\documentclass[referee]{raa}   
\usepackage{natbib}
\usepackage{enumerate}
\usepackage{float}
\usepackage{wrapfig}
\usepackage[section]{placeins}
\usepackage{enumitem}    
\usepackage{booktabs}
\usepackage{array}
\usepackage{threeparttable}

\usepackage{graphicx,times}
\usepackage{natbib}
\usepackage{amssymb,amsmath}
\bibpunct{(}{)}{;}{a}{}{,}

\usepackage{CJKutf8}

\usepackage[pagebackref=true]{hyperref}
\usepackage{booktabs}
\usepackage{adjustbox}
\usepackage{array}

\begin{document}
\begin{CJK*}{UTF8}{gbsn}

   \title{Chasing the formation history of the Galactic metal-poor disc}

 \volnopage{ {\bf 20XX} Vol.\ {\bf X} No. {\bf XX}, 000--000}
   \setcounter{page}{1}

   \author{Xiao-Kun Hou (侯晓琨)
   \inst{1,2}, Rui-Zhi Zhang\inst{1}, Hai-ning Li\inst{1}, Gang Zhao
      \inst{1,2}
   }

   \institute{ CAS Key Laboratory of Optical Astronomy, National Astronomical Observatories, Chinese Academy of Sciences, Beijing 100101, People's Republic of China;
   {\it gzhao@nao.cas.cn}\\
        \and
             School of Astronomy and Space Science, University of Chinese Academy of Sciences, Beijing 100049, People's Republic of China\\
\vs \no
   {\small Received 20XX Month Day; accepted 20XX Month Day}
}

\abstract{In our previous work, we identified $\sim100,000$ metal-poor stars ([Fe/H] $<$ -1.0) from the LAMOST Survey. This work estimates their chemical abundances and explores the origin and evolution of the Galactic metal-poor disc. Our chemo-dynamical analysis reveals four main populations within the metal-poor disc: (1) a primordial disc older than 12 Gyr with [Fe/H] $>$ -1.5; (2) debris stars from the progenitor galaxy of Gaia-Sausage-Enceladus (GSE), but now residing in the Galactic disc; (3) the metal-poor tail of the metal-rich, high-$\alpha$ disc formed 10-12 Gyr ago, with metallicity lower limit extending to -2.0; (4) the metal-poor tail of the metal-rich, low-$\alpha$ disc younger than 8 Gyr, reaching a lower metallicity limit of -1.8. These results reveal the presence of a primordial disc and show that both high-$\alpha$ and low-$\alpha$ discs reach lower metallicities than previously thought. Analysis of merger debris reveals that Wukong, with extremely low metallicity, likely originate from merger events distinct from GSE. Additionally, three new substructures are identified: ShangGu-1, characterized by unusual [Fe/H]-eccentricity correlations; ShangGu-2, possibly heated disc stars; and ShangGu-3, which can be divided into four subgroups based on differing orbital directions, with two aligning with the previously known Nyx and Nyx-2. 
}

   \authorrunning{X.-K. Hou et al. }            
   \titlerunning{The Galactic metal-poor disc}  
   \maketitle

%
\section{Introduction}           
According to galaxy formation models, the Milky Way grow through mergers with dwarf galaxies, living behind many merger remnants. Metal-poor stars ([Fe/H] $<$ -1.0), typically old and formed in chemically primitive environments, serve as excellent tracers of this process. However, their rarity makes their identification both challenging and essential. The LAMOST Survey \citep{zhao_stellar_2006,zhao_lamost_2012} has collected tens of millions of low-resolution stellar spectra, providing an unprecedented opportunity to search for metal-poor stars. \cite{li_catalog_2018} identified 10,000 very metal-poor stars ([Fe/H] $<$ -2.0) in LAMOST DR3, we further expanded this number to over 30,000 \citep{hou_very_2024} utilizing the more update DR9. During our search for very metal-poor stars, we also discovered approximately 100,000 metal-poor stars in total, most of them have reliable atmospheric parameters and velocities, making them ideal for investigating the assembly process of the Milky Way. Estimating elemental abundances is crucial, as chemical patterns reveal a star’s formation environment and help distinguish whether it formed in-situ or was accreted from dwarf galaxies. To achieve this, we use machine learning and template fitting to determine the abundances of various elements for these metal-poor stars.

The bimodal metallicity distribution between the Galactic halo and disc reflects their distinct formation histories: the metal-poor halo comprises primarily ancient stars formed during the early chaotic assembly phase, while the metal-rich disc ([Fe/H] $>$ -1.0) is dominated by younger stars shaped by post-merger secular evolution \citep{bland-hawthorn_galaxy_2016}. Recent discoveries of a substantial metal-poor disc population \citep[e.g.,][]{sestito_pristine_2020,bellazzini_metal-poor_2024,nepal_discovery_2024} challenge this view. Three principal hypotheses attempt to resolve this paradox. One hypothesis suggests that metal-poor disc stars are remnants of dwarf galaxies that merged with the Milky Way \citep{malhan_shiva_2024}. Another proposes that they are merely the metal-poor tail of the disc population, which has been overlooked due to their smaller numbers compared to the metal-rich disc stars \citep{naidu_evidence_2020}. A more novel perspective argues that these stars originate from the Milky Way's ancient primordial disc, distinct from the commonly recognized high-$\alpha$ and low-$\alpha$ discs \citep{xiang_formation_2024}. Observations of high-redshift galaxies also suggest that disc-like structures can emerge even in the early stages of galaxy evolution \citep[e.g.,][]{rowland_rebels_25_2024,baker_core_2024,fujimoto_primordial_2024}. However, some researchers argue that the concept of a primordial disc is unnecessary. For instance, \cite{dillamore_stellar_2023} proposed that resonances with the Galactic bar could lead to some halo stars acquiring disc-like orbits. Similarly, \cite{li_exploring_2024} suggested that bulge stars could also evolve into disc-like orbits through the influence of the decelerating Galactic bar.

Beyond the origin of the Galactic metal-poor disc, many substructures in the Galactic stellar halo trace past mergers that shaped the Milky Way. One of the earliest and most prominent examples is the Sagittarius dwarf galaxy, discovered by \cite{ibata_dwarf_1994}, which provided definitive evidence of an ongoing merger event. Subsequently, \cite{helmi_debris_1999} identified a kinematically coherent stream, suggesting the remnants of a disrupted dwarf galaxy. With the advent of Gaia data, more relics came to light: \cite{belokurov_co-formation_2018,helmi_merger_2018} pinpointed the Gaia-Sausage-Enceladus (GSE) merger as a major contributor to the inner halo. Numerous other substructures, such as Sequoia \citep{myeong_evidence_2019}, Thamnos \citep{koppelman_multiple_2019}, and Wukong \citep{naidu_evidence_2020}, have also been discovered. However, their precise origins often remain controversial. For instance, \cite{horta_chemical_2023} argued that Wukong might share a common progenitor with GSE, while other studies dispute that connection. These findings underscore the Milky Way's complex merger history and the need for more detailed chemical and kinematic data to distinguish in-situ from accreted populations, refine our understanding of known substructures, and discover any yet-undetected mergers.

The paper is structured as follows: Section \ref{sec2} describes the data and methods used in this study. Section \ref{sec3} presents the main results. Section \ref{sec4} analyzes these results in detail, discussing the formation process of the Galactic metal-poor disc, the possible origins of merger remnants, and properties of three newly discovered substructures, while comparing our findings with previous studies. Finally, Section \ref{sec5} summarizes the key discoveries.

\section{Data and Methods}\label{sec2}
In our previous work \citep{hou_very_2024}, we identified $\sim100,000$ metal-poor ([Fe/H] $<$ -1.0) stars from LAMOST low-resolution spectra, which are valuable for studying the Milky Way's formation history. In this study, we supplemented these stars with chemical abundances and kinematic properties to explore the formation process of the Galactic metal-poor disc. We estimated multi-element abundances using two methods: machine learning and template fitting. Machine learning provided abundances for 17 elements (C, Na, Mg, Si, Ca, Sc, Ti, Cr, Mn, Fe, Co, Ni, Zn, Sr, Ba, La and Eu) and three atmospheric parameters: the effective temperature ($T_\mathrm{eff}$), surface gravity ($\rm{log}\,\textsl{g}$), and microturbulence velocity ($v_{mic}$), while template fitting estimated abundances for four elements (Fe, Mg, Ca and C). Using a subset of these stars, we explored some key questions about the Milky Way. This section begins by detailing the process of estimating abundances for all metal-poor stars, followed by the methods used to calculate their kinematic parameters. Next, we describe the criteria for selecting sample stars for further study. Finally, we outline the classification of these stars into the kinematically cold disc, kinematically hot disc, and various well-known merger remnants.

\subsection{Estimating abundances}\label{2.1}
\subsubsection{Machine learning}\label{sec2.1.1}
We developed a neural network to predict chemical abundances and atmospheric parameters. The model is based on the Bayesian convolutional neural network framework from astroNN \citep{leung_deep_2019}. To account for the influence of model structure on predicted values, this framework incorporates a dropout mechanism during prediction, which randomly deactivates some nodes. The loss function used in the model is defined as:
\begin{equation}
Loss_i =  
\begin{cases}   
\frac{1}{2} (\hat{y}_i - y_i)^2 e^{-s_i} + \frac{1}{2} s_i & \text{for } y_i \neq \text{Nan} \\
0 & \text{for } y_i = \text{Nan}  
\end{cases}  
\end{equation}
Here $y_i$ and $\hat{y_i}$ are the label and predicted value for $i$th reference star, respectively, and $s_{i}=ln[\sigma_{known,i}^{2}+\sigma_{predicted,i}^{2}]$, where $\sigma_{known,i}^{2}$ and $\sigma_{predicted,i}^{2}$ are the uncertainties of the label and prediction. This loss function serves two purposes: it handles missing labels during training by ignoring them and incorporates both label and prediction uncertainties, enabling the model to output predictions along with their associated errors. The model input comprises normalized LAMOST low-resolution spectra within the wavelength range of 3900 to 5500 Å. This range was chosen because it includes key spectral features such as the Ca II H and K lines (∼3933 Å and ∼3968 Å), the CH G-band (∼4300 Å), and the Mg I triplet (∼5167–5184 Å). Notably, this differs from the wavelength range used in our previous work \citep{hou_very_2024}.

The reference set for the model includes 385 stars from \cite{li_four-hundred_2022} and 293 stars from the overlap between JINAbase \citep{abohalima_jinabasedatabase_2018} and LAMOST. The stars from \cite{li_four-hundred_2022} are metal-poor, initially identified using LAMOST low-resolution spectra and later observed with high-resolution spectroscopy for precise chemical abundances. JINAbase is a database compiling metal-poor stars and their chemical abundances from various studies. We incorporated metal-poor stars from these two distinct sources into our reference set to ensure that the metallicity coverage closely aligns with that of our target stars. Among the target stars, approximately 23.3\% have metallicities [Fe/H] $>$ -1.5. However, within the 385 stars from \cite{li_four-hundred_2022}, only 2 stars meet this criterion, representing just about 0.5\% of the sample. In contrast, among the 293 overlapping stars from JINAbase and LAMOST, 57 stars (approximately 19.5\%) have [Fe/H] $>$ -1.5. Moreover, the highest metallicity recorded in \cite{li_four-hundred_2022} is -0.63, which still does not sufficiently cover the metallicity range of the target stars. These statistics highlight the necessity of supplementing \cite{li_four-hundred_2022} with stars from the JINAbase–LAMOST overlap to achieve a more comprehensive coverage. To further illustrate this point, we included histograms (Fig. \ref{figA1}) showing the $T_{\rm{eff}}$, $\rm{log}\,\textsl{g}$ and metallicity distributions of the target sample, as well as those of \cite{li_four-hundred_2022}, the JINAbase–LAMOST overlap, and their combined reference set. While the overlapping stars from JINAbase and LAMOST span a broader metallicity range, their abundances and atmospheric parameters are drawn from diverse studies, which introduces some inconsistency. Therefore, we integrated these stars with those from \cite{li_four-hundred_2022} to build a reference set that balances both breadth of metallicity coverage and data uniformity.

The comparison of common stars between \cite{li_four-hundred_2022} and JINAbase, shown in Fig. \ref{figA2}, revealed systematic differences in $\rm{log}\,\textsl{g}$ and the abundances of Na, Mg, C, Si, Sc, Ti, Sr, and Ba, as their values deviate from a strict 1:1 relation. For each parameter ($\rm{log}\,\textsl{g}$ and the abundance of each element), we fitted a linear relationship between values from JINAbase and \cite{li_four-hundred_2022}:
\begin{equation}
    y = a \times x + b
\end{equation}
Here, $x$ represents the JINAbase value for a given parameter ($\rm{log}\,\textsl{g}$ or [X/H], with X representing Na, Mg, C, Si, Sc, Ti, Sr, or Ba), and $y$ corresponds to the respective value from \cite{li_four-hundred_2022}. The red dotted lines in Fig. \ref{figA2} show these linear fits, and the corresponding parameters $a$ and $b$ for each relationship are provided in Table \ref{tabA1}. These linear functions were then used to calibrate JINAbase values to the scale of \cite{li_four-hundred_2022}, which provides a more consistent dataset with uniformly analyzed abundances, while $\rm{log}\,\textsl{g}$ is derived from precise Gaia parallaxes. After this calibration, the reference stars were then divided into a training set and a test set in a 9:1 ratio. The model was trained on the training set, and its reliability was evaluated by comparing predicted values with labels in both sets, as shown in Fig. \ref{figA3}. To assess consistency, we calculated the mean and standard deviation of the differences between predicted values and labels. Overall, the predictions show good agreement with the labels for all atmospheric parameters and chemical abundances. We further analyzed the relationship between the absolute difference between predicted values and labels and the prediction errors, as shown in Fig. \ref{figA4}. The results indicate a positive correlation for most parameters and abundances, with larger prediction errors aligning with greater discrepancies between predicted and label values.

Beyond comparing predictions with labels within the reference set, we also performed an external comparison. \cite{li_stellar_2022} used a machine learning approach to estimate atmospheric parameters and chemical abundances for a large subset of stars from LAMOST DR8. By cross-matching our catalogue with theirs, we identified $\sim20,000$ common stars. Fig. \ref{figA5} compares our estimations of $T_\mathrm{eff}$, $\rm{log}\,\textsl{g}$, [Fe/H], and the abundances of C, Mg, Si, Ca, Mn, and Ni with their corresponding values. The results demonstrate strong consistency for $T_\mathrm{eff}$ and $\rm{log}\,\textsl{g}$. For [Fe/H] and the abundances of C, Mg, Ca, Mn, and Ni, our values are systematically lower but still exhibit overall agreement. These systematic differences likely arise from the use of different reference stars, as \cite{li_stellar_2022} adopted stars common to LAMOST and APOGEE \citep{majewski_apache_2017} as their reference set. Overall, these findings indicate the robustness of our estimations.

During the spectral normalization process, we employed polynomial fitting to derive a pseudo-continuum and obtained the normalized spectrum by dividing the observed spectrum by this pseudo-continuum. However, our analysis identified several poor-quality spectra, including cases with lost flux or significant cosmic-ray contamination, which adversely affected the quality of the normalized spectra. To evaluate the normalization quality, we calculated a new pseudo-continuum from the normalized spectrum. If this pseudo-continuum was smooth and had a mean value close to 1.0, the normalized spectrum was considered high quality and assigned a flag of 0; otherwise, it was flagged as 1. Finally, we applied the model to previously identified metal-poor stars, and the predicted abundances along with their associated uncertainties are compiled into a catalogue, available via the link provided at the end of the article.

\subsubsection{Template fitting}\label{sec2.1.2}
Although the neural network model performed well on the reference set, we employed an independent approach to further verify our measurements. Specifically, we used template fitting to re-estimate the abundances of Fe, Mg, Ca, and C, chosen for their strong absorption lines in LAMOST low-resolution spectra. The wavelength ranges used for fitting the abundance of each element are listed in Table \ref{tabA2}. For other elements, factors such as low abundances, weak absorption features, severe line blending and noise make template fitting impractical.

To identify the best-fitting template for each observed spectrum, we applied a Bayesian method. In this approach, the abundance of Fe, Mg, Ca, or C is assumed to follow a Gaussian prior distribution:
\begin{equation}
    P(\theta) = \frac{1}{\sqrt{2\pi\sigma_\theta^2}} \exp\left(-\frac{(\theta - \mu_\theta)^2}{2\sigma_\theta^2}\right)
\end{equation}
where $\theta$ represents the abundance, and $\mu_{\theta}$ and $\sigma_{\theta}$ are the mean and standard deviation derived from the neural network model's predictions and uncertainties. Next, we constructed a likelihood function to quantify the agreement between the template and observed spectrum:
\begin{equation}
    P(S_{\text{obs}} | \theta) = \exp\left(-\frac{1}{2} \chi^2(\theta)\right), \quad   
\end{equation}
where $\chi^2(\theta)$ is calculated as:
\begin{equation}
    \chi^2(\theta) = \sum_i \frac{\left(S_{\text{obs},i} - S_{\text{template},i}(\theta)\right)^2}{\sigma_i^2}  
\end{equation}
Here, $S_{\text{obs},i}$ and $S_{\text{template},i}$ are the normalized fluxes of the observed spectrum and template at the $i$th pixel, and $\sigma_{i}$ is the standard deviation of the observed flux. Smaller $\chi^2(\theta)$ values indicate better agreement between the template and observed spectrum, leading to a higher likelihood. Finally, the posterior distribution is obtained by combining the prior distribution and the likelihood function:
\begin{equation}
    P(\theta | S_{\text{obs}}) =  \frac{1}{\sqrt{2\pi\sigma_\theta^2}} \exp\left(-\frac{(\theta - \mu_0)^2}{2\sigma_\theta^2}\right) \cdot \exp\left(-\frac{1}{2} \chi^2(\theta)\right)
\end{equation}
The process of generating template involve several steps. First, we use KMOD, a publicly available IDL-based code \footnote{http://leda.as.utexas.edu/stools} widely adopted in lot of investigations \citep[e.g.,][]{meszaros_interpolation_2013,yong_most_2013}, to interpolate the ATLAS9 atmospheric model grid \citep{castelli_new_2004}. This interpolation is performed to obtain the target atmospheric model corresponding to the desired $T_\mathrm{eff}$, $\rm{log}\,\textsl{g}$, and [Fe/H]. Once the atmospheric model is established, we employ the SPECTRUM \citep{gray_calibration_1994} program to generate template based on the sampled chemical abundances.

Abundances were estimated as the mean value of samples drawn from the posterior distribution. Their errors arise from two sources: (1) posterior distribution uncertainty, defined as half the difference between the 16th and 84th quantiles of the posterior samples, and (2) the propagated uncertainties from the atmospheric parameters $T_{\mathrm{eff}}$, $\log g$, and [Fe/H]. Since our abundance measurements are based on template fitting, where the stellar parameters are fixed and the chemical abundance of the template is varied to match the observed spectrum, any uncertainty in the stellar parameters can directly affect the derived abundances. To account for this, we varied each parameter by its $\pm 1 \sigma$ uncertainty and re-estimated the abundance. The difference in abundance caused by this variation was taken as the contribution to the abundance error from that parameter. This process was performed separately for $T_{\mathrm{eff}}$, $\log g$, and [Fe/H]. The final abundance uncertainty was calculated as the square root of the sum of the squares of the posterior uncertainty and the individual contributions from the three parameters. For iron, only the posterior uncertainty and the contributions from $T_{\mathrm{eff}}$ and $\log g$ were considered.

We compared the Fe, Mg, Ca, and C abundances estimated from template fitting and machine learning, as shown in Fig. \ref{figA6}, \ref{figA7}, \ref{figA8}, and \ref{figA9}. Each figure includes two panels: the left shows stars with $\rm{flag}=0$, while the right shows stars after error screening. For the error screening, we required that the errors from both machine learning and template fitting must not exceed 0.2 for Fe, Mg, and Ca, while a less stringent threshold of 0.3 was applied for C. After error filtering, the absolute mean differences between two methods for Fe, Mg, and Ca are 0.01, 0.03, and 0.1, respectively, with standard deviations of 0.21, 0.21, and 0.24, indicating excellent consistency. However, significant discrepancies were observed for C, with template fitting yielding higher abundances. We attribute this discrepancy to the neural network’s reference set, which included labels from both JINAbase and \cite{li_stellar_2022}. As described in Section \ref{sec2.1.1}, [C/H] from these two sources do not follow a strict 1:1 relation. To address this, we corrected JINAbase values to the scale of \cite{li_stellar_2022} using a linear function derived from stars common to both sources. However, due to the small number of the common stars, this calibration may be unreliable. To assess the accuracy of [Fe/H] and the abundances of C, Mg, and Ca obtained by template fitting, we compared our estimations to those from \cite{li_stellar_2022}, as shown in Fig. \ref{figA10}. The comparison shows good agreement but our values are systematically lower than those reported by \cite{li_stellar_2022}. This offset aligns with the findings in Section \ref{sec2.1.1}, where we compared abundances obtained by machine learning to those from \cite{li_stellar_2022}. We estimated the abundances of Fe, Mg, Ca, and C for all spectra with $\rm{flag}=0$ using template fitting and compiled the results together with those from machine learning in a single catalogue.

\subsection{Computing Phase-space Quantities}\label{sec2.2}
The Integrals of Motion (IoMs) of stars from the same stellar system would remain coherent as the Galaxy evolved, which can be used to identify the Galactic substructures. To compute the positions, velocities, IoMs, and orbital parameters for further classification, we first selected stars with precise astrometric measurements in our sample.

We cross-matched our metal-poor stars with \textit{Gaia} DR3 catalogues \citep{Gaia2023} and the \texttt{StarHorse} photo-astrometric distance catalogues \citep{Anders2022}. 
Stars with relative distance uncertainties of \citet{Anders2022} greater than 20\% or \textit{Gaia} RUWE greater than 1.4 are removed since their astrometric measurements are unreliable \citep{Lindegren2021}. 
For stars with multi-observation results in LAMOST, we only keep the record with the highest signal-to-noise ratio.

We computed positions and velocities with coordinates and proper motions from \textit{Gaia} DR3, photo-astrometric distance from \citet{Anders2022}, and radial velocities from LAMOST DR10 or \textit{Gaia} DR3.
we prioritized the radial velocity measurement from LAMOST DR10; otherwise, we adopted the value from Gaia DR3.
We assumed that the sun is located 8.21 kpc from the Galactic centre \citep{McMillan2017} and 20.8 pc above the Galactic plane \citep{Bennett19}. We adopted the motion of the Local Standard of Rest (LSR) as $V_{\mathrm{LSR}}=233.1\;\mathrm{km\;s^{-1}}$ \citep{McMillan2017} and the solar peculiar motion as $(U_{\bigodot},V_{\bigodot},W_{\bigodot})=(11.1, 12.24, 7.25)\;\mathrm{km\;s^{-1}}$ \citep{Schonrich2010}.

We further computed IoMs (e.g., energy, actions) and orbital parameters (e.g., eccentricity, maximum height to the Galactic plane) using the software \texttt{Agama} \citep{Agama} with the axisymmetric potential model of the Milky Way \citep{McMillan2017}. 
We assumed a Gaussian uncertainties for the observables and ran a 10,000 times Monte Carlo (MC) simulation for each star to estimate the uncertainties of the dynamical parameters.
The value and the uncertainty of each quantity are represented by the median value and half the difference between the 16th and 84th quantiles.

In total, we obtained phase-space quantities for $\sim74,000$ stars.

\subsection{Stars for studying the Milky Way}\label{sec2.3}
\begin{figure}
   \centering
    \includegraphics[width=0.6\textwidth]{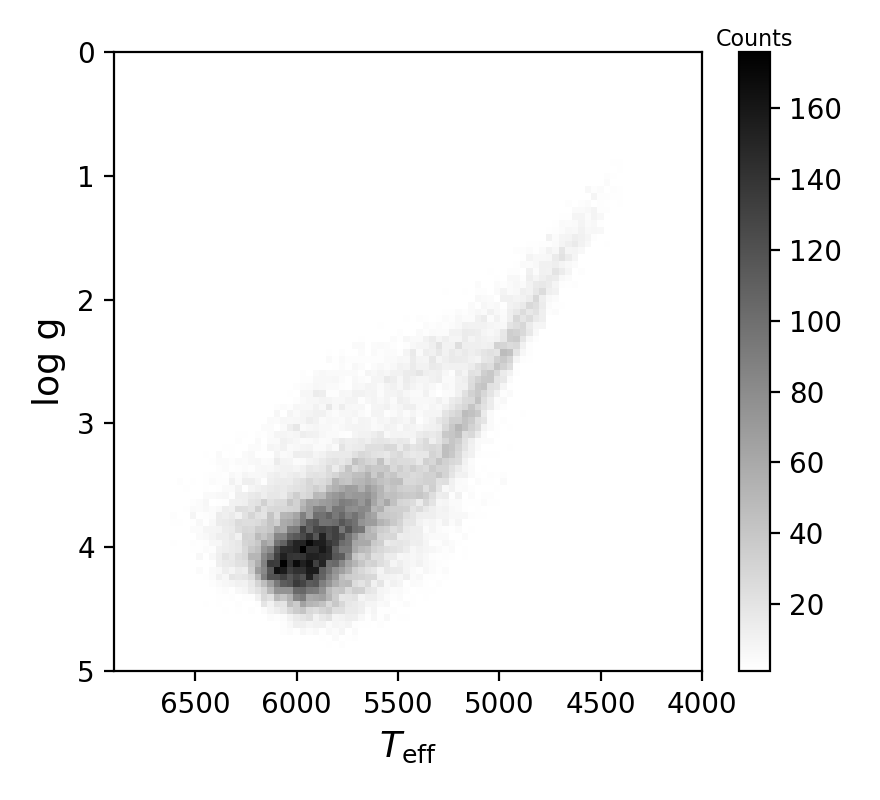}
    \caption{Kiel diagram of the selected stars for studying the Milky Way, with grayscale indicating the star count density.}
    \label{fig1}
\end{figure}

To study the Milky Way, we selected 46,178 stars from the $\sim74,000$ stars with phase-space quantities computed in Section \ref{sec2.2}, based on the precision of their kinematic parameters:
\begin{enumerate}[label=(\roman*), leftmargin=0em, align=left]  
\setlength{\itemindent}{1em} 
    \item $J_{r} > 0\, \rm{km\, s^{-1}}$
    \item $J_{z} > 0\, \rm{km\, s^{-1}}$
    \item $J_{\phi,\rm{err}} < 180\, \rm{km\, s^{-1}}$
    \item $E_{\rm{err}}/E<0.3$
    \item $J_{r,\rm{err}}/J_{r}<0.3$
    \item $J_{z,\rm{err}}/J_{z}<0.3$ 
\end{enumerate}  
Here, $E$ represents orbital energy, while $J_{\phi}$, $J_{r}$, and $J_{z}$ are the azimuthal, radial, and vertical actions, with $E_{\rm{err}}$, $J_{\phi,\rm{err}}$, $J_{r,\rm{err}}$, and $J_{z,\rm{err}}$ as their respective uncertainties. As shown in Fig. \ref{fig1}, our sample spans diverse evolutionary stages, with the majority of stars concentrated at the main-sequence turnoff, and others distributed across the subgiant, giant, and asymptotic giant branch (AGB) phases. Thanks to the exceptionally high observational efficiency of the LAMOST survey, we have assembled such a remarkably large sample of metal-poor stars, enabling detailed studies of the Galactic evolution, particularly regarding the Galactic metal-poor disc. Previous investigations into the Galactic metal-poor disc have predominantly relied on photometric surveys such as Gaia and Pristine \citep[e.g.,][]{bellazzini_metal-poor_2024,sestito_pristine_2020,fernandez-alvar_pristine_2021,hong_candidate_2024}. Although these photometric surveys offer metallicity measurements for hundreds of millions of stars, their chemical abundance information typically remains limited to iron. Consequently, researchers often supplement these datasets with high-resolution spectroscopic surveys like GALAH and APOGEE \citep[e.g.][]{malhan_shiva_2024,fiorentin_icarus_2021,fiorentin_icarus_2024,carter_ancient_2021,fernandez-alvar_metal-poor_2024}. However, integrating photometric and spectroscopic data substantially reduces the available sample size. Leveraging LAMOST’s extensive dataset, we can maintain a large stellar sample while simultaneously exploring multi-element abundance patterns, thus providing deeper insights into the composition and kinematics of the Galactic metal-poor disc.

For our analysis, we primarily use the abundances of Fe, Mg, and Ca obtained by machine learning, applying strict criteria to ensure the reliability: both the machine learning and template fitting uncertainties must be below 0.2, and the absolute difference between predicted values from two methods must not exceed 0.2. When these stringent criteria result in insufficient stars for study, we relax the standards to only require machine learning uncertainties below 0.2, explicitly noting when this applies. Additionally, we include the abundances of Sc, which are measured only through machine learning, with uncertainty thresholds of 0.15.

Additionally, we incorporated stellar ages from \cite{xiang_time-resolved_2022}, who estimated the ages of over 200,000 LAMOST subgiant stars using the isochrone fitting method. To ensure the subgiant phase, we selected stars with $T_{\rm{eff}}$ and $\rm{log}\,\textsl{g}$ errors below 100 K and 0.2, respectively. We further required ages less than 15 Gyr, age uncertainties within 1 Gyr, and relative age uncertainties under 0.2. The filtered stars were cross-matched with our sample, integrating their age data for further analysis. 

In conclusion, our sample contains 46,178 stars with chemical abundances, spatial positions, orbital parameters, and integrals of motion, including 2517 subgiant stars with precise age estimates.

\subsection{Classification of sample stars}\label{sec2.4}
Based on their kinematic properties, we classified the 46,178 sample stars into two main categories: the kinematically cold disc and kinematically hot disc, as well as six well-known merger remnants: Gaia-Sausage-Enceladus (GSE), Pontus, Thamnos, Sequoia, Helmi Stream, and Wukong. The detailed classification criteria are provided in Table \ref{tab1}. However, since the Galactic discs and remnants lack clear kinematic boundaries, the initial classification may result in mutual contamination, potentially affecting subsequent analysis. To mitigate this, we applied the HDBSCAN algorithm \citep{campello_hierarchical_2015} to cluster the stars in a space defined by energy ($E$) and actions ($J_{\phi}$, $J_{r}$, and $J_{z}$). This approach identified 6,549 stars grouped into 222 Dynamically Tagged Groups (DTGs), following the naming convention of \cite{yuan_dynamical_2020}. HDBSCAN is well-suited for identifying groups in data with varying densities, and it does not require a pre-defined number of clusters. The reason why not all stars in our sample were clustered is that HDBSCAN is designed to find dense regions of data while discarding sparse regions as noise. This feature ensures that only stars with sufficient density are grouped into meaningful clusters, while stars that are too isolated or do not form a significant structure are excluded. For our analysis, we carefully selected the parameters for HDBSCAN to ensure that the clusters formed were both robust and meaningful. Specifically, we set the \textbf{min\_cluster\_size} to 15, meaning that a group had to consist of at least 15 stars to be considered a valid cluster. We also used a \textbf{min\_samples} value of 6, which defines the number of neighboring points required to form a dense region. Additionally, the \textbf{metric} was set to 'euclidean' to measure the similarity between stars in a straightforward manner, and the \textbf{cluster\_selection\_method} was set to 'leaf', which ensures that the final clusters are chosen based on density and not by a hierarchical tree structure. These parameter choices were carefully optimized and tested to capture the true dynamical structures of the stellar populations, while minimizing the impact of statistical fluctuations and noise. Each DTG was assigned to one of the previously mentioned structures or remnants based on the mean kinematic parameters of the stars within the group. Specifically, the mean kinematic values of the member stars were used as the representative parameters for the entire DTG, ensuring that the classification reflected the overall dynamical behavior of the DTG. Based on this approach, 194 DTGs were successfully associated with known structures, while 28 DTGs did not align with the kinematic criteria of any known structure or remnant listed in Table \ref{tab1}. These 28 DTGs were further analyzed and grouped into six distinct categories. The criteria for this grouping, along with a detailed analysis of these six groups, are provided in Section \ref{sec3.3}. Due to the large number of DTGs, the kinematic parameters of all stars, along with a summary table of the DTG properties, have been made available through a link at the end of the article.
\begin{table}
    \centering
    \caption{Classification criteria for dividing sample stars and DTGs into the kinematically cold disc, kinematically hot disc, and six merger remnants.}
    \setlength{\tabcolsep}{2mm}{
    \begin{tabular}{lc}
     \toprule
        Name & Criteria \\
    \midrule
        Kinematically cold disc & $J_{\phi}>1.66\,(\times10^{3}\,\rm{kpc\,km\,s^{-1}})$; $e<0.2$; $|V_{z}|<15\,(\rm{km\,s^{-1}})$\\
        Kinematically hot disc & $J_{\phi}<1.66\,(\times10^{3}\,\rm{kpc\,km\,s^{-1}})$; $0.2<e<0.6$; $|V_{z}|<25\,(\rm{km\,s^{-1}})$;\\     &$-200<V_{\phi}<-90\,(\rm{km\,s^{-1}})$;\\
        GSE & $e>0.7$; $|V_{\phi}|<100\,(\rm{km\,s^{-1}})$\\
        Pontus & $-1.72<E<-1.56\,(\times10^{5}\,\rm{km^{2}\,s^{-2}})$; $-470<J_{\phi}<5\,(\rm{kpc\,km\,s^{-1}})$;\\  &$0.245<J_{r}<0.725\,(\times10^{3}\,\rm{kpc\,km\,s^{-1}})$;\\ 
            &$0.115<J_{z}<0.545\,(\times10^{3}\,\rm{kpc\,km\,s^{-1}})$; $0.39<L_{\bot}<0.865\,(\times10^{3}\,\rm{kpc\,km\,s^{-1}})$;\\
            &$0.5<e<0.8$; $8<r_{\rm{apo}}<13\,(\rm{kpc})$; $1<r_{\rm{peri}}<3\,(\rm{kpc})$; \\
        Thamnos & $-1.8<E<-1.6\,(\times10^{5}\,\rm{km^{2}\,s^{-2}})$; $e<0.7$; $J_{\phi}<0\,(\rm{kpc\,km\,s^{-1}})$\\
        Sequoia & $E>-1.6\,(\times10^{5}\,\rm{km^{2}\,s^{-2}})$; $J_{\phi}<-0.7\,(\times10^{3}\,\rm{kpc\,km\,s^{-1}})$; $\eta<-0.15$\\
        Helmi Streams & $0.75<J_{\phi}<1.7\,(\times10^{3}\,\rm{kpc\,km\,s^{-1}})$; $1.6<L_{\bot}<3.2\,(\times10^{3}\,\rm{kpc\,km\,s^{-1}})$\\
        Wukong & $0<J_{\phi}<1\,(\times10^{3}\,\rm{kpc\,km\,s^{-1}})$; $E<-1.15\,(\times10^{5}\,\rm{km^{2}\,s^{-2}})$;\\ &$(J_{z}-J_{r})/J_{\rm{tot}}>0.3$; $90^{\rm{o}}<\rm{arccos}$$(L_{z}/L)<120^{\rm{o}}$\\
    \bottomrule
    \end{tabular}}
    \begin{tablenotes}
        \item {\textbf{Note.} $E$ denotes orbital energy. $J_{\phi}$, $J_{r}$, and $J_{z}$ are the azimuthal, radial, and vertical components of action ($\textbf{J}$), with $J_{\rm{tot}}$ being their quadrature sum. $V_{\phi}$ and $V_{z}$ represent azimuthal and vertical velocities. $e$ and $\eta$ stand for eccentricity and circularity, while $r_{\rm{apo}}$ and $r_{\rm{peri}}$ are the apogalaction and perigalaction radii. $L_{z}$ is the azimuthal angular momentum component and $L_{\bot}$ ($L_{\bot}=\sqrt{L_{x}^{2}+L_{y}^{2}}$)is the perpendicular component.}
    \end{tablenotes}
    \label{tab1}
\end{table}

\cite{yoshii_density_1982} and \cite{gilmore_new_1983} found that the spatial distribution of stars in the Galactic disc can be well fitted by two exponential components. One component exhibits a larger scale height, while the other has a smaller scale height, leading to the morphological classification of the disc into 'thick disc' and 'thin disc'. With the advancement of spectroscopic surveys, the chemical abundances of an increasing number of stars have been measured. \cite{bovy_spatial_2012} used data from the Sloan Digital Sky Survey (SDSS), proposed that the Galactic disc can also be divided into two components based on $\alpha$ abundance: high-$\alpha$ disc and low-$\alpha$ disc. Stars in the high-$\alpha$ disc are generally older and more metal-poor, while those in the low-$\alpha$ disc tend to be younger and more metal-rich. Furthermore, \cite{bovy_spatial_2012} and \cite{hayden_chemical_2015} found significant geometric differences between the high-$\alpha$ and low-$\alpha$ discs. High-$\alpha$ disc stars are predominantly located within $\sim$10 kpc from the Galactic center and tend to lie at vertical distances greater than 0.5 kpc from the midplane. In contrast, low-$\alpha$ disc stars remain prominent out to $\sim$15 kpc and are mainly confined within 0.5 kpc of the midplane. \cite{bensby_exploring_2014} geometrically classified stars into the thin and thick discs and found that most thick disc stars belong to the high-$\alpha$ disc, although the correspondence is not perfect. In summary, the Galactic disc can be classified not only based on geometry (thick and thin discs), but also in terms of chemical abundances (high-$\alpha$ and low-$\alpha$ discs). Most studies of the Galactic disc have focused on stars with metallicities $[\mathrm{Fe}/\mathrm{H}] > -1$, where the disc populations are most prominent. As noted earlier, the high-$\alpha$ disc is generally associated with the geometric thick disc, while the low-$\alpha$ disc corresponds roughly to the geometric thin disc. This approximate mapping has led to the commonly held view that the Milky Way disc is composed of two distinct stellar populations, both chemically and kinematically. However, the correspondence between chemical and geometric definitions is not perfect. Recent studies \citep[e.g.,][]{martig_red_2016,lian_unveiling_2025} have shown that some stars with geometric thick disc kinematics are not $\alpha$-enhanced and may even be relatively young. For example, at Galactocentric distances of 10–14 kpc, \cite{lian_unveiling_2025} identified stars with thick disc-like orbital properties but $[\alpha/\mathrm{Fe}] < 0.2$ and ages around 6.6 Gyr. These findings suggest that the geometric thick disc includes a chemically and temporally diverse stellar population that challenges the traditional paradigm.

This work primarily examines the properties of the Galactic disc at the metal-poor end, demonstrating that the situation becomes increasingly complex at lower metallicities ([Fe/H] $<$ –1.0). The kinematically hot and cold discs do not represent single stellar populations, but rather include multiple distinct components. Therefore, in this work, the kinematically defined hot and cold discs do not directly correspond to the traditional metal-rich, high-$\alpha$ and low-$\alpha$ disc populations, and they differ from the thick and thin discs typically observed at the metal-rich end of the Galactic disc. To better characterize these components in the metal-poor disc, we classify stars based on their kinematic properties and then examine the distributions of age and chemical abundances across different groups. Similarly, we classify all stars with very high eccentricities as belonging to the Gaia-Sausage-Enceladus (GSE). In this paper, GSE specifically refers to these high-eccentricity stars, rather than to the progenitor dwarf galaxy itself. It is important to note that these classifications are not intended to define physically distinct stellar populations, but rather serve as a practical framework to facilitate a clearer analysis of the structure and evolution of the Galactic metal-poor disc.

\section{Results}\label{sec3}
\begin{figure*}
   \centering
    \includegraphics[width=1.0\textwidth]{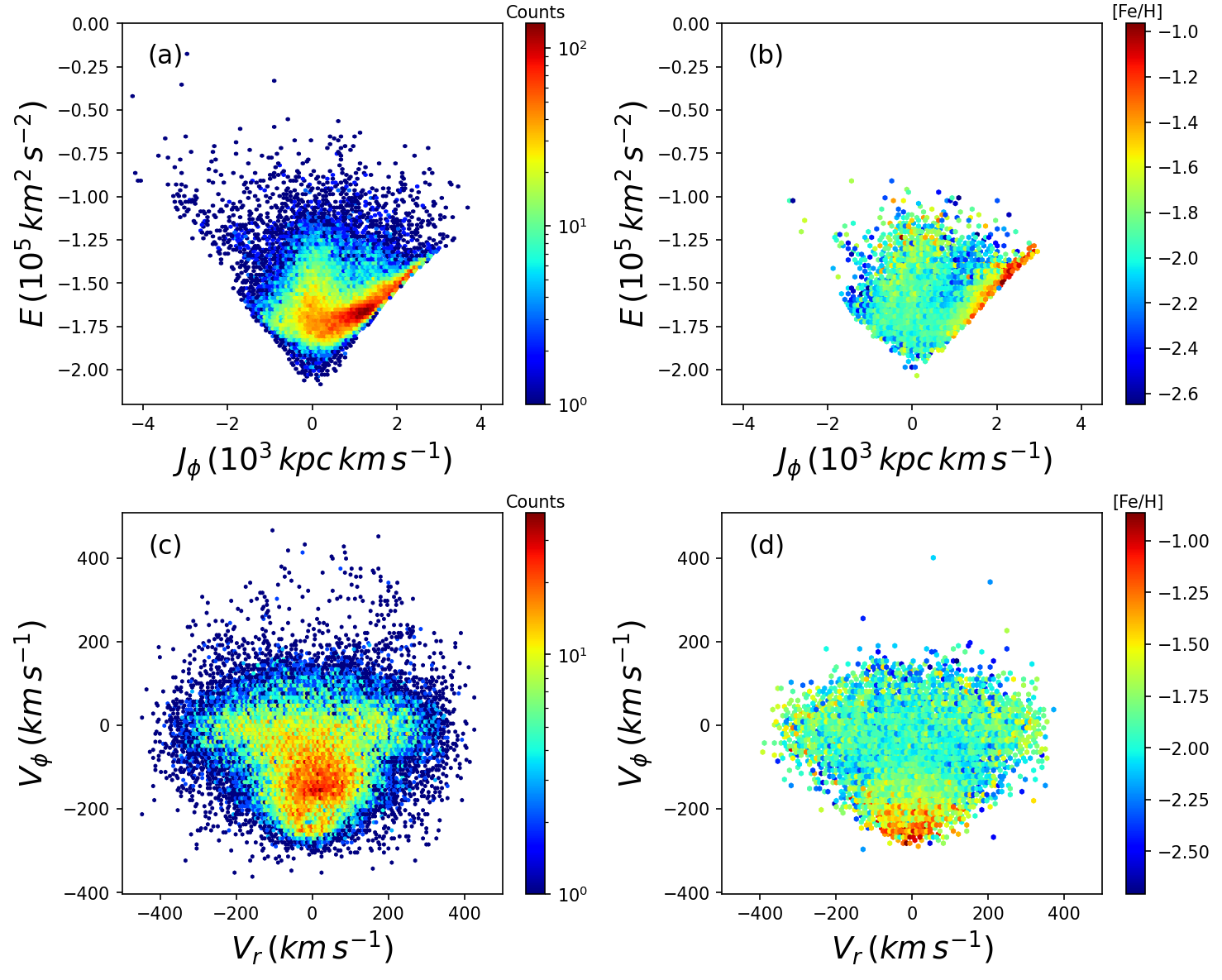}
    \caption{Panels (a) and (b) show the distributions of sample stars in the $E-J_{\phi}$ plane, and panels (c) and (d) in the $V_{\phi}-V_{r}$ plane. Colors in panels (a) and (c) represent stellar number density, while in panels (b) and (d), they indicate metallicity. The number density plots (a and c) include all 46,178 sample stars, whereas the metallicity plots include only stars with sufficient [Fe/H] precision. See Section \ref{sec2.3} for details.}
    \label{fig2}
\end{figure*}

\begin{figure*}
   \centering
    \includegraphics[width=1.0\textwidth]{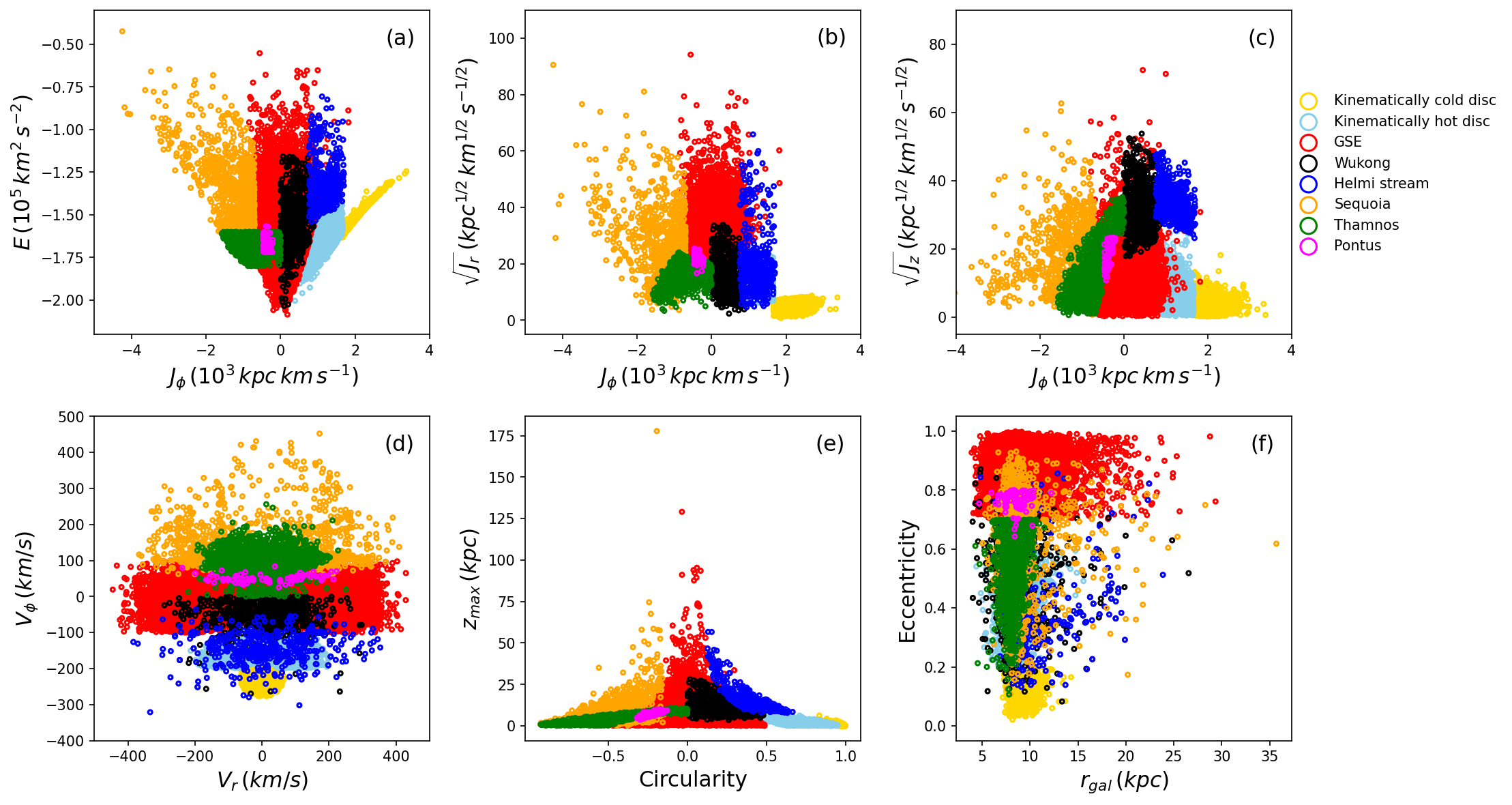}
    \caption{Kinematic space distributions of sample stars, with colors indicating different structures or remnants. See Section \ref{sec2.4} for classification details.}
    \label{fig3}
\end{figure*}

\begin{figure*}
   \centering
    \includegraphics[width=1.0\textwidth]{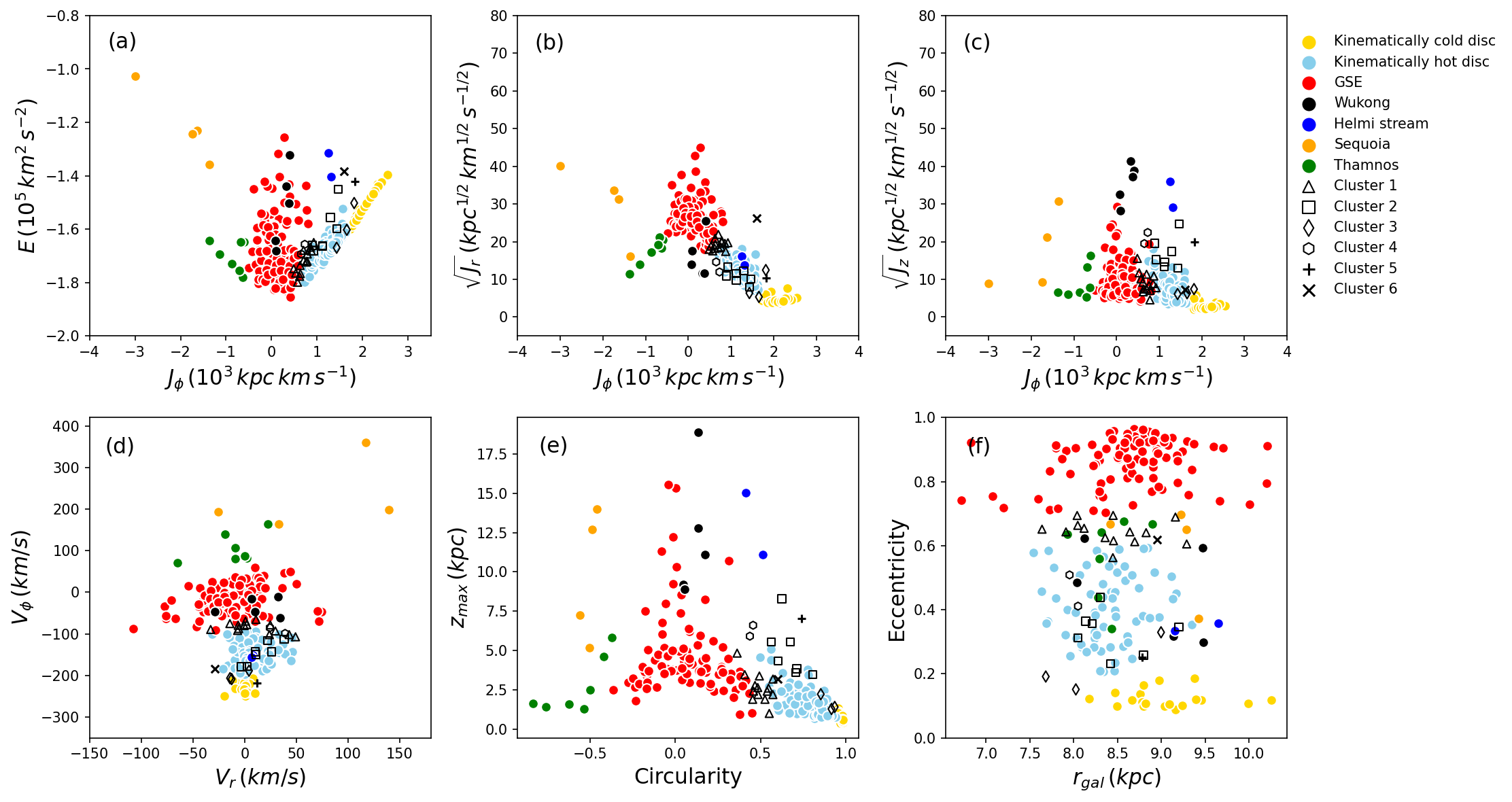}
    \caption{Similar to Fig. \ref{fig3}, but for the DTGs. Circles represent a successfully associated DTGs, while other shapes indicate those that do not align with known structures or remnants. The unclassified DTGS are divided into six distinct groups, distinguished by shapes such as squares and triangles.}
    \label{fig4}
\end{figure*}

We provide an overview of the chemical and kinematic properties of the sample stars. Fig. \ref{fig2} illustrates their distributions in the $E-J_{\phi}$ and $V_{\phi}-V_{r}$ planes. In panels (a) and (c), colors represent stellar number density, while in panels (b) and (d), they indicate metallicity. It is important to note that the stellar samples shown in the left and right panels differ slightly due to quality cuts. Panels (a) and (c) include all stars with reliable orbital parameters, as selected by the criteria described in Section \ref{sec2.3}. In contrast, panels (b) and (d) only include stars with high-quality [Fe/H] measurements: we require both the machine learning and template fitting estimates of [Fe/H] to have uncertainties smaller than 0.2, and the absolute difference between the two methods must also be less than 0.2. These stricter conditions reduce the number of stars shown in the metallicity panels. Panel (c) clearly highlights the GSE, characterized by nearly no net rotation ($V_{\phi}\approx0$), and a wide radial velocity range ($|V_{r}|$ from 0 to $400\, \rm{km\,s^{-1}}$). Kinematically hot disc is also evident with $V_{\phi}$ around $-180\, \rm{km\,s^{-1}}$ and $V_{r}$ near 0, while stars with $V_{\phi}$ under $-200\, \rm{km\,s^{-1}}$, correspond to kinematically cold disc. In panel (a), the number density shows a smooth, hook-like transition between GSE, kinematcally hot disc, and kinematically cold disc, with the highest density in kinematically disc region. Panels (b) and (d) reveal kinematically cold disc has the highest metallicity, generally above -1.25.
As discussed in Section \ref{sec2.4}, we classified these sample stars into the kinemtically cold disc, hot disc, and six dwarf galaxy debris. Fig. \ref{fig3} illustrate their distributions in kinematic space, while Fig. \ref{fig4} displays the distributions of DTGs. For 28 unclassified DTGs, we grouped them into six groups (Cluster 1 to 6), marking them with hollow triangles, squares, diamonds, hexagons, plus signs, and crosses.

Motivated by the unclear boundaries between GSE, the kinematically hot disc, and the cold disc in the $E-J_{\phi}$ plane, we investigated the origin of the Milky Way's metal-poor disc by comparing these three structures, as detailed in Section \ref{sec3.1}. And we analyzed the properties of GSE and other merger remnants in Section \ref{sec3.2}. Finally, Section \ref{sec3.3} details the DTGs that could not be associated.

\subsection{GSE vs Galactic discs}\label{sec3.1}
\begin{figure*}
   \centering
    \includegraphics[width=1.0\textwidth]{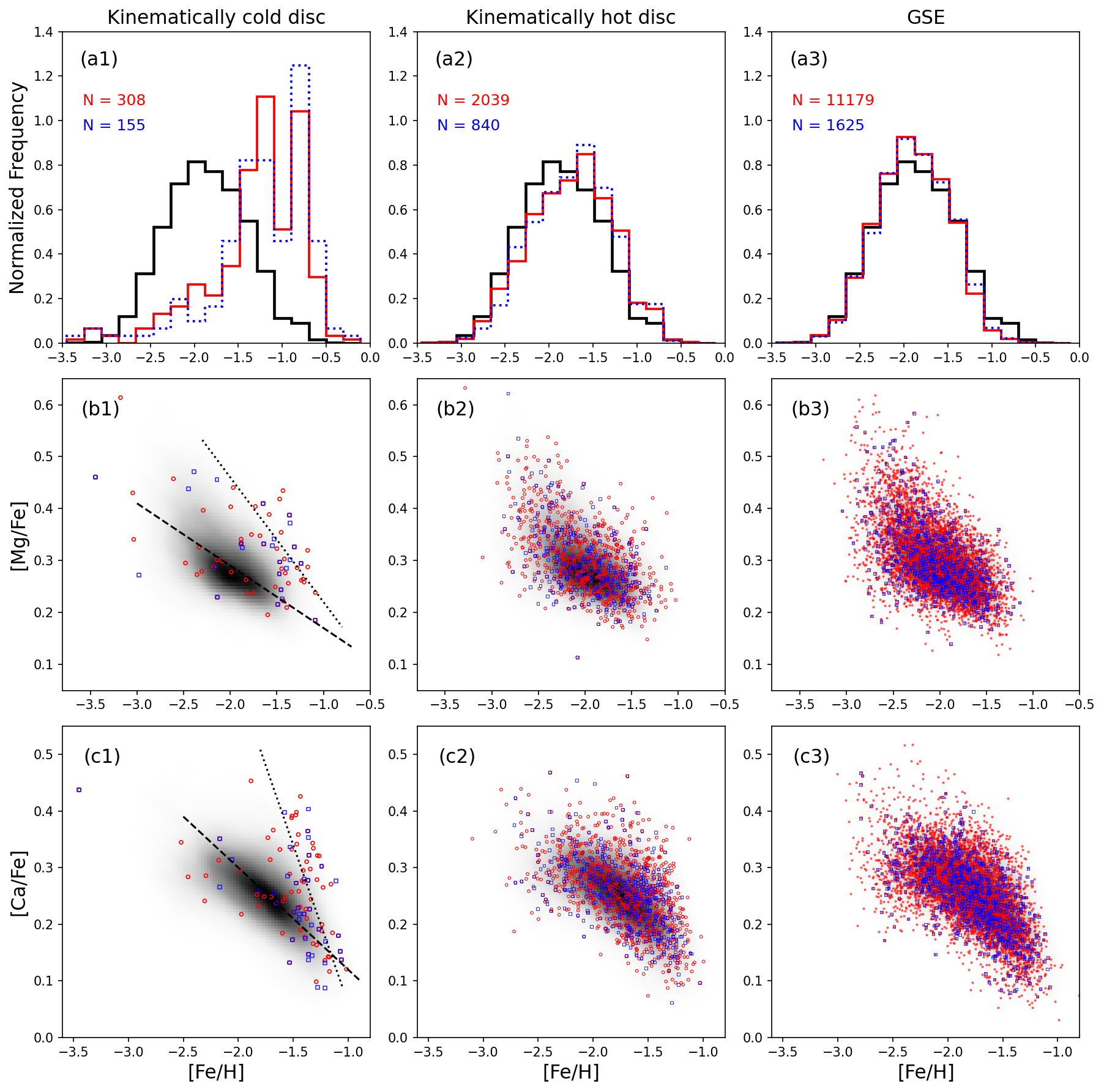}
    \caption{The first row shows the normalized metallicity histograms of the kinematcially cold disc, hot disc, and GSE, while the second and third rows display [Mg/Fe]-[Fe/H] and [Ca/Fe]-[Fe/H] distributions, respectively. Black represents the entire sample, red shows stars classified directly by their kinematic properties, and blue denotes stars identified through clustering in phase space and subsequently classified based on their collective kinematic features. Panels (b1) and (c1) include dashed and dotted lines to emphasize two distinct [$\alpha$/Fe]-[Fe/H] trends for the kinematically cold disc. In the histograms, red numbers indicate the count of stars selected directly, while blue numbers represent those selected indirectly.}
    \label{fig5}
\end{figure*}

First, we compare the normalized metallicity histograms of the kinematically cold disc, kinematically hot disc, and GSE, along with their distributions in the [Mg/Fe]-[Fe/H] and [Ca/Fe]-[Fe/H] planes as shown in Fig. \ref{fig5}. As stated in Section \ref{sec2.4}, we adopted two strategies to select the kinematically cold disc, kinematically hot disc, and GSE stars. The first classifies stars directly based on their kinematic properties, while the second identifies compact stellar groups in phase space and classifies them according to their collective kinematic features. In the comparison, samples from the first and second methods are shown in red and blue, respectively, while the entire sample is displayed in black as a background to highlight the differences among the kinematically cold disc, kinematically hot disc, and GSE. Regardless of the classification methods, the kinematically cold disc shows the highest metallicity, GSE the lowest, and the kinematically hot disc lies in between. Moving from the kinematically cold disc to the kinematcially hot disc and then to GSE, rotational velocity ($|V_{\phi}|$) decreasing from over 200 $\rm{km\,s^{-1}}$ to nearly 0, a trend mirrored by a progressive decline in metallicity. The kinematcially hot disc and GSE show unimodal metallicity histograms, peaking at approximately -1.5 and -2.0, respectively, while the kinematically cold disc exhibits three peaks at -0.8, -1.2, and -2.0, respectively. 

The [$\alpha$/Fe]-[Fe/H] relationship is a key indicator of the star formation and chemical enrichment history. We examines the variation of [Mg/Fe] and [Ca/Fe], two typical $\alpha$-elements, with metallicity. For both the entire sample and individual structures, [$\alpha$/Fe] generally decreases as [Fe/H] increases. Panels (b2), (b3), (c2), and (c3) in Fig. \ref{fig5} show that the kinematically hot disc and GSE share a single trend consistent with the total sample, while panels (b1) and (c1) reveal two distinct declining trends for the kinematically cold disc. One group of the kinematically cold disc stars align with the overall trend (dashed line), while another exhibits a steeper decline (dotted line), reflecting different formation histories. These results remain consistent even when using stars belonging to DTGs, confirming the robustness of the analysis.

\begin{figure*}
    \centering
    \includegraphics[width=1.0\textwidth]{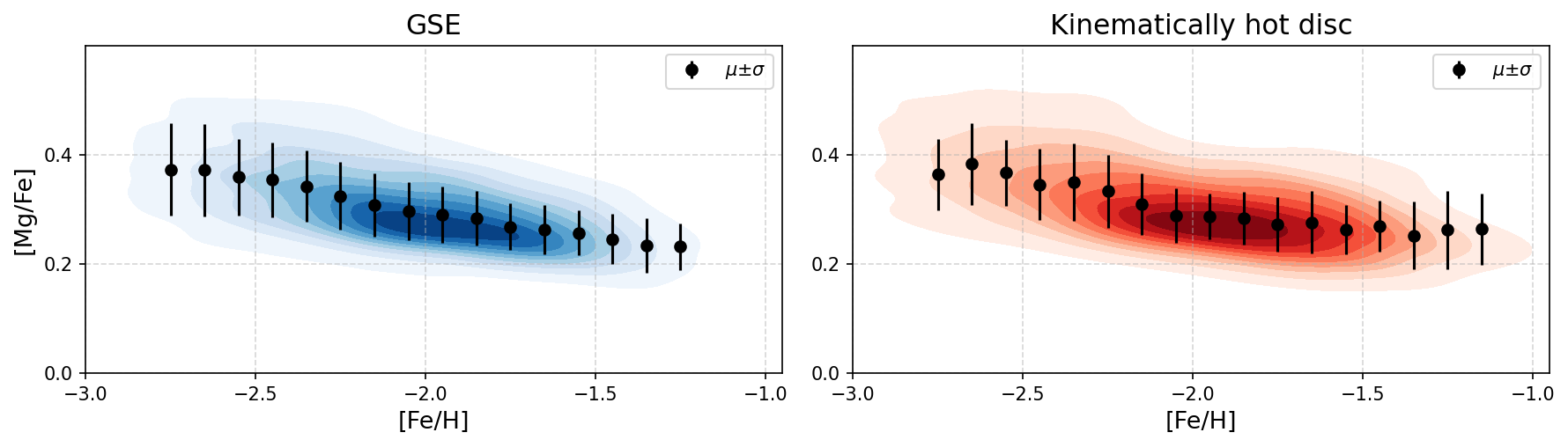}
    \caption{Two-dimensional density maps of the [Mg/Fe]–[Fe/H] distribution for GSE and the kinematically hot disc. The color shading represents stellar number density, with darker regions indicating higher densities. Stars are binned in 0.1 intervals in [Fe/H] from –2.8 to –1.2 for GSE, and from –2.8 to –1.1 for the kinematically hot disc. In each bin, the mean [Mg/Fe] and 1$\sigma$ dispersion are shown as black points with vertical error bars. The stellar samples used in both panels are identical to those shown in Fig. \ref{fig5}.}
    \label{fig6}
\end{figure*}

\cite{nissen_two_2010} analyzed high-resolution spectra of 94 stars, including 16 thick-disc stars, and identified two distinct sequences of halo stars in the [$\alpha$/Fe]–[Fe/H] plane: a high-$\alpha$ sequence and a low-$\alpha$ sequence. They found that the high-$\alpha$ halo stars occupy the same region as the thick-disc stars in [$\alpha$/Fe]-[Fe/H] plane. Based on this overlap, they proposed that the high-$\alpha$ sequence likely originates from the heated Galactic disc. In contrast, the low-$\alpha$ sequence was interpreted as having been accreted from external dwarf galaxies. It is important to note that all stars in their sample have [Fe/H] $>$ –1.6, and that the separation between the high-$\alpha$ and low-$\alpha$ sequences becomes increasingly pronounced at higher metallicities. \cite{horta_chemical_2023}, using APOGEE data, analyzed the [$\alpha$/Fe] distribution of GSE and identified its $\alpha$-knee at [Fe/H] $\approx$ –1.2. They reported that GSE stars have lower [$\alpha$/Fe] than the disc stars for [Fe/H] $>$ –1.2, corresponding to the “shin” region beyond the knee. However, their analysis does not clearly define the metal-poor disc, leaving it uncertain whether GSE and the disc are distinguishable in [$\alpha$/Fe] at lower metallicities. \cite{naidu_evidence_2020} introduced the concept of the Metal-Weak Thick disc (MWTD), which they defined as stars with [Fe/H] $<$ -0.8 and [$\alpha$/Fe] $>$ 0.25. In their study, many GSE stars also show [$\alpha$/Fe] values exceeding 0.25, as seen in their Figure 16. \cite{zhang_four-hundred_2024} compared GSE and MWTD in very metal-poor regime ([Fe/H] $<$ –2) and found no significant difference in [$\alpha$/Fe] between the two populations. Although none of these studies explicitly define a metallicity threshold where GSE and the disc diverge in [$\alpha$/Fe], they collectively suggest that a possible difference apparent at [Fe/H] $>$ –1.6. In our study, the initial scatter plot shown in Fig. \ref{fig5} displays overlapping trends between GSE and the kinematically hot disc in [$\alpha$/Fe]–[Fe/H] diagram. However, Fig. \ref{fig6} provides a more detailed view through two-dimensional density map. In this figure, the color shading reflects stellar number density, with darker regions indicating higher concentrations of stars. For the GSE population, we bin stars in 0.1 intervals in [Fe/H] from –2.8 to –1.2, and plot the mean [Mg/Fe] and its 1$\sigma$ dispersion as black points with vertical error bars. The same procedure is applied to the kinematically hot disc, with stars binned from [Fe/H] = –2.8 to –1.1. This consistent binning enables a direct comparison between the two populations across their overlapping metallicity ranges. The plot reveals that GSE exhibits a clear decreasing in [Mg/Fe] at [Fe/H] $>$ –1.6, whereas the kinematically hot disc displays a much flatter [Mg/Fe]–[Fe/H] relation over the same range. This suggests a chemical distinction that is not clearly visible in the scatter plot. It is also important to note that stars with [Fe/H] $>$ –1.6 make up only a small fraction of our sample. This underrepresentation of higher-metallicity stars may contribute to the difficulty in identifying [$\alpha$/Fe] difference between GSE and the disc. Finally, we emphasize that the kinematically hot disc in our study is not chemically homogeneous. As discussed in Section \ref{sec4}, it contains multiple components.

\begin{figure}
   \centering
    \includegraphics[width=0.6\textwidth]{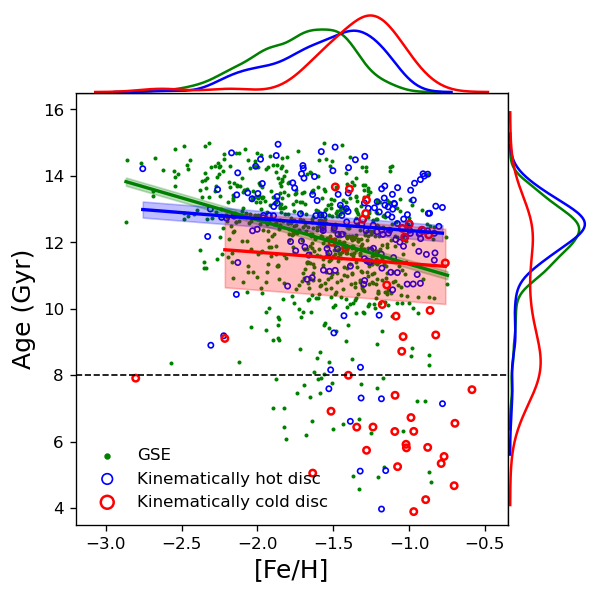}
    \caption{Stellar age-metallicity relation for the kinematcially cold disc, hot disc, and GSE, represented by red, blue, and green, respectively. Solid lines indicate linear fits of this relationship, with shaded regions showing 1$\sigma$ dispersion. All fits utilize stars older than 8 Gyr. A dashed line marks an age of 8 Gyr, and marginal distributions for age and [Fe/H] are also included.}
    \label{fig7}
\end{figure}

We then incorporated stellar ages into our analysis, but the limited number of stars with age data prevents the use of strict precision constraints on iron and magnesium abundances, as this would reduce the sample size. Therefore, we applied relatively loose constraints (see Section \ref{sec2.3} for details). Fig. \ref{fig7} shows the distributions of the kinematically cold disc, hot disc, and GSE in the age-[Fe/H] plane. Most kinematically hot disc and GSE stars are older than 8 Gyr, while kinematcially cold stars are about 50\% older and 50\% younger than 8 Gyr. For stars older than 8 Gyr, linear fits to the age-metallicity relationship reveal similar slopes for the kinematically cold and hot discs, both smaller than the slope for GSE. Marginal distributions further show that the metallicity peak of the kinematically hot disc corresponds to a higher metallicity than that of GSE, while the age peak of the kinematically hot disc corresponds to an older age compared to GSE.
The stars were divided into four age groups: $<$ 8 Gyr, 8-10 Gyr, 10-12 Gyr, and $>$ 12 Gyr. The normalized metallicity histograms of the kinematically cold disc, kinematically disc, and GSE were re-examined, as shown in Fig. \ref{fig8}, revealing the following patterns:
\begin{figure*}
   \centering
    \includegraphics[width=1.0\textwidth]{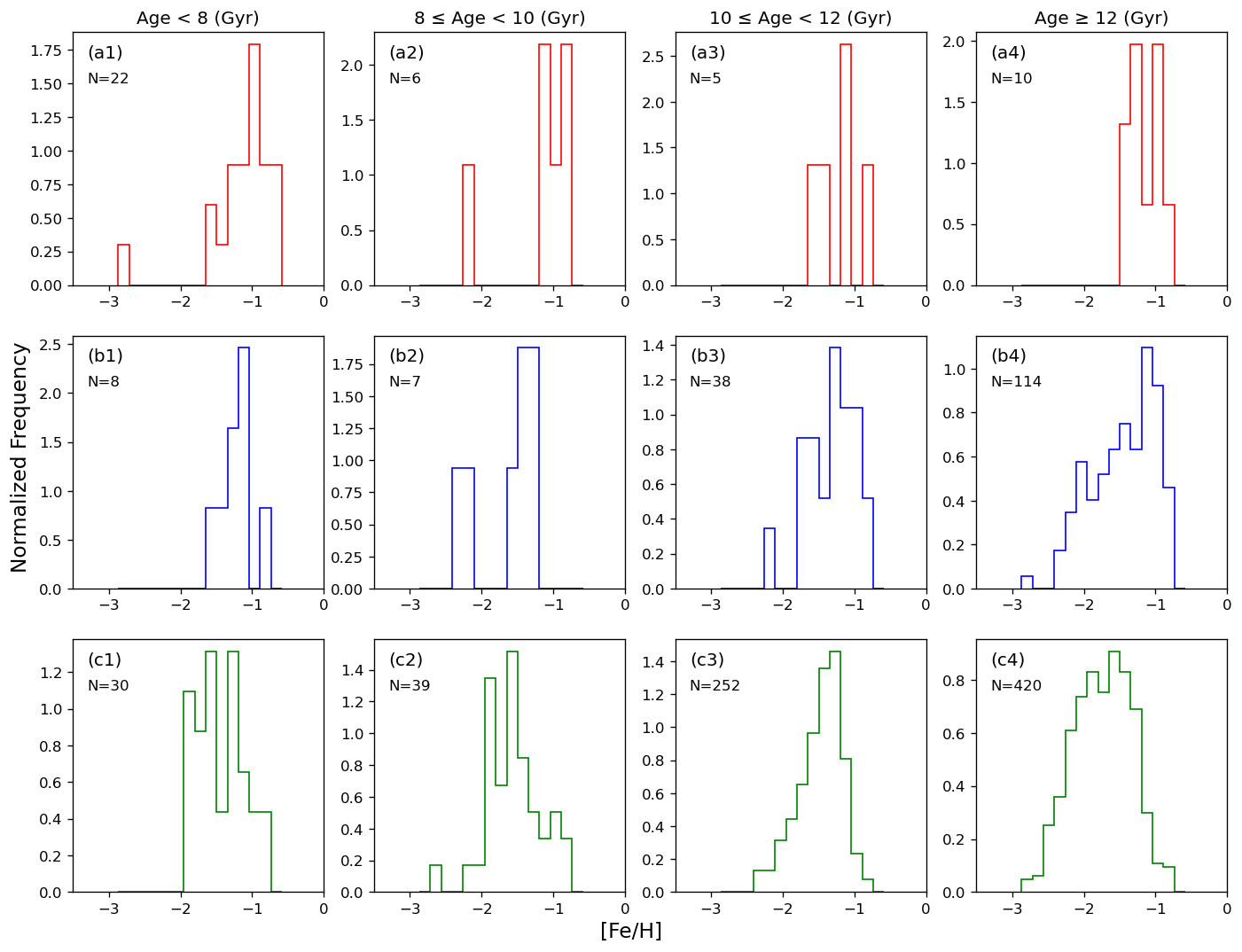}
    \caption{Normalized metallicity histograms for the kinematically cold disc, hot disc, and GSE across four age groups: $<$ 8 Gyr, 8-10 Gyr, 10-12 Gyr, and $>$ 12 Gyr. The three rows from top to bottom represent the kinematically cold disc, hot disc, and GSE, respectively.}
    \label{fig8}
\end{figure*}
\begin{enumerate}[label=(\roman*), leftmargin=0em, align=left] 
\setlength{\itemindent}{1em} 
    \item For all age groups, metallicity generally follows the order: kinematically cold disc $>$ kinematically hot disc $>$ GSE. This pattern is most pronounced for stars older than 12 Gyr, where the kinematically hot disc's metallicity histogram appears to combine characteristics of both the kinematically cold disc and GSE in specific proportions. For stars aged 10-12 Gyr, the differences between the three structures become less distinct, with all histograms peaking around -1.2.
    
    \item In the kinematically cold disc, stars older than 12 Gyr generally have [Fe/H] $>$ -1.5, while stars with [Fe/H] $<$ -1.5 first appear in the 10-12 Gyr group, and some in the 8-10 Gyr group even have [Fe/H] $<$ -2.0. For stars younger than 8 Gyr, nearly all have [Fe/H] $>$ -1.7 (except for one star with [Fe/H] = -2.8), and the histogram peaks at a [Fe/H] value around -1.0. Although the sample size is limited, this observation suggests a pattern of metallicity change in the kinematically cold disc.
    
    \item The kinematically hot disc shows a clear trend in metallicity evolution over time. For stars older than 12 Gyr, the proportion of stars with [Fe/H] $<$ -2.0 is relatively high, approximately 13.2\%. In the 10-12 Gyr group, this proportion drops significantly to only 5.3\%. From the older than 12 Gyr group to stars younger than 8 Gyr, the median metallicities are -1.4, -1.27, -1.5, and -1.25, while the mean metallicities are -1.45, -1.32, -1.64, and -1.23, respectively. 
    
    \item  For GSE stars older than 12 Gyr, the metallicity histogram peaks around -1.5, with a significant proportion having [Fe/H] $<$ -2.0. Specifically, out of the 420 stars in the older than 12 Gyr group, 112 stars have [Fe/H] $<$ -2, accounting for approximately 26.7\%. In the 10-12 Gyr group, the peak [Fe/H] value shifts to -1.2, with only 17 stars out of 252 having [Fe/H] $<$ -2.0, which is about 6.7\%. In the 8-10 Gyr group, the peak [Fe/H] value decreases again to -1.5. The median metallicities for these groups are -1.75, -1.42, and -1.56, while the mean metallicities are -1.74, -1.47, and -1.56, respectively. This trend suggests that the metallicity of GSE initially increased after its formation and then decreased, with this process concluding around 8 Gyr ago. The progenitor of GSE is believed to have merged with the Milky Way around 10 Gyr ago, making the observation of GSE stars younger than 8 Gyr unusual. These younger stars, which exhibit significant radial velocities and high eccentricities, may not originate from the GSE merger itself. They were likely included in our sample because they were selected using kinematic criteria associated with GSE membership.
    
    \item Across GSE and the kinematically hot disc samples, stars aged 8-10 Gyr are less numerous compared to those aged 10-12 Gyr and over 12 Gyr.
\end{enumerate}  

We also re-examined the [Mg/Fe]-[Fe/H] distributions under the same age groupings with relaxed error constraints on Mg and Fe abundances. Fig. \ref{fig9} presents these distributions, with the kinematically cold disc, hot disc, and GSE displayed from top to bottom. Each panel in the Fig. \ref{fig9} includes one auxiliary line, except panel (c1), which contains two. The auxiliary lines are of two types: dashed and dotted. Dashed lines correspond to the linear fit of the [Mg/Fe]-[Fe/H] relationship for kinematically cold stars older than 12 Gyr, while dotted lines represent the fit for kinematically cold stars younger than 8 Gyr. The equations for the lines are:
\begin{align}
    \rm{[Mg/Fe]}&=-0.12\times\rm{[Fe/H]}+0.05 \\
    \rm{[Mg/Fe]}&=-0.24\times\rm{[Fe/H]}-0.02 
\end{align}
From Fig. \ref{fig9}, we can observe the following patterns:
\begin{figure*}
   \centering
    \includegraphics[width=1.0\textwidth]{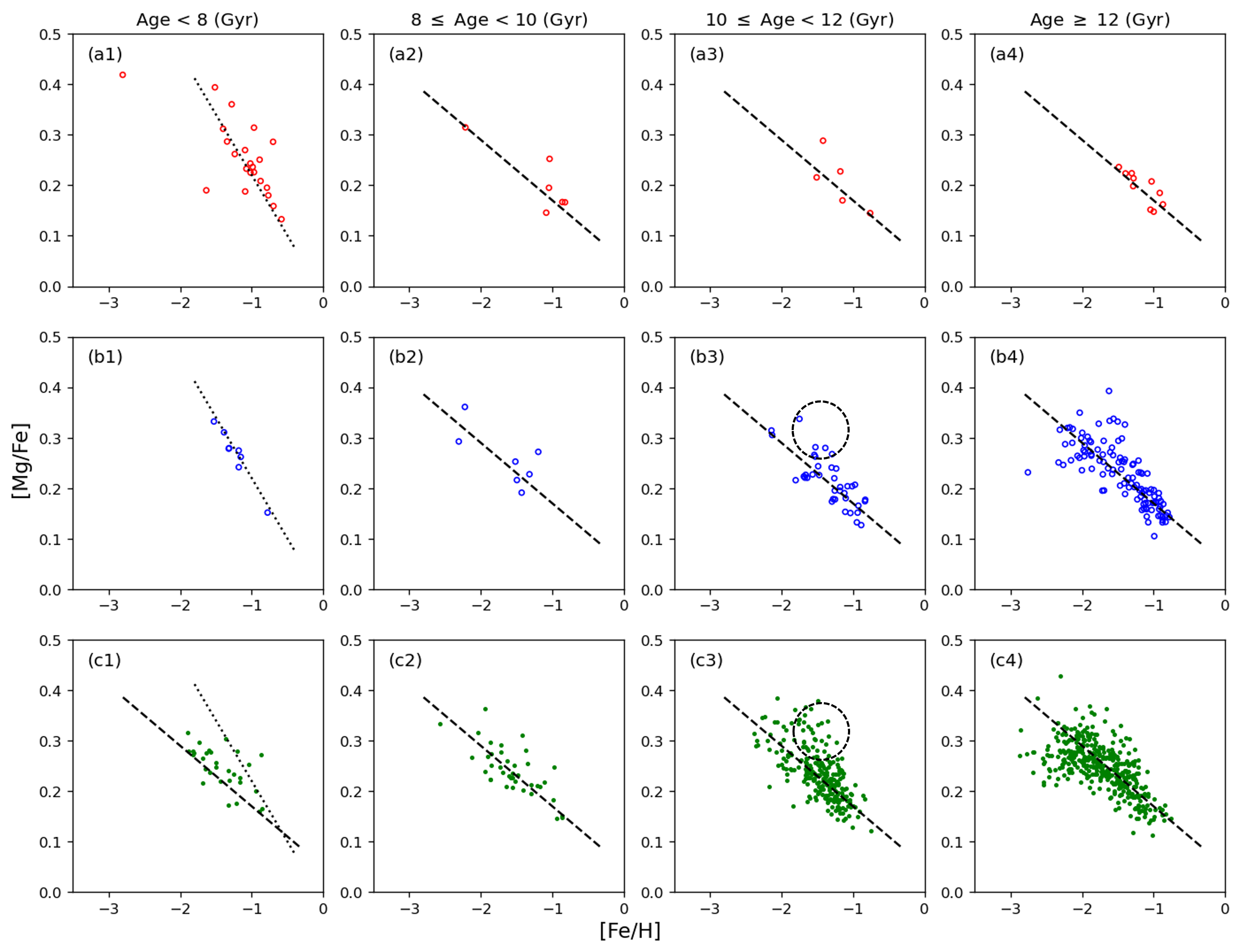}
    \caption{[Mg/Fe]-[Fe/H] distributions for the kinematically cold disc, hot disc, and GSE (shown in top, middle, and bottom rows) across four age groups: $<$ 8 Gyr, 8-10 Gyr, 10-12 Gyr, and $>$ 12 Gyr. Each panel contains one ore two auxiliary lines of two types: dashed lines derived from fitting kinematically cold disc stars older than 12 Gyr, and dotted lines from kinematically cold disc stars younger than 8 Gyr. Panels (b3) and (c3) include dashed circles highlighting Mg-enhanced stars, which are defined as stars with [Mg/Fe] values significantly higher than predicted by the dashed line relationship at a given [Fe/H].}
    \label{fig9}
\end{figure*}

\begin{enumerate}[label=(\roman*), leftmargin=0em, align=left] 
\setlength{\itemindent}{1em} 
    \item For the kinematically cold and hot discs, stars older than 8 Gyr show significantly different [Mg/Fe]-[Fe/H] relationships compared to younger stars. Stars younger than 8 Gyr exhibit a steeper slope, with [Mg/Fe] decreasing more rapidly as [Fe/H] increases.
    
    \item The kinematically hot disc and GSE exhibit similar distributions across three age groups: 8-10 Gyr, 10-12 Gyr, and $>$ 12 Gyr. In the oldest group ($>$ 12 Gyr), both exhibit an $\alpha$-knee at [Fe/H] $\approx$ -1.3. Dashed lines represent the expected [Mg/Fe]-[Fe/H] relationship, with Mg-enhanced stars defined as those showing significantly higher [Mg/Fe] values than predicted at a given [Fe/H]. In the 10-12 Gyr group, many stars in both the kinematically hot disc and GSE show such Mg enhancements, as shown in panels (b3) and (c3) of Fig. \ref{fig9}. This enhancement is less pronounced in the other age groups.

    \item  For stars younger than 8 Gyr, the number of stars following the dotted [Mg/Fe]-[Fe/H] relationship decreases progressively from kinematically cold disc to kinematically hot disc to GSE. While the kinematically cold disc contains the most stars conforming to this relationship, such stars are nearly absent in GSE.
\end{enumerate}

In Fig. \ref{fig7}, we find that both the kinematically hot and cold disc samples contain a subset of metal-poor stars with ages younger than 8 Gyr. This is unexpected, given that younger stars are generally associated with higher [Fe/H] in the context of Galactic chemical evolution. Building on this, Fig. \ref{fig5} reveals that within the kinematically cold disc sample, there are two distinct sequences in the [$\alpha$/Fe]–[Fe/H] plane: a high-[$\alpha$/Fe] sequence and a low-[$\alpha$/Fe] sequence. When stellar ages are incorporated, as shown in panels (a1) and (b1) of Fig. \ref{fig9}, we find that disc stars on the high-[$\alpha$/Fe] sequence are precisely those younger than 8 Gyr. This finding is particularly surprising, as it contradicts the widely accepted trend that [$\alpha$/Fe] decreases with time in the Galactic disc. One possible explanation is that binary stellar evolution might lead to underestimated ages for some genuinely old stars. To investigate this possibility, we examined whether stars shown in (a1) and (b1) of Fig. \ref{fig9} might be unresolved binaries. \cite{penoyre_astrometric_2022} noted that Gaia astrometric measurements with RUWE larger than 1.4 could indicate binarity. However, in our study, stars with RUWE $>$ 1.4 were already excluded in Section \ref{sec2.2}. Additionally, \cite{mowlavi_gaia_2023} released a catalogue of eclipsing-binary candidates based on Gaia DR3, and after cross-matching, we found no overlap between stars in our high-[$\alpha$/Fe] sequence and their eclipsing-binary catalogue. Furthermore, 14 stars from (a1) and (b1) in Fig. \ref{fig9} have been repeatedly observed by LAMOST low-resolution survey. The multiple line-of-sight velocity measurements and their corresponding uncertainties for these stars are listed in Table \ref{tab2}. We used these line-of-sight velocity variations to statistically evaluate the possibility of binarity. To quantify whether a star exhibits significant radial velocity variations, indicative of a binary system, we employed the $\chi^2$ test. Specifically, we computed:
\begin{equation}
\chi^2 = \sum_{i=1}^{n} \frac{(v_i - \bar{v})^2}{\sigma_i^2}
\end{equation}
where $v_i$ is the $i$th line-of-sight velocity measurement, $ \bar{v}$ is the mean velocity, $\sigma_i$ is the uncertainty of the $i$th measurement, and $n$ is the number of observations. Under the null hypothesis—that the star is single and maintains a constant line-of-sight velocity, the calculated $\chi^2$ statistic follows a chi-square distribution with $n - 1$ degrees of freedom. For each star, we calculated the associated $p$-value, which represents the probability of observing a $\chi^2$ value as large as (or larger than) the measured one under the null hypothesis. A low $p$-value (typically $p < 0.05$) indicates statistically significant velocity variations and suggests the presence of binarity. The computed $\chi^2$ values and their corresponding $p$-values are listed in Table \ref{tab2}. Based on these $p$-values, none of the stars exhibit velocity variations significant enough to be identified as binaries.In summary, current evidence does not support the hypothesis that these high-$\alpha$, young stars are unresolved binaries.
\begin{table}
    \centering
    \caption{Line-of-sight velocities and corresponding uncertainties for 14 metal-poor, high-$\alpha$ sequence stars younger than 8 Gyr, each observed multiple times by the LAMOST survey.}
    \setlength{\tabcolsep}{2.1mm}{
    \begin{tabular}{cccccc}
    \toprule
    uid &  velocity list & velocity error list & $\chi^2$ & $p$-value \\
    \midrule
    G14028202809079 &8.31; 16.96; 8.94; 13.68  &3.48; 5.28; 6.47; 5.49 & 2.316 & 0.509 \\
    G15971821115630 &45.87; 37.86 & 7.63; 2.39 & 3.084 & 0.079 \\
    G16220243843745 &-7.20; -7.12 & 4.04; 3.84 & 0.000 & 0.989 \\
    G16225978524450 &-1.78; -3.19 & 3.83; 4.69 & 0.056 & 0.812 \\
    G16237508594175 &2.23; -4.19 & 5.37; 6.58 & 0.595 & 0.440 \\
    G16263861478746 &-10.79; -5.38 & 3.30; 4.74 & 0.998 & 0.317 \\
    G16310961074871 &-26.86; -29.30; -32.14; -28.23 & 5.45; 5.99; 3.12; 4.38 & 1.146 & 0.766 \\
    G16424162144631 &-24.13; -26.35; -16.38; & 4.36; 4.76; 4.79 & 3.135 & 0.679\\
                    &-21.11; -22.34; -14.24 & 2.89; 4.52; 15.19 & & \\
    G16496513210296 &40.96; 33.03 & 2.83; 5.04 & 2.582 & 0.108 \\
    G16498871828495 &-24.25; -18.38 & 7.05; 6.30 & 0.390 & 0.532 \\
    G17297423527435 &11.77; 13.41 &3.25; 6.40 & 0.080 & 0.777 \\
    G17567156788740 &-6.75; -20.85 &4.50; 8.77 & 3.101 & 0.078 \\
    G15215012847084 &-129.00; -124.24 &4.74; 7.98 & 0.341 & 0.559 \\
    G17266455315544 &-28.94; -34.52 &3.97; 3.72 & 1.056 & 0.304 \\
    \bottomrule
    \end{tabular}}
     \begin{tablenotes}
        \item {\textbf{Note.} uid is the unique source ID in LAMOST. The “velocity list” and “velocity error list” entries for each star are paired in order.}
    \end{tablenotes}
    \label{tab2}
\end{table}

\begin{figure*}
    \centering
    \includegraphics[width=1.0\textwidth]{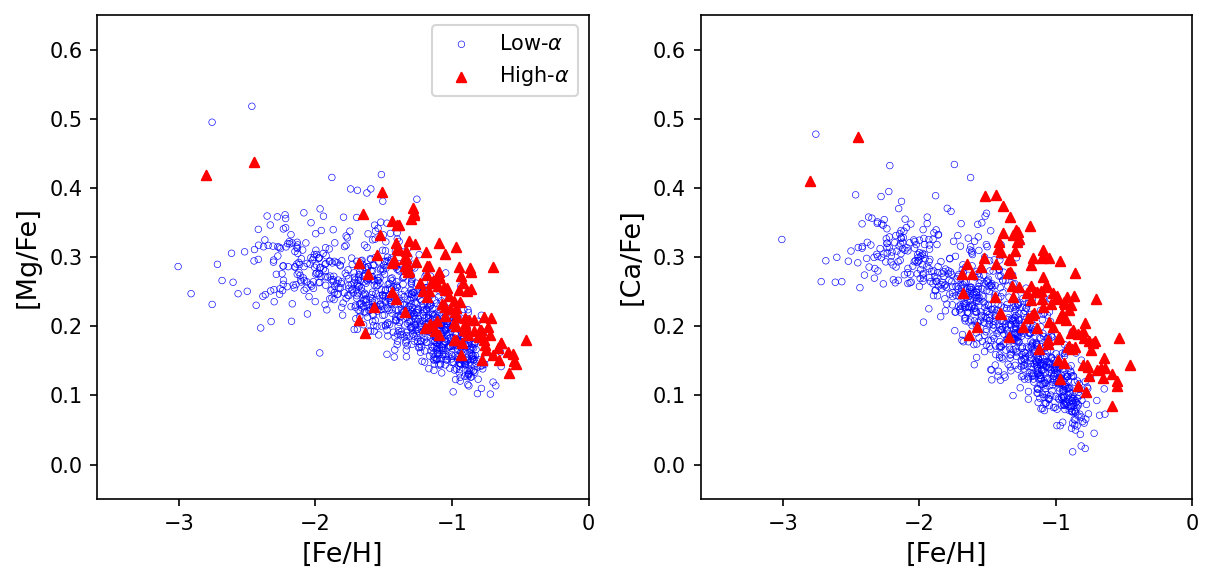}
    \caption{Distributions of [Mg/Fe] and [Ca/Fe] as a function of [Fe/H] for two stellar populations: high-$\alpha$ (red triangles) and low-$\alpha$ (blue circles). These groups are defined based on stellar age, with high-$\alpha$ stars being younger than 8 Gyr and low-$\alpha$ stars older. Only stars with [Mg/H] and [Fe/H] uncertainties less than 0.2 (from machine-learning estimates) are included.}
    \label{fig10}
\end{figure*}

\begin{figure*}
    \centering
    \includegraphics[width=1.0\textwidth]{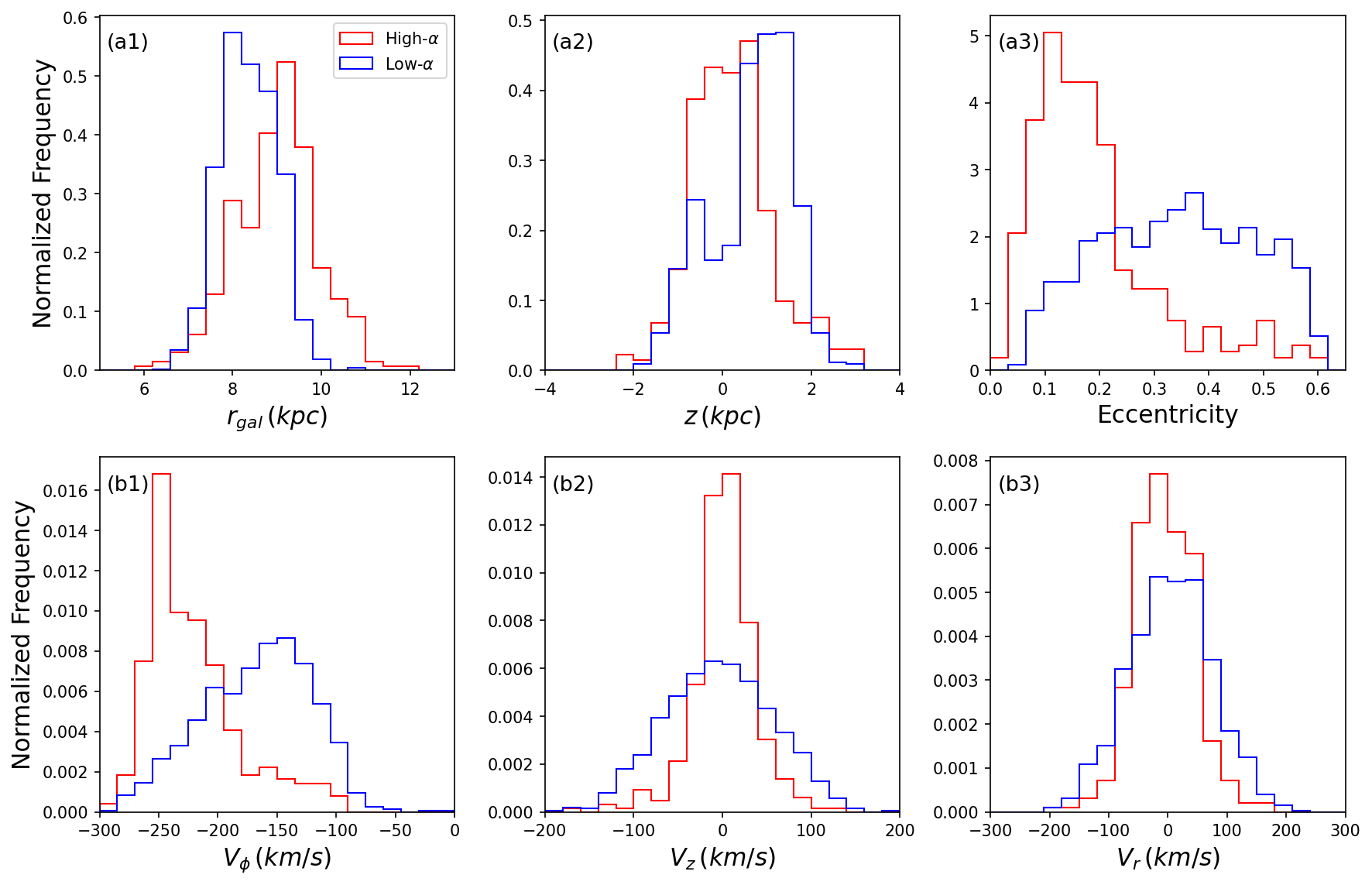}
    \caption{Normalized histograms showing spatial and kinematic properties of high-$\alpha$ (red) and low-$\alpha$ (blue) disc stars. Top panels: Galactocentric radius $r_{\text{gal}}$ (a1), vertical distance from the mid-plane $z$ (a2), and orbital eccentricity (a3). Bottom panels: azimuthal velocity $V_\phi$ (b1), vertical velocity $V_z$ (b2), and radial velocity $V_r$ (b3). The two populations are selected from stars with reliable age estimates, excluding those with $e>0.6$ or retrograde motion ($V_\phi > 0$). High-$\alpha$ stars are defined as those younger than 8 Gyr; the rest are classified as low-$\alpha$.}
    \label{fig11}
\end{figure*}

As previously discussed, our kinematically cold disc sample reveals two distinct populations: metal-poor, high-$\alpha$ and metal-poor, low-$\alpha$ stars (Fig. \ref{fig5}). When incorporating stellar age information, we find that the metal-poor, high-$\alpha$ population predominantly comprises stars younger than 8 Gyr (Fig. \ref{fig9}). Notably, age information also reveals the presence of metal-poor, high-$\alpha$ stars within the kinematically hot disc, although they constitute smaller fraction compared to the kinematically cold disc. From the 2,517 stars with reliable age estimates, we applied two selection criteria to isolate disc stars. First, we excluded stars with orbital eccentricities greater than 0.6 to eliminate those with halo-like orbits. Second, we removed stars exhibiting retrograde motion by filtering out those with $V_{\phi}>0$. After these selections, 1,396 stars remained. Among them, 329 stars with ages younger than 8 Gyr were classified as high-$\alpha$ population, with the remainder designated were as low-$\alpha$ population. We examined the abundance trends of the two groups. Fig. \ref{fig10} shows the [Mg/Fe]–[Fe/H] and [Ca/Fe]–[Fe/H] distributions, including only stars with machine-learning–estimated [Mg/H] and [Fe/H] having uncertainties less than 0.2. As shown in Fig. \ref{fig10}, the younger group clearly exhibits higher [Mg/Fe] and [Ca/Fe] ratios compared to the older group. A detailed comparison between the high-$\alpha$ and low-$\alpha$ populations was subsequently performed, focusing on their spatial distribution and kinematic properties. As shown in Fig. \ref{fig11}, the normalized histograms highlight clear and significant differences between the two groups:
\begin{enumerate}[label=(\roman*), leftmargin=0em, align=left] 
\setlength{\itemindent}{1em} 
    \item High-$\alpha$ stars are preferentially found at larger Galactocentric radii, with a distribution peak near 9.5 kpc and a substantial fraction extending beyond 10 kpc. In contrast, low-$\alpha$ stars are more centrally concentrated, peaking around 8.2 kpc, with almost no stars located beyond 10 kpc (Fig. \ref{fig11} (a1)).
    
    \item High-$\alpha$ stars are more tightly confined to the Galactic mid-plane, with majority located within $\pm$1.5 kpc and a distribution peak near $z\approx0.5$ kpc. On the other hand, low-$\alpha$ exhibit a broader vertical distribution, peaking slightly higher around $z\approx1.0$ kpc (Fig. \ref{fig11} (a2)).

    \item The eccentricity distribution of high-$\alpha$ stars is relatively narrow, peaking at $e\approx0.12$, and more than 90\% of the stars have $e<0.4$, indicating near-cicular orbits. Incontrast, low-$\alpha$ stars show a flatter and broader distribution peaking around $e\approx0.4$, with a notable fraction extending to $e>0.5$ (Fig. \ref{fig11} (a3)).
    
    \item  High-$\alpha$ stars demonstrate higher absolute azimuthal velocities, with a sharp distribution peak at $V_{\phi}\approx-250\, \rm{km}\,\rm{s^{-1}}$, suggesting faster rotation. In contrast, low-$\alpha$ stars peak around $-150\, \rm{km}\,\rm{s^{-1}}$, indicating slower rotational support and a kinematically hotter component (Fig. \ref{fig11} (b1)).
    
    \item  In both radial and vertical velocities, high-$\alpha$ stars exhibit narrower and more centrally concentrated distributions. Their $V_{r}$ and $V_{z}$ values are tightly clustered around $0\,\rm{km}\,{s^{-1}}$, with standard deviations significantly smaller than those of low-$\alpha$ stars. Low-$\alpha$ stars show broader velocity spreads in both components, reaching up to $\pm200 \rm{km}\,\rm{s^-1}$, further reinforcing their dynamically hotter nature (Fig. \ref{fig11} (b2) and (b3).
\end{enumerate}

\cite{han_insights_2020} investigated the Galactic disc using a sample of G- and K-type dwarfs, classifying stars into thin and thick disc populations based on their [$\alpha$/Fe]. Their sample primarily consists of stars with [Fe/H] $>$ –1.0. In their classification, stars with lower [$\alpha$/Fe] are assigned to the thin disc, while those with higher [$\alpha$/Fe] are identified as thick disc members. They found that thin disc stars (metal-rich and low-$\alpha$) generally have orbital eccentricities below 0.4, with a peak near 0.2, whereas thick disc stars (metal-rich and high-$\alpha$) exhibit a broader eccentricity distribution, extending up to 0.8 (see their Figure 8). In terms of eccentricity, our sample of young, metal-poor, high-$\alpha$ disc stars shows kinematic behavior similar to the thin disc population (metal-rich and low-$\alpha$) described by \cite{han_insights_2020}, while our old, metal-poor, low-$\alpha$ stars more closely resemble their thick disc (metal-rich and high-$\alpha$). A similar chemo-dynamic comparison was presented by \cite{lee_formation_2011}, who analyzed G-type dwarfs from the SEGUE survey to characterize the Galactic thin and thick discs. In their study, thin disc stars (metal-rich and low-$\alpha$) predominantly exhibit azimuthal velocities $|V_\phi| > 200 \, \mathrm{km\,s^{-1}}$, while the thick disc stars (metal-rich and high-$\alpha$) tend to show lower $|V_\phi|$ values. Spatially, their thin disc population (metal-rich and low-$\alpha$) is more radially extended and located farther from the Galactic center. Vertically, their thin disc stars (metal-rich and low-$\alpha$) are primarily confined within 1.5 kpc of the Galactic mid-plane, whereas thick disc (metal-rich and high-$\alpha$) stars concentrated between 0.5 and 2 kpc, peaking around 1 kpc. Based on these characteristics, our old, metal-poor, low-$\alpha$ population shows similarities with the thick disc (metal-rich and high-$\alpha$) defined by \cite{lee_formation_2011}, while our young, metal-poor and high-$\alpha$ stars resembles their thin disc (metal-rich and low-$\alpha$). However, we note an important distinction: a subset of our young, metal-poor, high-$\alpha$ stars are located at significant vertical distances (Fig. \ref{fig11} (a2)), with several exceeding 1.5 kpc and some reaching up to 3 kpc in height. In summary, combining insights from our sample and prior studies \citep[e.g.,][]{lee_formation_2011,han_insights_2020}, we find that the chemo-dynamic properties of disc stars vary markedly across the metallicity regime. At higher metallicities, high-$\alpha$ stars are dynamically hotter and they are found closer to the Galactic center, possess higher orbital eccentricities, lower rotational velocities, and extend to greater vertical heights. In contrast, low-$\alpha$ stars at high metallicity tend to be dynamically colder, with lower eccentricities and higher azimuthal velocities. Interestingly, this trend reverses at lower metallicities: high-$\alpha$ stars in this regime exhibit kinematically colder properties, while low-$\alpha$ stars appear dynamically hotter.

A number of recent studies have identified a peculiar population of stars exhibiting relatively high [$\alpha$/Fe] despite their young ages. Using asteroseismic data to determine stellar ages and spectroscopic measurements to derive chemical abundances, \cite{martig_young_2015,chiappini_young_2015,wu_mass_2018} independently reported the existence of such stars. \cite{chiappini_young_2015} found that these high-$\alpha$ stars are younger than 8 Gyr, while \cite{martig_young_2015}, even under conservative assumptions intended to overestimate stellar ages, identified high-$\alpha$ stars with ages below 6 Gyr. Both studies showed that these young high-$\alpha$ stars occupy the same region in the [Mg/Fe]–[Fe/H] diagram as the old, metal-rich, high-$\alpha$ disc stars. In terms of metallicity, the young high-$\alpha$ stars studied by \cite{martig_young_2015} have [Fe/H] $>$ –0.5, while those in \cite{chiappini_young_2015} have [Fe/H] $>$ –0.7. \cite{chiappini_young_2015} further noted that these stars are primarily located within 6 kpc of the Galactic center and proposed that their formation may be influenced by the Galactic bar. In contrast, \cite{martig_young_2015} found that the guiding radii and radial velocity distributions of these stars are similar to those of the canonical thick disc, implying that their spatial and kinematic properties are consistent with those old, high-$\alpha$ stars. In our study, we also identify a population of high-$\alpha$ stars with ages younger than 8 Gyr. However, in contrast to the young, high-$\alpha$ stars reported by \cite{martig_young_2015,chiappini_young_2015}, our high-$\alpha$ stars are metal-poor. Furthermore, our analysis reveals that our young, metal-poor, high-$\alpha$ stars exhibit kinematic properties more closely aligned with the thin disc: they are found at larger Galactocentric radii and exhibit dynamically colder motions. These results suggest that our young, metal-poor, high-$\alpha$ stars are distinct from previously identified metal-rich, young, high-$\alpha$ populations, both chemically and kinematically, and may represent a different evolutionary pathway or formation environment.

\subsection{GSE vs other merger remnants}\label{sec3.2}
Gaia's observations have uncovered numerous dwarf galaxy remnants in the Galactic halo, with stars in each remnant typically exhibiting similar kinematic and chemical properties. However, the origins of these remnants remain unclear. Some researchers suggest that certain remnants share a common progenitor, while others argue they originate from different merger events. This study aims to provide new evidence on this topic, focusing on GSE, Pontus, Thamnos, Sequoia, Helmi Streams, and Wukong. We excluded other remnants like Sagittarius, Cetus, and Heracles. Despite its size, Sagittarius's great distance from the Sun limits the number of its stars in our sample. Cetus is mainly located in the southern celestial hemisphere, while Heracles lies near the Galactic center, where LAMOST lacks observational coverage.

We compared the normalized metallicity histograms of GSE, Pontus, Thamnos, Sequoia, Helmi Streams, and Wukong, as shown in Fig \ref{fig12}. The key findings are as follows:
\begin{figure*}
   \centering
    \includegraphics[width=1.0\textwidth]{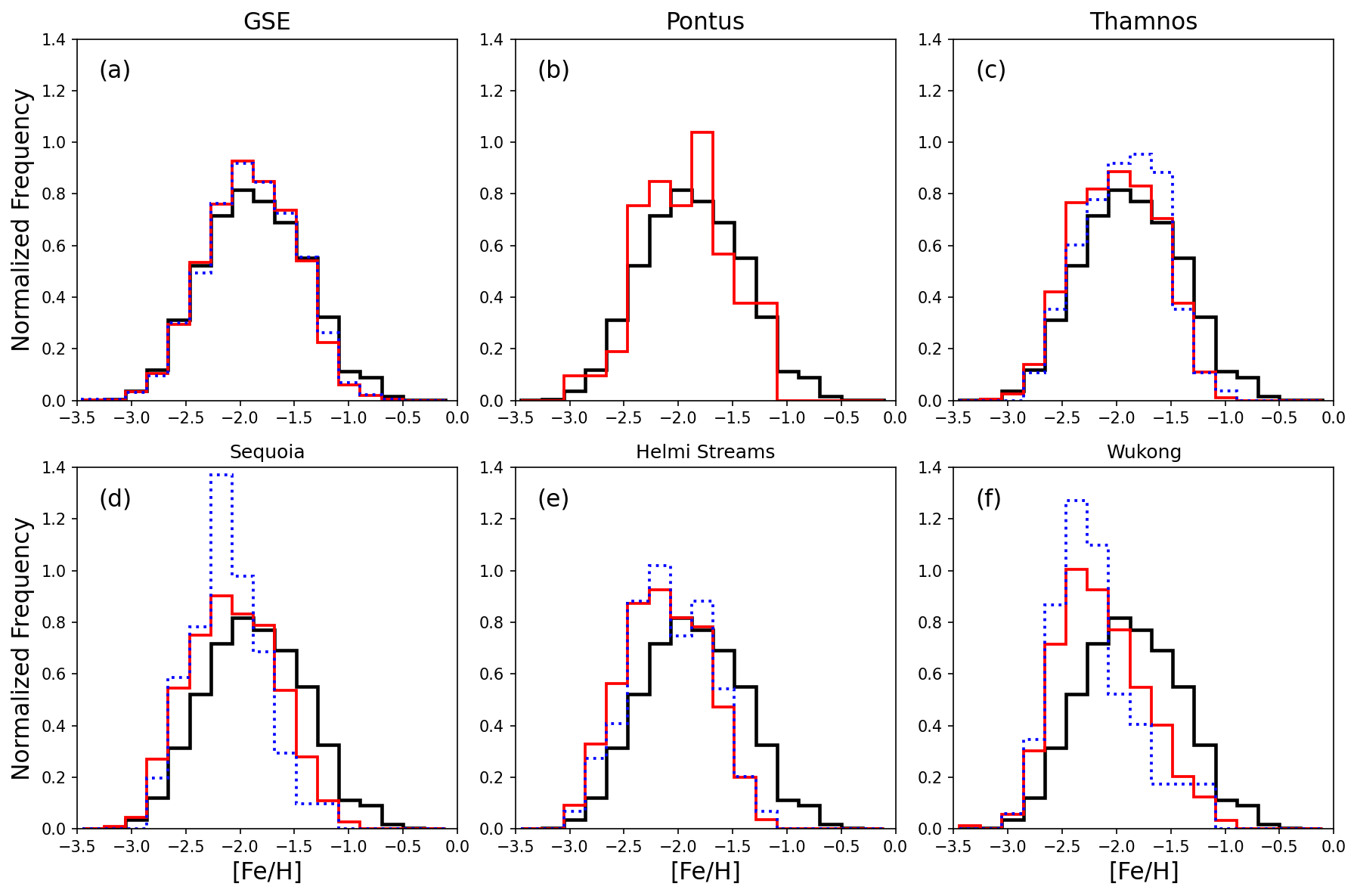}
    \caption{Normalized metallicity histograms of GSE, Pontus, Thamnos, Sequoia, Helmi Streams, and Wukong. As in Fig. \ref{fig5}, each panel shows: the black solid line for all sample stars, the red solid line for stars directly classified by their kinematic properties, and the blue dashed line for stars identified through clustering in phase space and subsequently attributed to that remnant.}
    \label{fig12}
\end{figure*}

\begin{enumerate}[label=(\roman*), leftmargin=0em, align=left] 
\setlength{\itemindent}{1em} 
    \item Pontus contains proportionally fewer stars with [Fe/H] $>$ -1.7, and more with -2.5 $<$ [Fe/H] $<$ -2.0 compared to GSE.
    
    \item Thamnos peaks at [Fe/H] $\approx$ -2.0, and contains proportionally fewer stars with [Fe/H] $>$ -1.5 and a higher fraction of stars in the range -2.7 $<$ [Fe/H] $<$ -2.0 compared to GSE.
    
    \item Sequoia, Helmi Streams, and Wukong have metallicity histograms peaking at lower [Fe/H] values than GSE and contain more stars with [Fe/H] $<$ -2.0, with Wukong exhibiting the lowest metallicities.
\end{enumerate}

\subsection{Unclassified DTGs}\label{sec3.3}
According to our classification criteria for the Galactic discs and dwarf galaxy remnants, 28 DTGs could not be assigned to either category. These 28 DTGs were further analyzed and divided into six groups based on their kinematic characteristics, labeled Cluster 1 to 6. Stars within each group exhibit similar kinematic properties, as shown by their distributions in the kinematic space in Fig. \ref{fig4}. Specifically, Cluster 1 to 3 likely correspond to known structures but were excluded from the main classification due to the strict criteria applied, while Cluster 4 to 6 likely represent unknown substructures. The detailed kinematic characteristics of each group are provided in the following paragraphs. Additionally, a table listing the specific DTGs within each group can be found at the end of the article via a provided link.

\begin{figure*}
   \centering
    \includegraphics[width=1.0\textwidth]{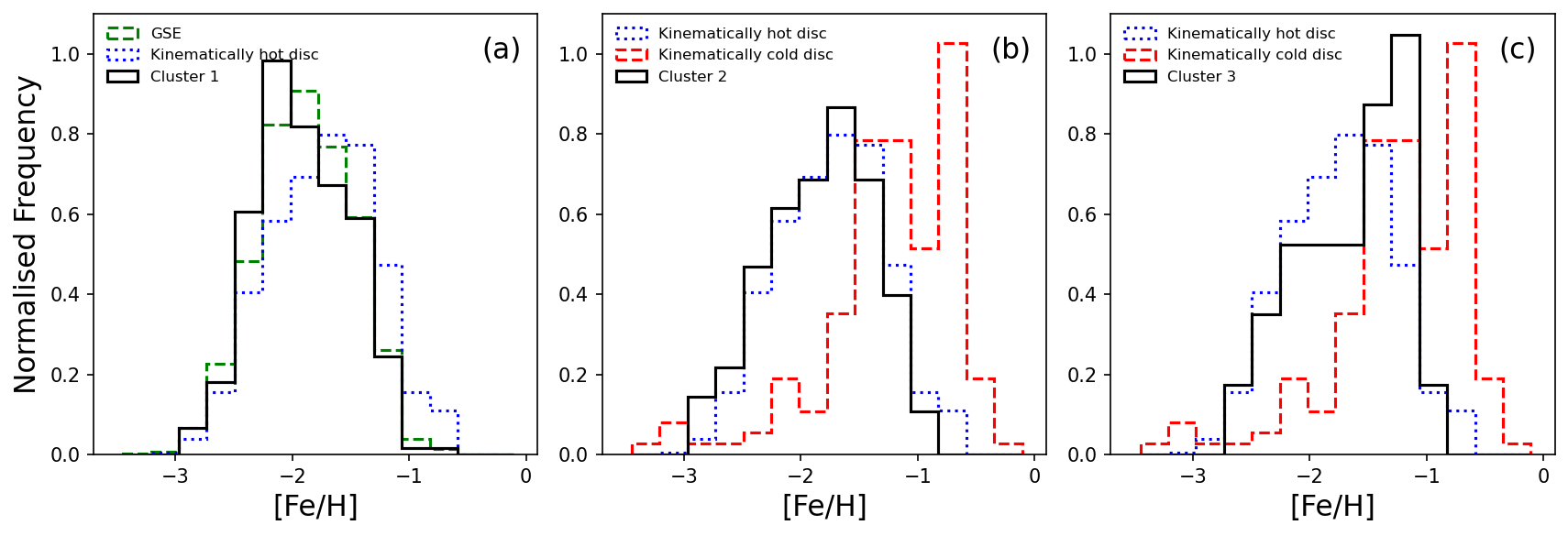}
    \caption{(a) Comparison of normalized metallicity histograms between GSE, the kinematically hot disc, and Cluster 1. (b) Comparison of normalized metallicity histograms between the kinematically hot disc, the kinematically cold disc, and Cluster 2. (c) Comparison of normalized metallicity histograms between the kinematically hot disc, the kinematically cold disc, and Cluster 3. GSE, kinematically hot disc, and kinematically cold disc stars shown here were first identified through phase-space clustering and then classified accordingly.}
    \label{fig13}
\end{figure*}

Cluster 1, containing 14 DTGs, exhibits consistent kinematic properties, such as $V_{\phi}$ around -100 $\rm{km\,s^{-1}}$ and eccentricities near 0.65. These DTGs are located at the periphery of the region occupied by GSE and share a similar metallicity histogram, as shown in Fig. \ref{fig13}. However, they were excluded from GSE due to our arbitrary eccentricity criterion ($e>0.7$). Given their kinematics and metallicity, we conclude that Cluster 1 is part of GSE. Cluster 2, comprising 7 DTGs, overlaps with the kinematically hot disc in kinematic space. The DTGs in this cluster have $|V_{z}|$ values slightly above 25 $\rm{km\,s^{-1}}$, exceeding the kinematically hot disc criterion of $|V_{z}|$ $<$ 25 $\rm{km\,s^{-1}}$. Nevertheless, their metallicity histogram, illustrated in panel (b) of Fig. \ref{fig13}, closely resembles that of the kinematically hot disc. Therefore, we conclude that Cluster 2 is part of the kinematically hot disc. Cluster 3 consists of 3 DTGs with kinematic properties intermediate between those of the kinematically hot and cold discs. Two DTGs were excluded from the kinematically cold disc because, despite they have $J_{\phi}$ $>$ 1660 $\rm{kpc\,km\,s^{-1}}$ and $|V_{z}|$ $<$ 15 $\rm{km\,s^{-1}}$, their eccentricities are greater than 0.3. The third DTG, on the other hand, was excluded from the kinematically hot disc because, while its $J_{\phi}$ $<$ 1660 $\rm{kpc\,km\,s^{-1}}$, its eccentricity is lower than 0.2. These DTGs exhibit kinematic characteristics partially aligning with both discs, suggesting they belong to the transition region between the two. Their metallicity histogram, shown in panel (c) of Fig. \ref{fig13}, lies between that of the kinematically cold and hot discs, supporting this conclusion. Therefore, we infer that Cluster 3 represents the transition from the kinematically cold disc to the kinematically hot disc. It should be noted that the stars shown in Fig. \ref{fig13} for the kinematically hot disc, cold disc, and GSE were first identified through phase-space clustering, and subsequently assigned to each component based on their kinematic properties. For plotting purposes, we only include stars with [Fe/H] uncertainties less than 0.2 from both machine learning and template fitting methods, and with a difference between the two methods also smaller than 0.2.

For Clusters 4 to 6, Fig. \ref{fig14} shows the distribution of their stars in the $V_{\phi}$-$V_{r}$ and $V_{\phi}$-$V_{z}$ planes, alongside stars belonging to GSE, the kinematically hot disc, and the kinematcially cold disc. As in Fig. \ref{fig13}, the disc and GSE samples were first identified through phase-space clustering and then assigned to their respective components based on their kinematic properties. The following observations can be made:
\begin{figure*}
   \centering
    \includegraphics[width=1.0\textwidth]{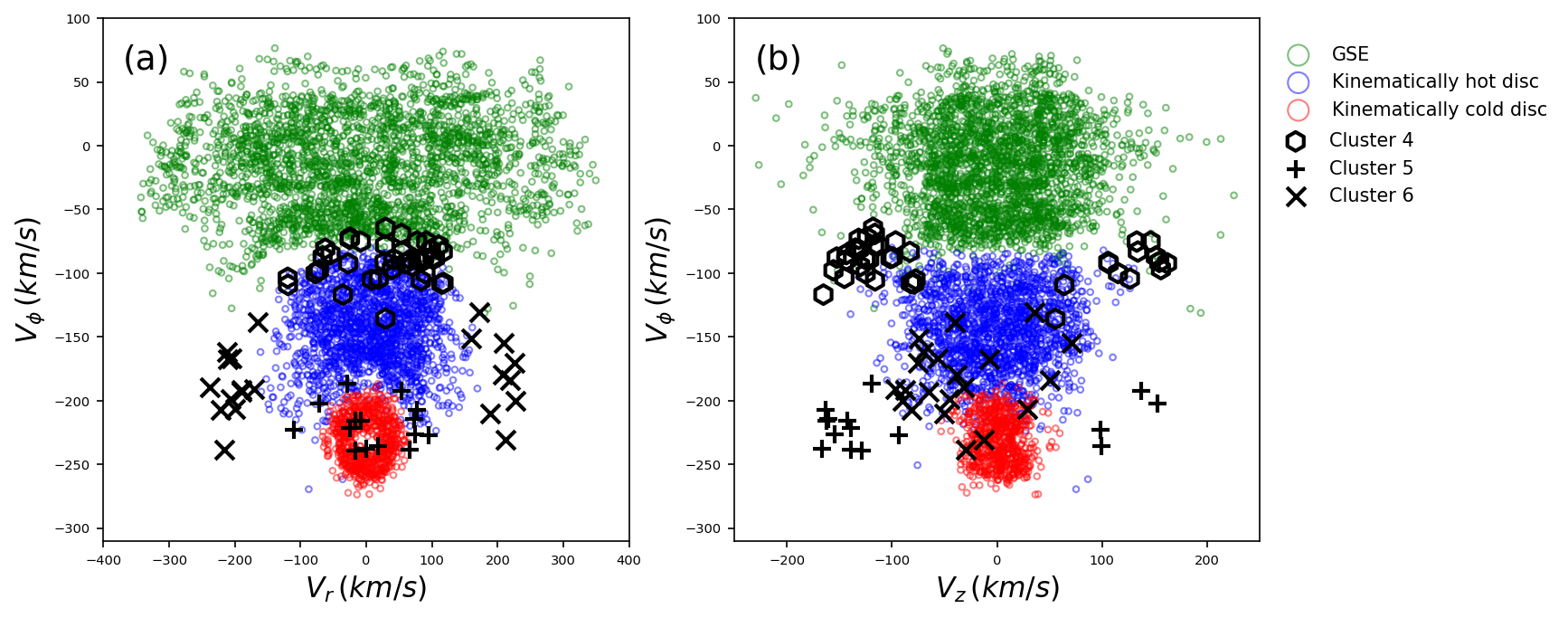}
    \caption{$V_{\phi}$-$V_{r}$ and $V_{\phi}$-$V_{z}$ distributions of stars in Cluster 4, 5, and 6, represented by Hexagons, plus signs and crosses, respectively. The distributions of GSE, the kinematically hot disc, and the kinematically cold disc are shown by green, blue, and red circles, respectively.}
    \label{fig14}
\end{figure*}
\begin{enumerate}[label=(\roman*), leftmargin=0em, align=left] 
\setlength{\itemindent}{1em} 
    \item Stars in Cluster 4 have $V_{\phi}$ near -100 $\rm{km\,s^{-1}}$, between GSE and the kinematically hot disc. Their $V_{r}$ dispersion is similar to the kinematically hot disc, but their $|V_{z}|$ values is significantly higher, mostly exceeding 100 $\rm{km\,s^{-1}}$.

    \item Stars in Cluster 5 have $V_{\phi}$ and $V_{r}$ similar to the kinematically cold disc, with $V_{\phi}$ mostly below -200 $\rm{km\,s^{-1}}$ and $|V_{r}|$ below 100 $\rm{km\,s^{-1}}$. However, their $|V_{z}|$ values are much higher than the kinematically cold disc, mostly exceeding 100 $\rm{km\,s^{-1}}$.

    \item Stars in Cluster 6 exhibit disc-like $V_{z}$ and $V_{\phi}$. Specifically, $|V_{z}|$ is mostly below 100 $\rm{km\,s^{-1}}$, consistent with the kinematically hot disc, while $V_{\phi}$ ranges from -240 to -130 $\rm{km\,s^{-1}}$, showing features of both the kinematically hot and cold discs. However, their $|V_{r}|$ is unusually large, mostly around 200 $\rm{km\,s^{-1}}$.
\end{enumerate}

\begin{figure*}
   \centering
    \includegraphics[width=1.0\textwidth]{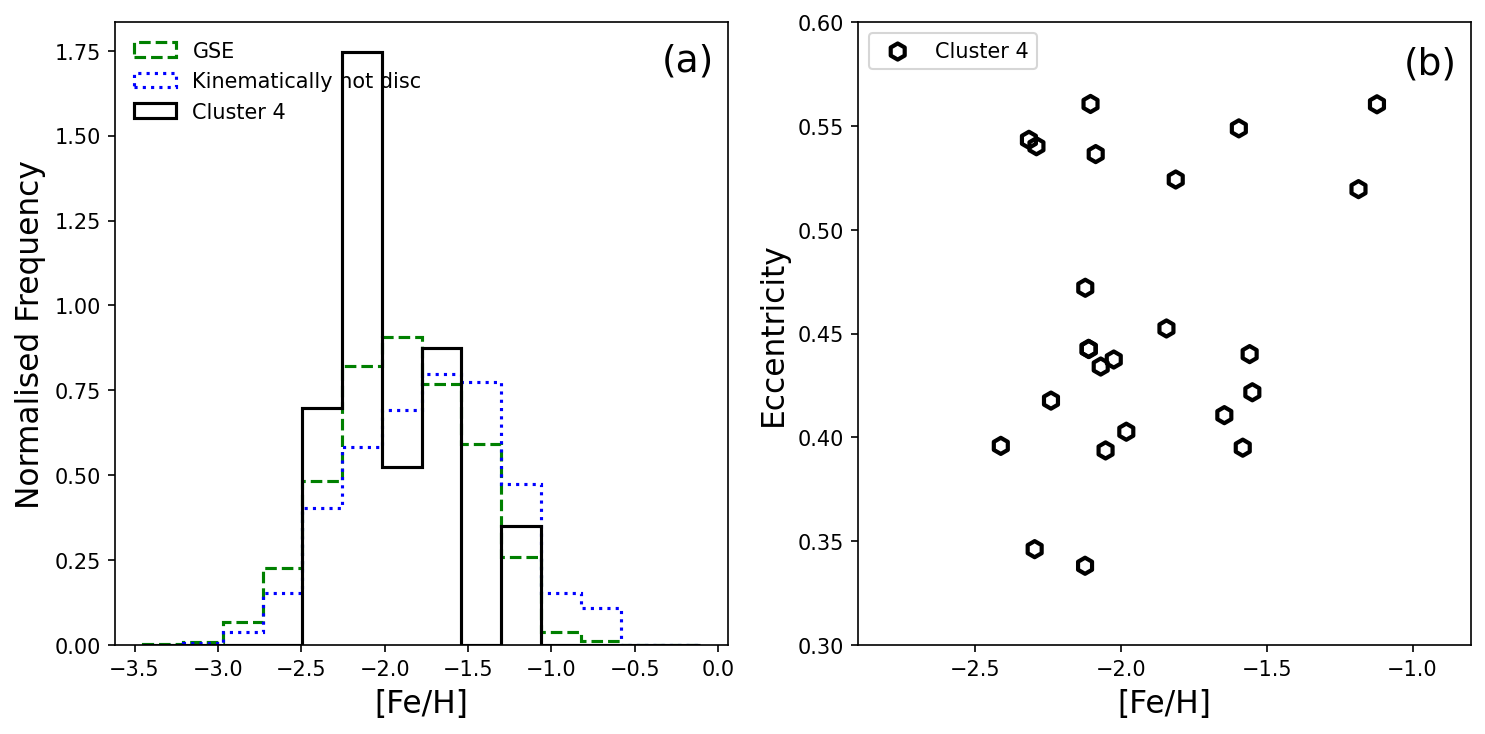}
    \caption{(a) Comparison of normalized metallicity histograms between GSE, the kinematically hot disc, and Cluster 4. (b) $e$-[Fe/H] distribution of stars in Cluster 4. Stars associated with GSE and the kinematically hot disc were identified through phase-space clustering and subsequently assigned to their respective components based on kinematic criteria. For both panels, only stars with [Fe/H] uncertainties smaller than 0.2 from both the machine learning and template fitting methods, and with a discrepancy between the two methods less than 0.2 are included.}
    \label{fig15}
\end{figure*}

\begin{figure}
   \centering
    \includegraphics[width=0.6\textwidth]{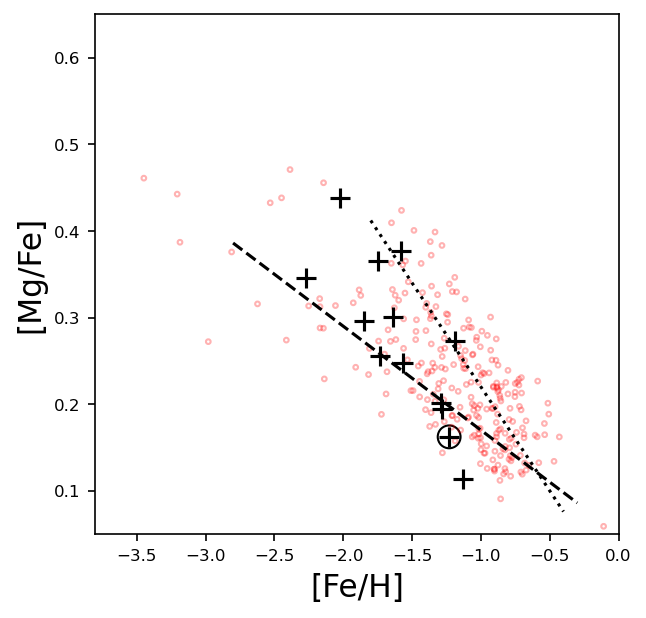}
    \caption{[Mg/Fe]-[Fe/H] distribution of stars in Cluster 5 and the kinematically cold disc, represented by plus signs and circles, respectively. Auxiliary lines from Fig. \ref{fig5} (b1) are also included to illustrate the two distinct trends. One star in Cluster 5 with available age information is marked by a black circle.}
    \label{fig16}
\end{figure}

\begin{figure*}
   \centering
    \includegraphics[width=1.0\textwidth]{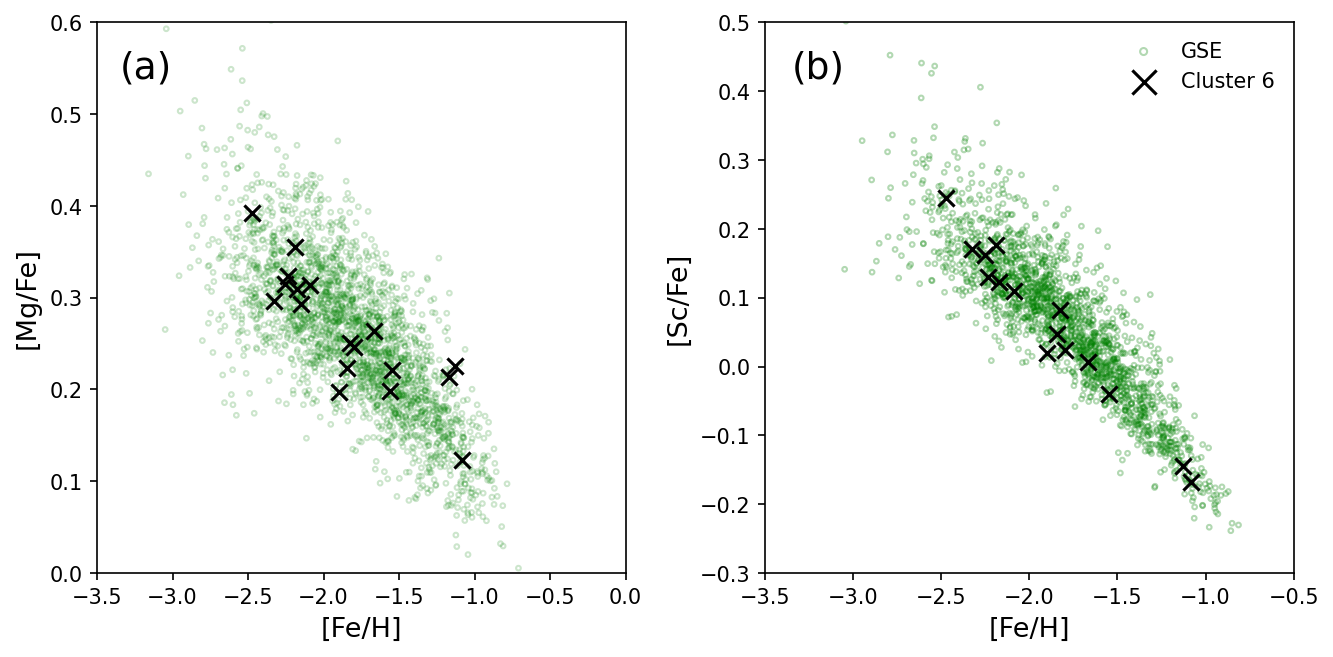}
    \caption{(a) [Mg/Fe]-[Fe/H] distribution of Cluster 6 and GSE. (b) [Sc/Fe]-[Fe/H] distribution of Cluster 6 and GSE. Each cross or circle represents a star.}
    \label{fig17}
\end{figure*}

The kinematics of stars in Cluster 4 are intermediate between GSE sample and the kinematically hot disc sample, but their metallicity is generally lower than both, as shown in Fig. \ref{fig15} (a). It is important to note that the metallicity histogram for Cluster 4 stars is based on a sample of 24 stars. The mean and median metallicities for Cluster 4 are [Fe/H] = -1.93 and -2.06, respectively, while the mean and median metallicities for GSE are [Fe/H] = -1.89 and -1.89, and for the kinematically hot disc, they are [Fe/H] = -1.73 and -1.7, respectively. Additionally, Fig. \ref{fig15} (b) reveals a positive correlation between eccentricity and [Fe/H] for Cluster 4 stars. For Cluster 5, [Mg/Fe]-[Fe/H] distributions reveal two distinct sequences resembling those of kinematically cold disc sample, as shown in Fig. \ref{fig16}. Stars in Cluster 6 show large $|V_{r}|$, similar to GSE, prompting a comparison of their [Mg/Fe]-[Fe/H] and [Sc/Fe]-[Fe/H] distributions with GSE in Fig. \ref{fig17}. Cluster 6 shows an $\alpha$-knee at [Fe/H] $\approx$ -1.3, consistent with GSE. In the [Sc/Fe]-[Fe/H] plane, both Cluster 6 and GSE exhibit linear trends, but Cluster 6 has smaller [Sc/Fe] dispersion at a given [Fe/H]. In Figs \ref{fig16} and \ref{fig17}, the disc and GSE samples are drawn from their respective DTGs. A relaxed error screening is applied to Mg, Fe, and Sc abundances in these figures: stars are retained only if the machine learning–derived uncertainties in [Mg/H] and [Fe/H] are less than 0.2, and the uncertainty in [Sc/H] is less than 0.15.

\section{Discussion}\label{sec4}
\subsection{Formation history of the Galactic metal-poor disc}\label{sec4.1}
\subsubsection{Formation process}\label{sec4.1.1}
We suggest a possible formation scenario that can account for the above observed phenomena, and we summarize it in four stages:
\begin{enumerate}[label=(\roman*), leftmargin=0em, align=left] 
\setlength{\itemindent}{1em} 
    \item Primordial phase: Fig. \ref{fig7} and Fig. \ref{fig8} indicates that the kinematically cold disc sample contains stars older than 12 Gyr, which we identify as the Galactic primordial disc. Although we believe the merger between the progenitor of GSE and the Milky Way caused significant contamination of the disc by stars from GSE's progenitor, we argue that the kinematically cold disc sample older than 12 Gyr are not the result of this contamination. Since these stars have higher metallicity than GSE stars of the same age, and lack the distinct $\alpha$-knee at [Fe/H] $\approx$ -1.3. \cite{helmi_merger_2018} noted that the mass of GSE's progenitor is about one-fourth of the Milky Way's, explaining why the primordial disc has higher metallicity, and smaller [Mg/Fe] dispersion at a given [Fe/H]. The Milky Way's larger mass led to a higher star formation rate and faster metallicity enrichment at early stage, whereas the smaller mass of GSE's progenitor and slower star formation caused its metallicity to enrich more slowly. Additionally, the Milky Way’s greater mass ensured more uniform gas distribution, whereas the progenitor of GSE likely experienced multiple star formation events, resulting in less uniform gas and greater [Mg/Fe] dispersion. In fact, based on our sample, there should also be primordial Galactic disc stars present in our kinematically hot disc sample. As shown in Fig. \ref{fig7}, the kinematically hot disc contains some members older than 12 Gyr. However, these kinematically hot disc stars with ages greater than 12 Gyr span a wide range of metallicities, some have [Fe/H] $>$ -1.5, while others are more metal-poor, even with [Fe/H] $<$ -2.5. In Fig. \ref{fig8}, if we consider the metallicity histogram in panel (a4) as representative of primordial disc stars, and panel (c4) as that of stars from the progenitor of GSE, then the histogram in panel (b4) can be well described as a combination of (a4) and (c4) in certain proportions (see Section \ref{sec4.1.3} for details). This indicates the presence of primordial disc stars in our kinematically hot disc sample. Therefore, we suggest that the primordial Galactic disc may not have been “thin”.
    
    \item Merger phase: We infer that the merger between the GSE progenitor and the Milky Way occurred approximately 10–12 Gyr ago. This inference is mainly based on the evolution of metallicity and [Mg/Fe] in the GSE sample, along with the metallicity trends observed in the kinematically hot disc. Fig. \ref{fig18} presents the metallicity histograms for both the kinematically hot disc and GSE samples in the 10–12 Gyr and $>$12 Gyr groups. Unlike in Fig. \ref{fig8}, Fig. \ref{fig18} includes soame key statistical indicators such as the mean and median metallicities. Furthermore, we apply kernel density estimation (KDE) to model the metallicity distribution of the four subsamples. The KDE fits are shown as dashed lines, with the corresponding peak [Fe/H] values annotated.

    The first line of observational evidence is that the 10–12 Gyr GSE sample exhibits higher metallicities compared to the older ($>$12 Gyr) GSE population. This trend is consistently reflected in the median, mean, and peak values of the [Fe/H] distributions. Moreover, the $>$12 Gyr GSE sample contains a significantly larger fraction of very metal-poor stars ([Fe/H] $<$ –2.0), whereas such stars are rare in the 10–12 Gyr age group.

    The second piece of evidence is that stars in the 10–12 Gyr GSE sample show higher [Mg/Fe] compared to those in the $>$12 Gyr sample. As shown in Fig. \ref{fig9}, [Mg/Fe] decreases with increasing [Fe/H] among stars older than 12 Gyr, with a noticeable $\alpha$-knee appearing at [Fe/H] $\approx$ –1.3. In contrast, the 10–12 Gyr GSE group does not show a distinct $\alpha$-knee. Instead, for [Fe/H] $<$ –1.5, these younger stars display elevated [Mg/Fe] relative to their older counterparts. In panels (c3) and (c4) of Fig. \ref{fig9}, the $>$12 Gyr stars are roughly symmetrically distributed around the dashed reference line, whereas the majority of stars in the 10–12 Gyr group lie above the line.

    The third line of evidence comes from the kinematically hot disc. As discussed in Section \ref{sec3.1}, it superficially appears that the metallicity of the kinematically hot disc increases from the $>$12 Gyr group to the 10–12 Gyr group. This impression arises because the older group contains a substantial fraction of very metal-poor stars, while such stars are nearly absent in the 10–12 Gyr group. Additionally, as seen in Fig. \ref{fig18}, both the median and mean [Fe/H] values increase from the $>$12 Gyr group to the 10–12 Gyr group. However, we argue that this apparent metallicity increase does not reflect genuine chemical evolution. Instead, the $>$12 Gyr kinematically hot disc is likely a composite population, consisting of (1) stars formed in the early Milky Way, and (2) stars originally from the GSE progenitor that now exhibit disc-like kinematics. In Section \ref{sec4.1.3}, we demonstrate that the metallicity distribution of the $>$12 Gyr hot disc can be reproduced as a mixture of the $>$12 Gyr GSE and cold disc samples. This suggests that the $>$12 Gyr hot disc is not a chemically homogeneous population, but rather a blend of native and accreted stars. Because GSE progenitor was less massive than the Milky Way, its stars are generally more metal-poor than those formed in-situ. As a result, their presence biases interpretations of metallicity evolution in the native disc. Based on the KDE peak values in Fig. \ref{fig18}, the peak [Fe/H] for the $>$12 Gyr hot disc is –1.1, while its shifts to –1.24 for the 10–12 Gyr group. This implies that, contrary to the initial impression, the metallicity of native disc decreases over time. In Fig. \ref{fig19}, we compare the KDE-fitted metallicity distributions of the 10–12 Gyr hot disc with the two decomposed components of the $>$12 Gyr hot disc. Relative to the native component of the older hot disc, the 10–12 Gyr population exhibits lower median, mean, and peak metallicities, further supporting a decline in metallicity within the in-situ disc over time. This conclusion hinges on the population decomposition of the $>$12 Gyr kinematically hot disc, based on a comparative analysis of the metallicity distributions among the hot disc, cold disc, and GSE samples, as detailed in Section \ref{sec4.1.3}. Previous studies suggest that stars formed in the GSE progenitor typically have lower [Al/Fe], while stars born in the Milky Way show higher [Al/Fe] \citep[e.g.,][]{feuillet_selecting_2021,belokurov_dawn_2022}. However, due to the limited spectral resolution of LAMOST, we are unable to measure [Al/Fe] for the stars in our sample and thus cannot provide direct chemical evidence for this decomposition. Nevertheless, other studies have identified metal-poor stars in the Galactic disc with [Al/Fe] $<$ 0 \citep[e.g.,][]{malhan_shiva_2024,fiorentin_icarus_2024}. It is important to note that the metallicity statistics presented in Fig. \ref{fig18} and \ref{fig19} may show slightly discrepancies. This is because the values in Fig. \ref{fig18} are derived from the observed stellar samples, whereas those in Fig. \ref{fig19} are based on KDE-fitted functions.
\begin{figure*}
   \centering
    \includegraphics[width=1.0\textwidth]{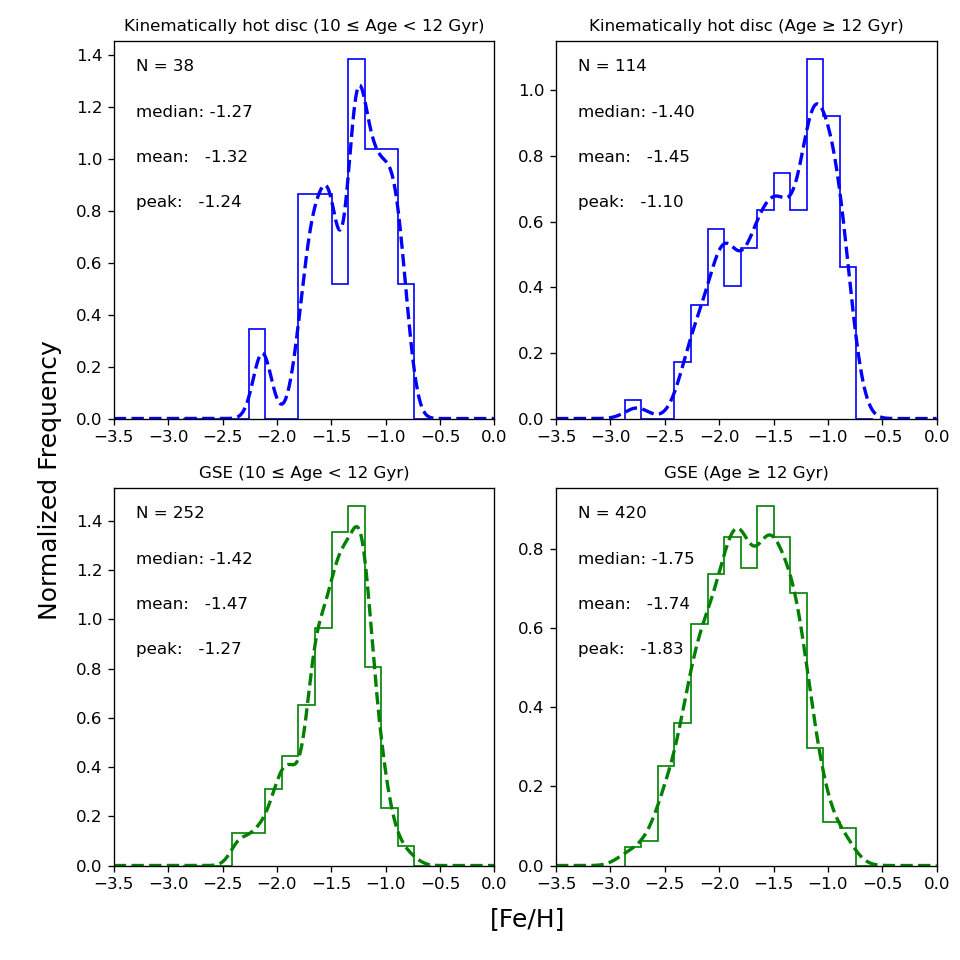}
    \caption{Normalized [Fe/H] histograms for the kinematically hot disc and Gaia-Sausage-Enceladus (GSE) subsamples, shown for the 10–12 Gyr and $>$12 Gyr age groups only. Each panel lists the number of stars (N), along with the median, mean, and KDE peak [Fe/H] values. Solid lines show the histograms; dashed lines indicate kernel density estimation (KDE) fits.}
    \label{fig18}
\end{figure*}
\begin{figure*}
   \centering
    \includegraphics[width=1.0\textwidth]{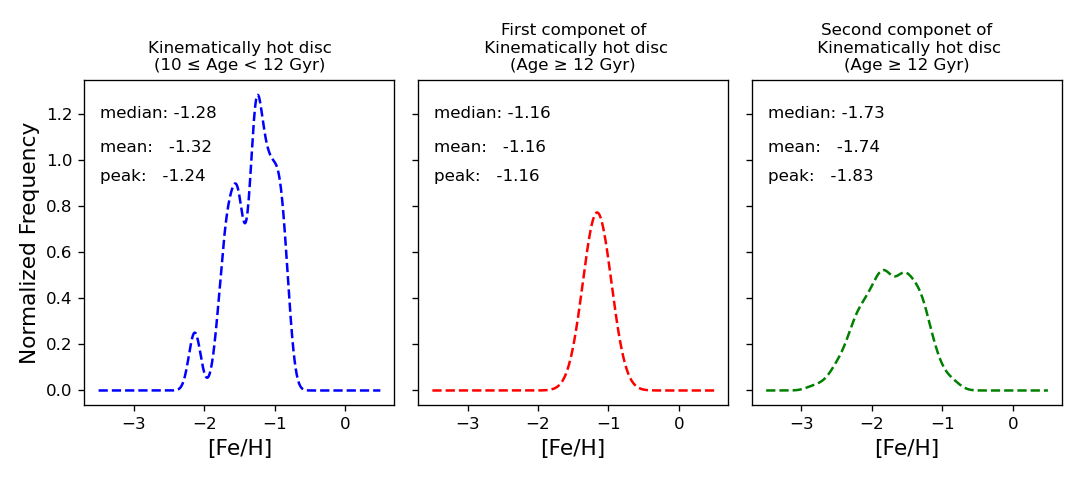}
    \caption{KDE-fitted [Fe/H] distributions for the kinematically hot disc with ages 10–12 Gyr (left) and the two decomposed components of the $>$12 Gyr kinematically hot disc (middle and right). The $>$12 Gyr hot disc is modeled as a composite population consisting of stars formed in the early Milky Way (middle panel) and stars originally from the GSE progenitor (right panel). The decomposition process is described in Section \ref{sec4.1.3}. Each panel reports the median, mean, and peak [Fe/H] values derived from the fitted KDE distributions.}
    \label{fig19}
\end{figure*}

    These phenomenons can be explained by the interaction between the GSE progenitor and the Milky Way. Since GSE progenitor had a lower mass than the Milky Way, its gas should been correspondingly more metal-poor. During their interaction, gas from both systems likely mixed,  producing an interstellar medium with intermediate metallicity, higher than that of the GSE progenitor, but lower than that of the Milky Way. Stars formed from this mixed gas would reflect these metallcity properties. In particular, stars with highly eccentric orbits would appear more metal-rich than older GSE stars. At the same time, disc stars formed from this mixed gas would be more metal-poor than the older, in-situ disc stars. The interaction likely triggered a burst of star formation, leading to creation of many massive stars. These massive stars, with their short lifespans, quickly ended their lives as Type II supernovae, releasing large amounts of $\alpha$ elements such as magnesium into the surrounding interstellar medium. This rapid enrichment would result in elevated [Mg/Fe] in stars formed during the merger event. It is important to note that [Mg/Fe] is also influenced by iron production, which predominantly come from Type Ia supernovae. In the early stages of galaxy evolution, before Type Ia supernovae had significantly enriched the environment with iron, [Mg/Fe] would naturally remain high. However, in our case, we observe that the 10–12 Gyr GSE stars are not only more metal-rich but also show higher [Mg/Fe] compared to the older ($>$12 Gyr) GSE population. If no starburst had occurred to enhance magnesium, one would expect the younger, more metal-rich GSE stars to exhibit lower [Mg/Fe] than their older counterparts. The fact that this is not observed indicates the need for an additional source of magnesium. The most plausible explanation is a burst of star formation triggered by the merger between GSE and the Milky Way.

    This timescale is also consistent with previous studies. \cite{helmi_merger_2018} estimated the GSE merger at 10 Gyr ago, and \cite{belokurov_co-formation_2018} suggested a range of 8–11 Gyr. Similarly, \cite{ciuca_chasing_2023} identified a population of stars aged 10-12 Gyr that are more [Mg/Fe]-rich than older stars, attributing this to starbust by GSE merger. \cite{aguado-agelet_cluster_2025} found two groups of GSE-associated globular cluster: one around $\sim$13 Gyr and [Fe/H] $\approx$ –1.6, likely formed before the merger, and another around $\sim$11.5 Gyr and [Fe/H] $\approx$ –1.1, likely formed during the merger process. \cite{vincenzo_fall_2019} also concluded, via chemical evolution modeling, that the merger occured about 10 Gyr ago.
    
    In conclusion, the GSE merger had a profound impact on the Milky Way. First, it introduced stars from the GSE progenitor into the Galactic disc. Our previous analysis shows that kinematically cold disc stars older than 12 Gyr are largely uncontaminated by GSE debris, as their [Fe/H] is much higher than that of GSE stars of the same age. Thus, the kinematically cold disc stars in this age range can represent the metallicity of the Galactic primordial disc. In contrast, the stars from the GSE progenitor that polluted the primordial disc are found mainly in our kinematically hot disc sample. Therefore, kinematically hot disc stars older than 12 Gyr include both primordial disc stars and GSE-origin stars. This explains why the metallicity distribution of kinematically hot disc stars older than 12 Gyr falls between that of the cold disc and the GSE at the same age. Second, the merger brought metal-poor gas into the Milky Way, triggering a starburst that formed a new generation of metal-poor disc stars. Because this mixed gas was more metal-poor than the original Galactic gas, kinematically hot disc stars formed 10–12 Gyr are more metal-poor than that of older in-situ disc stars.
    
    \item Quiescent phase: Following the GSE merger, the Milky Way’s disc entered a quiescent phase 8-10 Gyr ago, marked by slow gas accretion and reduced star formation. This is evidenced by the significantly fewer stars aged 8-10 Gyr compared to 10-12 Gyr, the absence of Mg-enhanced stars  in the 8-10 Gyr group.

    \item Latest formation phase: From 8 Gyr ago to the present, the formation of the Milky Way’s disc proceeded independently of the GSE merger. Stars younger than 8 Gyr in the kinematically cold disc and hot disc exhibit a steep decline in [Mg/Fe] with increasing [Fe/H], more pronounced than in other age groups and lacking an $\alpha$-knee. This pattern suggests an early phase of $\alpha$-element enhancement, followed by a gradual enrichment in iron over time. The missing $\alpha$-knee might occur at very low [Fe/H] (below -2.0). An outlier star in Fig. \ref{fig9} (a1) with [Fe/H] = -2.8 and [Mg/Fe] = 0.42 could suggest a knee between [Fe/H] = -3 and -2, but this single star might be unrelated to this latest formed disc.
\end{enumerate}
In summary, the metal-poor disc of the Milky Way primarily consists of four components: the primordial Galactic disc, stars originating from GSE's progenitor, stars formed from the mixed gas of GSE's progenitor and the ancient Milky Way, and stars formed within the last 8 Gyr. We believe these recently formed stars represent the metal-poor tail of the otherwise metal-rich, low-$\alpha$ disc, and we will explain this interpretation in detail in Section \ref{sec4.1.4}. This formation scenario explains the metallicity histogram of the kinematically cold disc shown in panel (a1) of Fig. \ref{fig5}, where three distinct peaks are observed. We attribute the prominent peak at [Fe/H] = -0.8 to yong, metal-poor, high-$\alpha$ stars, which we identify as the metal-poor tail of the traditional metal-rich, low-$\alpha$ disc, while the peak at [Fe/H] = -1.2 likely originates from the Galactic primordial disc. The less prominent peak at [Fe/H] = -2.0 suggests stars related to GSE merger.

\subsubsection{The primordial disc}\label{sec4.1.2}
The primordial disc formed over 12 Gyr ago with a metallicity lower limit of about -1.5. \cite{xiang_formation_2024} also identified a primordial disc but suggested a higher metallicity limit ([Fe/H] $>$ -0.5). This discrepancy likely stems from methodological differences: we directly applied a kinematic criterion to identify disc stars, while \cite{xiang_formation_2024} did not distinguish between disc and halo stars. Instead, they grouped stars by age and studied the relationship between age and morphology, defining morphology as the ratio of scale length to scale height. In their analysis, the primordial disc at [Fe/H] $<$ -0.5 is obscured by numerous GSE stars, which typically have [Fe/H] $<$ -1.0 and exhibit a halo-like geometry. \cite{xiang_formation_2024} observed disc-like structures in stars as old as 13 Gyr, with these features becoming more prominent in stars younger than 12.5 Gyr. This can be explained by our findings: disc older than 12 Gyr are primarily composed of the primordial disc stars and stars from GSE's progenitor, and Fig. \ref{fig7} suggests that our kinematically hot disc sample is slightly older than the GSE sample. Based on the above analysis, our kinematically hot disc sample consists of stars originating from both the progenitor of GSE and the Galactic primordial disc. Therefore, the Galactic primordial disc itself should be older than the GSE progenitor. At 13 Gyr, the relatively low presence of GSE stars allows for detection of weak disc-like features. However, between 12.5 and 13 Gyr, the increasing number of GSE stars significantly obscures the primordial disc, making the disc less apparent. 12.5 Gyr ago, the GSE merger triggered the growth of the disc, leading to the prominent disc-like structures observed by \cite{xiang_formation_2024}. These findings demonstrate consistency between our results and their observations. 

Several studies have provided indirect evidence for an ancient primordial disc in the Milky Way. Using common sources between APOGEE DR17 and Gaia DR3, \cite{belokurov_dawn_2022} identified Galactic in-situ stars with [Al/Fe] $>$ -0.1. These stars have [Fe/H] between -1.5 and -0.5. Stars with [Fe/H] $<$ -1.3 exhibit isotropic velocity distributions and large chemical abundance dispersion, characteristics of a local primordial halo they named 'Aurora'. Stars with -1.3 $<$ [Fe/H] $<$ -0.9 show increasing azimuthal velocity ($|V_{\phi}|$) and decreasing $V_{\phi}$ dispersion, indicating a transition from chaotic to ordered motion around redshift 2.5. Their finding of the lower [Fe/H] limit of -1.5 for early in-situ stars matches our lower limit for the primordial disc. While they observed disc-like features emerging at [Fe/H] = -1.3, slightly higher than -1.5, this difference likely stems from methodology. Their approach might ignore sparse disc-like stars between [Fe/H] -1.5 and -1.3. Furthermore, \cite{belokurov_biggest_2020} identified 'Splash', a local halo component with [Fe/H] $>$ -0.7, suggesting it originated from the heating of the Galactic primordial disc due to GSE merger. They believed Splash extends to [Fe/H] below -0.7, but most observed Splash stars have [Fe/H] $>$ -0.7, because at lower [Fe/H], Splash becomes difficult to distinguish from other halo populations such as debris from dwarf galaxies. In other words, the metallicity of the primordial disc could extend below -0.7, which is also consistent with our findings. \cite{gallart_uncovering_2019} discovered two distinct sequences in the color-magnitude diagram of the Galactic halo stars. Through analysis, they identified one sequence as GSE and the other as in-situ halo stars. Similar to the Splash \citep{belokurov_biggest_2020}, they proposed these in-situ stars come from the heated the Galactic primordial disc, due to the merger with GSE's progenitor. Their results show that both in-situ stars and GSE stars have ages primarily between 10-14 Gyr, consistent with our findings. Additionally, they found that in-situ stars generally have higher metallicities than GSE stars, with a lower metallicity limit of -1.7 for in-situ stars, which broadly agrees with our -1.5 lower limit. Although they did not directly identify the primordial disc, they suggested that the primordial disc was not completely heated, with some stars maintaining their disc-like motions. These studies indicate that ancient Galactic stars generally have [Fe/H] above -1.5, and suggest a possible primordial disc. Regarding the in-situ halo, some propose it contains both heated primordial disc stars and primordial halo stars \citep{belokurov_dawn_2022}, while others argue it formed entirely from the heated early disc \citep{matteo_milky_2019}. Whether a primordial halo exists remains an open question.

However, some works present different views. \cite{viswanathan_slow_2024} studied the Galactic disc's formation using high-$\alpha$ stars from Gaia DR3, attempting to exclude GSE stars. They found stars with [Fe/H] $<$ -1.5 have negligible rotation ($V_{\phi}\approx0$), attributing them to GSE. For stars with -1.5 $<$ [Fe/H] $<$ -1.0, they observed an average $V_{\phi}$ of -50 $\rm{km\,s^{-1}}$, which they identify as the Galactic earliest disc. When [Fe/H] exceeds -1.0, the average $V_{\phi}$ reaches -200 $\rm{km\,s^{-1}}$, representing the high-$\alpha$ disc. Based on the linear relationship between age and metallicity for high-$\alpha$ stars, they concluded the transition from the earliest disc to high-$\alpha$ disc was gradual. These conclusions differ from our results in two major aspects. First, regarding the earliest disc characteristics, they suggest a modest rotation ($V_{\phi}\approx-50\,\rm{km\,s^{-1}}$), while we find significant rotational motion ($V_{\phi}<-100\,\rm{km\,s^{-1}}$). We attribute this discrepancy to two factors: (a) their sample includes GSE stars. As shown in Fig. \ref{fig5}, GSE stars can have [Mg/Fe] lager than 0.3 in -1.5 $<$ [Fe/H] $<$ -1.0, resulting in a lower average $|V_{\phi}|$ due to the mixture of primordial disc and GSE stars; (b) our results show that although most stars from GSE's progenitor have highly eccentric orbits, some can exhibit disc-like orbits, making them particularly difficult to distinguish from in-situ disc stars through kinematics. Second, regarding the evolution from primordial disc to high-$\alpha$ disc, they propose a slow, gradual transition, while we argue for an abrupt change. This second point will be discussed in detail in Section \ref{sec4.1.3}.

Regarding the morphology of the primordial disc, our classification criteria for kinematically hot and cold discs show that the primordial disc exists in both. This suggests that the primordial disc of the Milky Way is not a very thin, kinematically cold disc.

\subsubsection{Impact of GSE merger}\label{sec4.1.3}
Gaia-Sausage-Enceladus (GSE) is a prominent merger remnant in the Galactic halo, notable for stars with nearly zero azimuthal velocity ($V_{\phi}$) and high radial velocity ($V_{r}$) \citep{belokurov_co-formation_2018,helmi_merger_2018}. This study reveals that the merger between the Milky Way and GSE's progenitor not only contributed numerous stars into the halo but also profoundly influenced the Galactic disc's evolution. The merger effected the disc in two major ways. First, it contaminated the primordial disc, as detailed in Section \ref{sec4.1.1}, introducing many stars from the GSE's progenitor into the Galactic primordial disc. While GSE stars are typically associated with highly eccentric orbits, \cite{koppelman_messy_2020}, through the merger simulation, showed that 9\% of stars from GSE's progenitor could exhibit eccentricities below 0.6, and some have $V_{\phi}$ reach -170 $\rm{km\,s^{-1}}$. This simulation-based result indicates that some stars from GSE's progenitor may have settled into disc-like orbits. Furthermore, \cite{vasiliev_radialization_2022} demonstrated that the orbits of satellite galaxies can become more circular during mergers under specific conditions.

The term GSE is often used inconsistently in the literature, sometimes referring to the dwarf galaxy that merged with the Milky Way and other times to the resulting remnant. As most studies associate this remnant with the Galactic halo, GSE often implies halo-like kinematic characteristics. However, this study shows that the merger also contributed stars to the disc. To clarify, this work defines GSE specifically as the stars with high eccentricities.

\begin{figure*}
   \centering
    \includegraphics[width=1.0\textwidth]{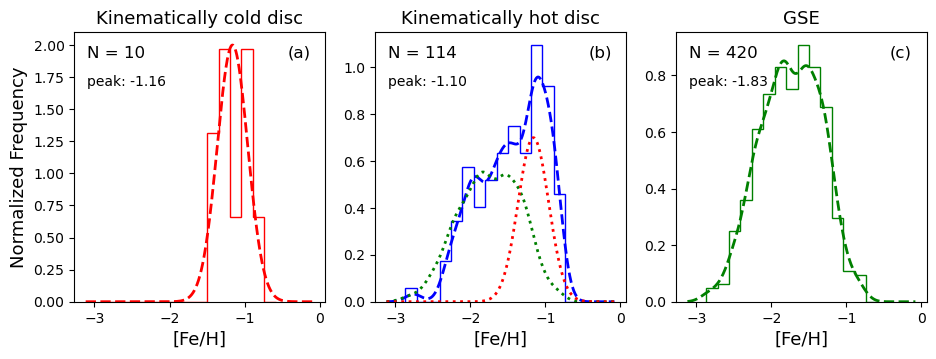}
    \caption{Solid lines show the normalized metallicity histograms for the kinematically cold disc (a), kinematically hot disc (b), and GSE (c) stars older than 12 Gyr, with the number of stars indicated in the top-left corner of each panel. Dashed lines represent the fitted distribution functions, with the peak [Fe/H] values indicated on the plots. In panel (b), the red dotted line corresponds to 0.31 times the kinematically cold disc fit, and the green dotted line corresponds to 0.69 times the GSE fit.
}
    \label{fig20}
\end{figure*}

Based on our analysis of Fig. \ref{fig8}, we identify that our kinematically cold disc stars older than 12 Gyr are entirely composed of the in-situ primordial disc stars. These stars exhibit significantly higher metallicities than the GSE stars of the same age, indicating that our kinematically cold disc could not have been contaminated by accreted stars from the GSE progenitor. In contrast, the kinematically hot disc stars older than 12 Gyr show a broad distribution in metallicity. Some stars are relatively metal-poor, with [Fe/H] values as low as –3.0, consistent with the chemical properties of the GSE sample of the same age. Meanwhile, others have comparatively higher metallicities around [Fe/H] $\approx$ –1.0, similar to those found in our cold disc of the same age. This wide metallicity spread suggests that the kinematically hot disc in this age range consists of two distinct stellar populations: (1) in-situ stars belonging to the Galactic primordial disc, and (2) stars accreted from the GSE progenitor. In summary, the Galactic primordial disc is present in both our kinematically cold and hot disc samples older than 12 Gyr. Crucially, our kinematically hot disc older than 12 Gyr is composed exclusively of two components: the Galactic primordial disc stars and accreted stars from the GSE progenitor. No additional stellar populations are needed to explain the observed chemical abundance patterns. To evaluate how strongly the merger contaminated the Galactic primordial disc, we modeled the metallicity distribution of our kinematically hot disc sample older than 12 Gyr as a combination of the distributions of our kinematically cold disc and GSE samples of the same age:
\begin{equation}
    f_{\rm{hot}}=\beta \times f_{\rm{cold}} + (1-\beta) \times f_{\rm{GSE}} 
\end{equation}
Here, $f_{\rm{hot}}$, $f_{\rm{cold}}$, and $f_{\rm{GSE}}$ are the metallicity distribution functions for the kinematically hot disc, cold disc, and GSE, respectively. These distribution functions were obtained by fitting the observed [Fe/H] histograms: we used kernel density estimation (KDE) for our kinematically hot disc sample and our GSE sample, and a Gaussian function for our kinematically cold disc sample. Fig. \ref{fig20} displays the observed [Fe/H] histograms alongside the corresponding fitted distribution functions. From this modeling, we derive a best-fit value of $\beta=0.39$, suggesting that 39\% of the [Fe/H] distribution of the kinematically hot disc matches that of the kinematically cold disc, while remaining 61\% aligns with our GSE sample. Given that the kinemaically hot disc sample contains 114 stars, we estimate that approximately 70 of them originated from the progenitor of GSE. Assuming that these GSE-origin stars now residing in the disc share the same metallicity distribution function as the GSE-origin stars currently found in the Galactic halo, and noting that our GSE sample with age older than 12 Gyr has 420 stars, we infer that 14.3\% of stars from the GSE progenitor were deposited into the disc, while the remaining 85.7\% staying in the Galactic halo. \cite{lane_stellar_2023} estimated the stellar mass of the GSE progenitor to be $1.45\times10^{8}$ $M_{\odot}$ (where $M_{\odot}$ is the solar mass) but their analysis did not account for GSE-origin disc-residing stars. Based on our infered fraction, we estimate the stellar mass of GSE-origin stars that ended up in the Galactic disc to be approximately $\sim2.42\times10^{7}$ $M_{\odot}$.

More importantly, the merger triggered the formation of the so-called metal-rich, high-$\alpha$ disc (thick disc). Three key observations support this: (1) a significant fraction of stars in our sample formed 10–12 Gyr ago, consistent with the metal-rich, high-$\alpha$ disc formation timescale proposed by many studies \citep[e.g.,][]{haywood_age_2013,xiang_time-resolved_2022}; (2) stars aged 10–12 Gyr show enhanced [Mg/Fe], indicating a violent starburst during this period; (3) for stars aged 10-12 Gyr, the majority exhibit kinematic properties consistent with our kinematically hot disc criterion, while the fraction meeting our kinematically cold disc criterion is relatively low. Previous studies also support this conclusion. \cite{belokurov_co-formation_2018} suggested that the merger likely induced the formation of the Galactic thick disc. Similarly, \cite{ciuca_chasing_2023} identified a teardrop-shaped region in the [Fe/H]-Age plane for stars aged 9–12 Gyr, termed the 'Dip'. Compared to stars older than 12 Gyr, these stars show a sharp drop in [Fe/H] and a steep rise in [Mg/Fe]. They concluded that the merger between the Milky Way and the GSE's progenitor brought in metal-poor gas, triggering a starburst that formed the high-$\alpha$ disc. Their findings align closely with ours.

\subsubsection{Metal-poor tail of metal-rich, low-$\alpha$ disc}\label{sec4.1.4}
We argue that disc stars with [Fe/H] $<$ -0.8 and age $<$ 8 Gyr represent the metal-poor tail of the metal-poor, low-$\alpha$ disc based on three key observations: First, their formation timing aligns with the metal-rich, low-$\alpha$ disc, which began forming 8 Gyr ago \citep{haywood_age_2013}. Second, their spatial distribution and kinematic properties matches metal-rich, low-$\alpha$ disc (thin disc) characteristics, as discussed in Section \ref{sec3.1}. Third, their [Mg/Fe]-[Fe/H] slope matches the metal-rich, low-$\alpha$ disc pattern. Extrapolating our fitted line yields [Mg/Fe] values of 0.11, 0.07, and 0.05 at [Fe/H] = -0.5, -0.2, and 0, respectively, consistent low-$\alpha$ characteristics at metal-rich end. Our findings differ somewhat from those of \cite{haywood_age_2013}, particularly regarding the early stages of low-$\alpha$ disc formation. \cite{haywood_age_2013} suggested that the inner low-$\alpha$ disc ($r_{\rm{gal}}<10\,\rm{kpc}$) initially had relatively high metallicity (-0.1 to 0.1) and low $\alpha$-abundance, inheriting the chemical composition of the metal-rich, high-$\alpha$ disc's late-stage gas, followed by a period of slow evolution. In contrast, our results indicate that the low-$\alpha$ disc began with much lower metallicity, possibly at least as low as -1.8, and experienced a brief period of rapid star formation, which significantly increased the [Mg/Fe]. Only after this phase did the low-$\alpha$ disc enter a slow evolutionary period, starting from [Fe/H] $\approx$ -1.8 and [Mg/Fe] $\approx$ 0.4. The mechanism that triggered the initial rapid star formation in the so-called metal-rich, low-$\alpha$ disc (thin disc) remains unclear. This suggests that there is no clear inheritance relationship between the so-called metal-rich, low-$\alpha$ and so-called metal-rich, high-$\alpha$ discs, but rather significant differences in their formation mechanisms. Some studies suggest that the thin disc's metallicity could be as low as -2.0 \citep{fernandez-alvar_pristine_2021,fernandez-alvar_metal-poor_2024}, but without age information, it is uncertain whether these stars are younger or older than 8 Gyr.

These stars exhibit relatively low [Fe/H] and elevated [Mg/Fe], yet they are remarkably young and display kinematically cold disc characteristics - a rather unusual combination. One possible explanation is that their ages have been underestimated due to observational or modeling uncertainties, such as unresolved binarity, inaccurate stellar parameters, or limitations in age-dating methods. In Section \ref{sec3.1}, we investigate whether these stars could be members of unresolved binary systems, but found insufficient evidence to support this scenario. Of course, this does not constitute definitive proof of their youth. We therefore strongly look forward to further observations, such as asteroseismic measurements and high-resolution spectroscopy, to better constrain the nature of these stars. Based on the currently available information, we propose that these stars may represent the metal-poor tail of the metal-rich, low-$\alpha$ disc population. In addition, compared to older stars, they exhibit higher [Mg/Fe] at a given [Fe/H], suggesting that they have experienced $\alpha$-enrichment. In our sample, the number of such young, metal-poor, high-$\alpha$ stars is small, and they do not appear to be associated with intense star formation activity. Consequently, the mechanism responsible for their $\alpha$-enrichment remains unclear.

\subsubsection{Comparative analysis of metal-poor disc}\label{sec4.1.5}
The origin of the Galactic metal-poor disc has been widely debated, with various scholars offering different perspectives. Here, we compare our findings with those in the literature.

\cite{naidu_evidence_2020} identified a group of stars, named Aleph, with $V_{\phi}$ $<$ -170 $\rm{km\,s^{-1}}$ and $|V_{r}|$ $<$ 70 $\rm{km\,s^{-1}}$. These stars have [Fe/H] $>$ -0.8, [$\alpha$/Fe] $<$ -0.27, and a wide $|z|$ distribution, reaching up to 10 kpc. \cite{horta_chemical_2023} analyzed Aleph's chemical abundances, finding its properties intermediate between metal-rich, high-$\alpha$ and metal-rich low-$\alpha$ discs but closer to the low-$\alpha$ disc. They suggested Aleph represents disc stars perturbed by the Sagittarius. We agree with this interpretation but since the low-$\alpha$ disc extends to [Fe/H] $\approx$ -1.8, Aleph likely includes lower-metallicity members, though these may be obscured by halo stars at large distances from the mid-plane.

\cite{fiorentin_icarus_2021} discovered Icarus, a group of metal-poor stars with average [Fe/H] $\approx$ -1.35 and disc-like kinematics. Initially, Icarus was thought to be a low-$\alpha$ structure with an average [Mg/Fe] $\approx$ -0.02. However, subsequent observations revealed the average [Mg/Fe] $\approx$ 0.27 \citep{fiorentin_icarus_2024}. Their detailed comparison of chemical abundances between Icarus and GSE revealed strong similarities, suggesting Icarus is the relic of a dwarf galaxy. While they did not explicitly propose that Icarus come from the progenitor of GSE, our work shows that some stars with disc-like kinematics belonging to GSE's progenitor, supporting the idea that Icarus and GSE originated from the same merger event.

\cite{naidu_evidence_2020} defined the Metal-Weak Thick Disc (MWTD) as stars with disc-like kinematics, [Fe/H] $<$ -0.8, and [$\alpha$/Fe] $>$ 0.25, suggesting it represents the metal-poor tail of the thick disc (the metal-rich high-$\alpha$ disc). Our analysis suggests MWTD is multi-component, including the metal-poor tail of the metal-rich, low-$\alpha$ disc (thin disc), stars from the progenitor of GSE, and the metal-poor tail of metal-rich high-$\alpha$ disc (thick disc).

\cite{malhan_shiva_2024} discovered two metal-poor structures in $E$–$L_{z}$ space: Shakti ($L_z \approx 1000~\mathrm{kpc\,km\,s^{-1}}$) and Shiva ($L_z \approx 300~\mathrm{kpc\,km\,s^{-1}}$), and found that some stars in these structures have [Al/Fe] $>$ 0 while others have [Al/Fe] $<$ 0. Since Galactic stars typically have [Al/Fe] $>$ 0 and GSE stars [Al/Fe] $<$ 0, and given these stars are highly clustered in $E$–$L_{z}$ space as well as their [Fe/H] $<$ –1, they concluded that Shakti and Shiva are relics of dwarf galaxies unrelated to GSE. However, we argue that these structures result from the mixing of GSE stars and the Galactic primordial disc stars. This mixing naturally explains the observed [Al/Fe] variations, with stars in both structures exhibiting [Al/Fe] $>$ 0 and [Al/Fe] $<$ 0.

\cite{nepal_discovery_2024} proposed that the existence of an 'old thin disc' in the Milky Way based on their study of old disc stars ([Fe/H] $<$ -1.0, $z_{max}$ $<$ 1 kpc). Unlike the common referred thin disc, which is low-$\alpha$ and formed within last 8 Gyr, their 'old thin disc' is ancient (13–14 Gyr old) and primarily consists of high-$\alpha$ stars. They suggested that metal-poor disc stars originated from this 'old thin disc', which was later heated during the GSE merger. They also identified low-$\alpha$ stars older than 10 Gyr within this disc, leading to the hypothesis that it underwent different evolutionary phases. However, we disagree with their assumption that all metal-poor disc stars originate from this 'old thin disc', as we believe their formation occurred in multiple stages, resulting in a diverse composition. Our proposed formation process explains several phenomena observed by \cite{nepal_discovery_2024}, such as the single age peak at 13.5 Gyr for stars with $|V_{\phi}|$ $>$ 180 $\rm{km\,s^{-1}}$ and two peaks at 12 Gyr and 13.5 Gyr for stars with $|V_{\phi}|$ $<$ 180 $\rm{km\,s^{-1}}$. This aligns with our hypothesis that contamination from GSE merger primarily affected the warmer part of the Galactic primordial disc, leaving the colder part largely intact. The single age peak for stars with $|V_{\phi}|$ $>$ 180 $\rm{km\,s^{-1}}$ corresponds to the less polluted primordial disc, while two age peaks at 12 Gyr and 13.5 Gyr for stars with $|V_{\phi}|$ $<$ 180 $\rm{km\,s^{-1}}$ reflects merger-polluted disc and the primordial disc, respectively. Additionally, \cite{nepal_discovery_2024} found that disc stars older than 13 Gyr span [Fe/H] from -2.5 to -0.5, with metal-rich stars dominating, consistent with our conclusion that the primordial disc has [Fe/H] $>$ -1.5 and was polluted by more metal-poor stars from GSE's progenitor. Furthermore, we did not observe low-$\alpha$ disc stars older than 10 Gyr and believe the formation of the low-$\alpha$ disc is unrelated to the ancient primordial disc.

\subsection{Properties of merger remnants}\label{sec4.2}
Pontus, identified by \cite{malhan_global_2022}, is a merger remnant with low orbital energy and slightly retrograde motion, exhibiting kinematics close to GSE and Thamnos. Their study reports that Pontus has a metallicity distribution function (MDF) spanning [Fe/H] from -2.3 to -1.3, with a median of [Fe/H] = -1.7, while GSE exhibits a broader metallicity range and a significant population of stars with [Fe/H] $>$ -1.5. \cite{malhan_global_2022} also suggested that Thamnos is more metal-rich, with [Fe/H] ranging from -1.4 to -1.1, and proposed that Pontus originated from a distinct merger event compared to GSE and Thamnos. In this work, we also find that Pontus has a lower proportion of stars with [Fe/H] $>$ -1.5 compared to GSE. However, we show that Thamnos can have [Fe/H] as low as -3.0, contradicting the claim that Thamnos is definitively more metal-rich. 

Wukong (also known as LMS-1), was first identified by \cite{yuan_low-mass_2020} and later independently discovered by \cite{naidu_evidence_2020}. This remnant exhibits mild prograde rotation and intermediate/high orbital energy. Its origin, whether as part of GSE or from a separate dwarf galaxy, remains uncertain. \cite{naidu_evidence_2020} and \cite{horta_chemical_2023} analyzed Wukong's chemical properties using stars from H3 and APOGEE surveys, respectively, and both constrained its metallicity to [Fe/H] $<$ -1.45. \cite{naidu_evidence_2020} found Wukong's [Fe/H] distribution similar to GSE's but with multiple peaks, unlike GSE's single peak. \cite{horta_chemical_2023} observed chemical similarities across multiple elements and suggested a shared origin for Wukong and GSE. In contrast, \cite{zhang_four-hundred_2024} and \cite{limberg_extending_2024} found that Wukong's [$\alpha$/Fe] forms a plateau at [Fe/H] $<$ -2.0, unlike GSE, where [$\alpha$/Fe] decreases with rising [Fe/H]. \cite{malhan_global_2022} showed Wukong is the most metal-poor accreted remnant, far poorer than GSE, and hosting three most metal-poor stellar streams: C-19 ([Fe/H] = -3.38), Sylgr ([Fe/H] = -2.92), and Phoenix ([Fe/H] = -2.70). Given these chemical and kinematic differences, \cite{malhan_global_2022} concluded that Wukong and GSE are likely products of separate merger events. Our study confirms Wukong as the most metal-poor structure among the six accreted remnants analyzed in this paper. Considering these chemical differences and their distinct kinematic properties, where GSE has most stars with eccentricities above 0.7 while Wukong's eccentricity distribution spans 0.2 to 0.8, and GSE has a net azimuthal velocity ($V_{\phi}$) close to 0 whereas Wukong has -50 $\rm{km\, s^{-1}}$, we tend to believe that Wukong and GSE originated from different progenitor galaxies.

Sequoia, a retrograde structure identified by \cite{myeong_evidence_2019} has slightly lower metallicity than GSE and moderate orbital eccentricity ($e\approx0.6$). \cite{myeong_evidence_2019} concluded that Sequoia and GSE originated from different dwarf galaxies. However, \cite{koppelman_messy_2020} and \cite{amarante_gastro_2022} suggested that a single merger event could produce both retrograde and high-eccentricity stars, implying a shared origin. \cite{naidu_evidence_2020} discovered that the metallicity distribution of Sequoia contains multiple peaks, and therefore suggested that Sequoia consists of three distinct components, with metallicity increasing in the order of I'itoi, Sequoia, and Arjuna. They believe that Arjuna belongs to GSE, while \cite{horta_chemical_2023} highlighted that different selection criteria could lead to varying conclusions. Our study found Sequoia has slightly lower metallicity than GSE, leaving the origin question unresolved. 

\subsection{New substructures}\label{sec4.3}
We identified three new kinematic substructures in Section \ref{sec3.3}: Cluster 4, Cluster 5, and Cluster 6. The stars in Cluster 4 have $V_{\phi}$ of approximately -100 $\rm{km\,s^{-1}}$, near the edge of the kinematically hot disc, but their $|V_z|$ is much higher, generally exceeding 100 $\rm{km\,s^{-1}}$. One Cluster 4 star was found to be 14.4 Gyr old. Cluster 4 is unlikely to be a disturbed remnant of the Galactic primordial disc, as its stars generally have metallicities below -1.5, whereas the primordial disc’s metallicity exceeds -1.5. Furthermore, Cluster 4 stars exhibit a unique positive correlation between metallicity and orbital eccentricity. These findings suggest that Cluster 4 is likely the remnant of an undiscovered dwarf galaxy, which we designate as ShangGu-1.

The stars in Cluster 5 have  $V_{\phi}$ values mostly below -200 $\rm{km\,s^{-1}}$, aligning with the rotational characteristics of the kinematically cold disc, but their $|V_z|$ is unusually high, generally exceeding 100 $\rm{km\,s^{-1}}$. Their metallicity ranges from -2.27 to -1.13, with an average of -1.58. Two stars have age information: one is 3.6 Gyr old, and the other 11.2 Gyr old. The [Fe/H] error of the star aged 3.6 Gyr is larger than 0.2, so it does not appear in Fig. \ref{fig16}. Meanwhile, the star aged 11.2 Gyr is marked with a black circle in Fig. \ref{fig16}. In the [Mg/Fe]-[Fe/H] distribution, Cluster 5 show two branches just like our kinematic cold disc sample. We suggest that Cluster 5 is composed of stars from the Milky Way’s metal-poor disc, which have been kinematically heated recently, thereby acquiring significant vertical motions. We designate Cluster 5 as ShangGu-2.

The stars in Cluster 6 exhibit disc-like rotation but unusual high radial velocity ($|V_r|$ $\approx$ 200 $\rm{km\,s^{-1}}$). Among the five stars with age data, the oldest is 13.8 Gyr, the youngest 12.4 Gyr, with an average age of 13.1 Gyr and 1-$\sigma$ dispersion of 0.33 Gyr. The average [Fe/H] of Cluster 6 is -2.06, close to the peak metallicity value of GSE in our sample. Based on its age, metallicity, and obvious radial motion, we suggest that Cluster 6 come from the progenitor galaxy of GSE. We designate stars with disc-like rotation and significant radial motion as ShangGu-3.

\begin{figure*}
   \centering
    \includegraphics[width=1.0\textwidth]{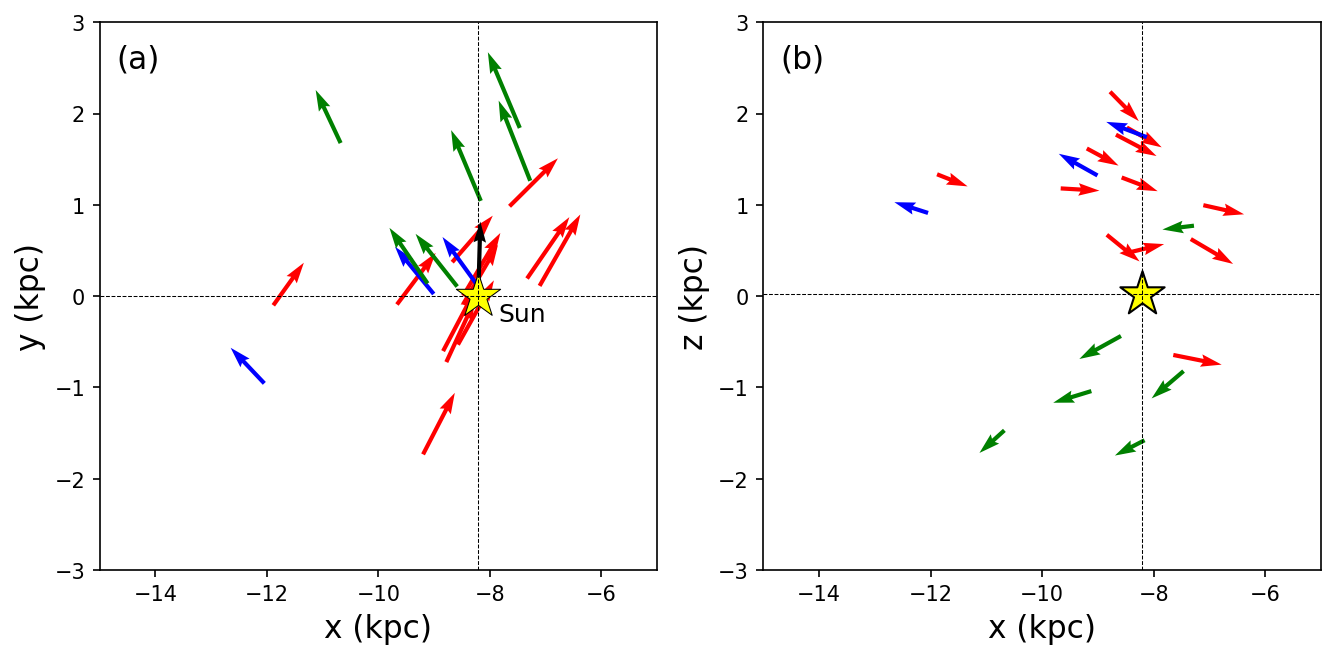}
    \caption{Velocities of ShangGu 3 stars in the Galactic x–y (a) and x–z (b) planes. The Galactic Centre is at (x, y, z) = (0, 0, 0) kpc while the Sun (indicated by the yellow star moving in the direction of the black arrow) is at (x, y, z) = (-8.21, 0, 0.02) kpc. These stars are divided into three subgroups, shown in red, blue, and green.}
    \label{fig21}
\end{figure*}

\begin{figure*}
   \centering
    \includegraphics[width=1.0\textwidth]{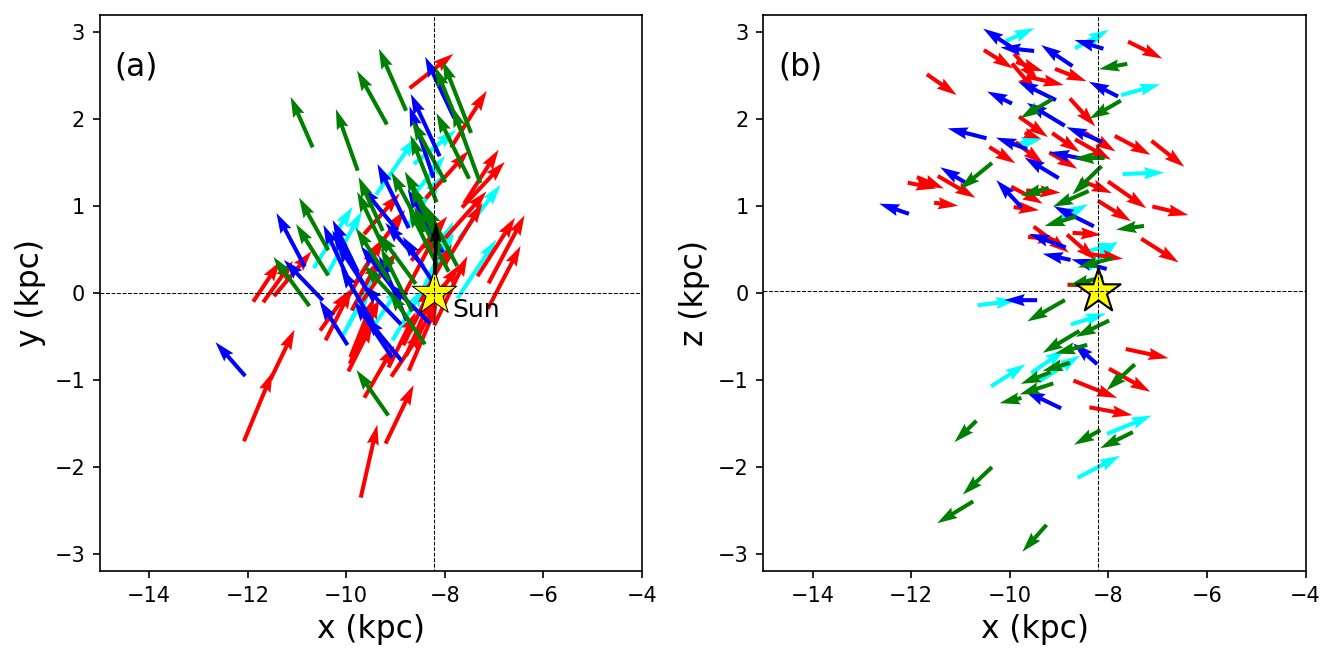}
    \caption{Similar to Fig. \ref{fig21}, but for ShangGu-3 stars selected using kinematic criteria in Section \ref{sec4.3}. Compared to Fig. \ref{fig21}, an additional subgroup with $V_{x}>0$ and $V_{z}>0$ is identified, marked by the cyan arrows.}
    \label{fig22}
\end{figure*}

ShangGu-3 shared kinematic similarities with the Nyx. \cite{necib_evidence_2020} identified Nyx with an average $V_{\phi}$ of -130 $\rm{km\,s^{-1}}$, an average $|V_r|$ of 134 $\rm{km\,s^{-1}}$, and a narrow [Fe/H] distribution averaging -0.55. Therefore, they suggested Nyx originated from a dwarf galaxy on a prograde orbit. \cite{horta_chemical_2023} later analyzed Nyx stars in APOGEE DR17, finding most have [Fe/H] $>$ -1.0, though some reach as low as -2.0. Their chemical similarity to the in-situ, metal-rich, high-$\alpha$ disc led \cite{horta_chemical_2023} to propose a local origin for Nyx. Fig. \ref{fig21} reveals ShangGu 3 comprises three distinct subgroups: one aligning with Nyx (positive $V_x$ and negative $V_z$), marked by red arrows; a second aligning with Nyx-2 (negative $V_x$ and $V_z$), marked by green arrows; and a third with negative $V_x$ and positive $V_z$, marked by blue arrows. Considering that the clustering algorithm might not have captured all ShangGu-3 members, we implemented additional kinematic criteria to directly select potential members:
\begin{enumerate}[label=(\roman*), leftmargin=0em, align=left]  
\setlength{\itemindent}{1em} 
    \item $-240 < V_{\phi} < -130\, \rm{km\, s^{-1}}$
    \item $150 < |V_{r}| < 240\, \rm{km\, s^{-1}}$
    \item $|V_{z}| < 100\, \rm{km\, s^{-1}}$
    \item $|z| < 3\, \rm{kpc}$
    \item $E > -1.41\, \times10^{5}\,\rm{km^{2}\,s^{-2}}$
\end{enumerate}  
This selection yielded 106 stars, whose motions in the x-y and y-z planes are shown in Fig. \ref{fig22}. Notably, this expanded sample reveals that ShangGu-3 actually consists of four distinct kinematic subgroups, rather than the three subgroups. Stars with positive $V_x$ an be further divided based on their $V_z$ values. This new finding suggests that Nyx and Nyx-2 are merely two subcomponents of a more complex structure, challenging the interpretation of Nyx as a coherent stellar stream from a merged dwarf galaxy \citep{necib_evidence_2020}.

\subsection{Future prospects}\label{sec4.4}
This study confirms the existence of the Galactic primordial disc, though its full scale remains uncertain due to our sample being limited to the solar vicinity. While we analyzed the impact of the GSE merger on the evolution of the Galactic disc, the effects of other merger events remain unclear. Many merger remnants have uncertain origins and lower masses than GSE, making their influence harder to detect. Our findings reveal that Galactic structures and merger remnants lack clear kinematic boundaries. For instance, primordial disc stars are found in both our kinematically cold and hot disc sample, while stars from the GSE's progenitor appear in both the halo and disc. Understanding the distribution of Galactic structures and merger remnants across physical, kinematic, and chemical spaces is crucial for uncovering the Milky Way’s assembly history. We plan to utilize deeper photometric surveys to address these issues in the future.

\section{Conclusions}\label{sec5}
LAMOST survey has collected tens of millions of low-resolution stellar spectra, from which we identified $\sim100,000$ metal-poor stars in our previous work. In this study, we estimated chemical abundances of these stars with two methods: machine learning and template fitting. The machine learning provided abundances for 17 elements, while template fitting estimated abundances of Fe, Mg, Ca, and C. By combining Gaia astrometry, we computed velocities, orbital parameters, and integrals of motion for over 70,000 stars, then selected about 46,000 with multiple-element abundances and precise kinematics to investigate the origin of the Galactic metal-poor disc and merger remnants. In addition, we supplemented age estimates for some subgiant stars in our sample.

We found that the evolution of the Galactic disc can be divided into four stages: 
\begin{enumerate}[label=(\roman*), leftmargin=0em, align=left] 
\setlength{\itemindent}{1em}
    \item  Primordial phase (over 12 Gyr ago): The Milky Way hosted a primordial disc with metallicity above -1.5.

    \item Merger phase (10-12 Gyr ago): The merger of the Gaia-Sausage-Enceladus (GSE) progenitor with the Milky Way significantly affected the Galactic disc in two major ways. First, stars originating from the GSE progenitor, many of which have metallicities [Fe/H] $<$ –1.5, were deposited into the primordial disc, contaminating it with an external stellar population. Second, the merger delivered a substantial amount of metal-poor gas to the Milky Way, which fueled a rapid episode of star formation. This event gave rise to a new disc component, with some stars reaching metallicities as low as [Fe/H] $\approx$ –2.0. We propose that this formation process is connected to the origin of the so-called metal-rich, high-$\alpha$ disc (commonly referred to as the thick disc), though our findings suggest it extends to lower metallicities than traditionally assumed.

    \item Quiescent phase (8-10 Gyr ago): After GSE merger, the Milky Way gradually accreted gas from its surroundings, leading to slow star formation.
    
    \item Latest formation phase (within the last 8 Gyr): This phase corresponds to the formation of the so-called metal-rich, low-$\alpha$ disc (commonly referred to as the thin disc). Contrary to the assumption that the low-$\alpha$ disc formed with [Fe/H] $>$ -0.7, we propose it was initially metal-poor, with metallicity as low as -1.8. This conclusion is supported by the identification of a population of young, metal-poor, high-$\alpha$ stars in our study. These stars exhibit kinematic properties highly consistent with those of the metal-rich, low-$\alpha$ disc, suggesting a shared evolutionary origin. Although current observational data do not indicate that these stars are part of unresolved binary systems, their apparently young ages remain uncertain and may be subject to measurement biases. Furthermore, relative to other metal-poor disc stars, they show enhanced $\alpha$-element abundances at a given [Fe/H], implying they have experienced $\alpha$-enrichment process. However, given their small numbers, it is unlikely that they are associated with episodes of intense star formation, and the underlying mechanism of their $\alpha$-enrichment remains unclear. We strongly look forward to further high-resolution spectroscopic observations to better constrain the nature of these stars.
\end{enumerate}

In summary, the Galactic disc formed through three major phases: the primordial phase, the GSE merger phase, and the post-merger phase. These processes resulted in four main metal-poor disc populations: stars from the Galactic primordial disc, stars from the GSE progenitor, and the metal-poor extensions of both the so-called metal-rich, high-$\alpha$ and low-$\alpha$ discs. Other populations, such as stars formed during the quiescent phase, are present but constitute a smaller fraction.

Regarding the merger relics, Wukong's much lower metallicity than GSE suggests they stemmed from separate merger events.

Additionally, we identified three new substructures: ShangGu-1, ShangGu-2, and ShangGu-3. ShangGu-1 exhibits mild rotation ($V_{\phi}$ $\approx$ -100 $\rm{km\,s^{-1}}$), strong vertical motion ($|V_z|$ $\gtrsim$ 100 $\rm{km\,s^{-1}}$), and a positive correlation between orbital eccentricity and metallicity. ShangGu-2 shows thin disc-like rotation ($V_{\phi}$ $\lesssim$ -200 $\rm{km\,s^{-1}}$) combined with high vertical velocity ($|V_z|$ $\gtrsim$ 100 $\rm{km\,s^{-1}}$). Its stellar population displays two distinct sequences in the [Mg/Fe]–[Fe/H] diagram, similar to those observed in our kinematically cold disc sample. These features suggest that ShangGu-2 is composed of metal-poor disc stars that have been dynamically heated. ShangGu-3 has prograde orbit (-240 $\lesssim$ $V_{\phi}$ $\lesssim$ -130 $\rm{km\,s^{-1}}$) with significant radial movements ($|V_{r}|$ $\approx$ 200 $\rm{km\,s^{-1}}$), and a mean stellar age of 13 Gyr. It can be divided into four subgroups based on orbital directions, and we propose it originated from the progenitor of GSE.

In conclusion, this work provides a detailed analysis of the Galactic metal-poor disc, proposing a metallicity floor of -1.5 for the primordial disc, and lowering the metallicity floors for high-$\alpha$ and low-$\alpha$ discs to -2.0 and -1.8, respectively. These findings are both novel and radical, and we look forward to future high-resolution observations to verify our views. Additionally, we argue that the merger between the progenitor of GSE and the Milky Way had a profound impact on the Milky Way, not only contributing to the Galaxy's halo but also significantly influencing the evolution of the Galactic disc. The value-added catalogues of chemical abundances and kinematic parameters for metal-poor stars used in this study are available at   \href{https://github.com/tubage886/VMP-StarsII}{https://github.com/tubage886/VMP-StarsII}.

\normalem
\begin{acknowledgements}
 This study is supported by the National Natural Science Foundation of China under grant Nos. 11988101, 12222305, and National Key R$\&$D Program of China No.2024YFA1611900. Guoshoujing Telescope (the Large Sky Area Multi-Object Fiber Spectroscopic Telescope LAMOST) is a National Major Scientific Project built by the Chinese Academy of Sciences. LAMOST is operated and managed by the National Astronomical Observatories, Chinese Academy of Sciences. This work has made use of data from the APOGEE (SDSS-IV) surveys. Funding for the Sloan Digital Sky Survey IV has been provided by the Alfred P. Sloan Foundation, the U.S. Department of Energy, Office of Science, and the participating institutions. This work has also made use of data from the European Space Agency (ESA) mission Gaia, processed by the Gaia Data Processing and Analysis Consortium (DPAC). Funding for the DPAC has been provided by national institutions, in particular the institutions participating in the Gaia Multilateral Agreement.
\end{acknowledgements}

\appendix                  

\section{Supplementary materials on chemical abundances}\label{ap1}
\begin{figure*}
    \centering
    \includegraphics[width=1.0\textwidth]{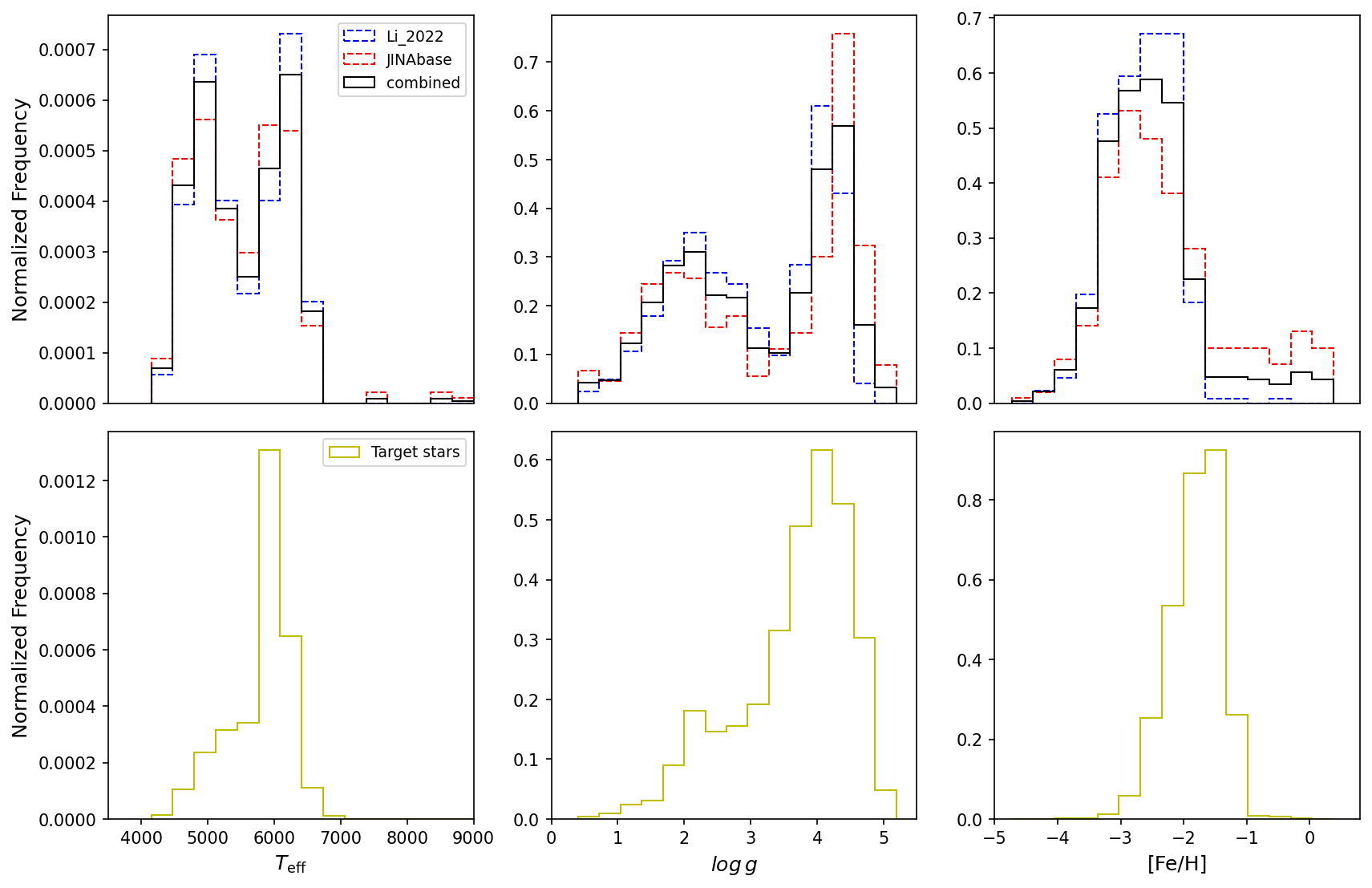}
    \caption{Normalized histograms of effective temperature ($T_{\mathrm{eff}}$), surface gravity ($\log g$), and metallicity ([Fe/H]) for the reference set and target stars. The top panels show the distributions for the stars from \cite{li_four-hundred_2022} (blue dashed lines), the JINAbase–LAMOST overlapping sample (red dashed lines), and the combined reference set (solid black lines). The bottom panels show the corresponding distributions for the target stars (yellow lines). The target stars are metal-poor candidates identified by \citep{hou_very_2024} using LAMOST spectra, and the stellar parameters used to construct this figure are taken from \citep{hou_very_2024}.}
    \label{figA1}
\end{figure*}

\begin{figure*}
    \centering
    \includegraphics[width=1.0\textwidth]{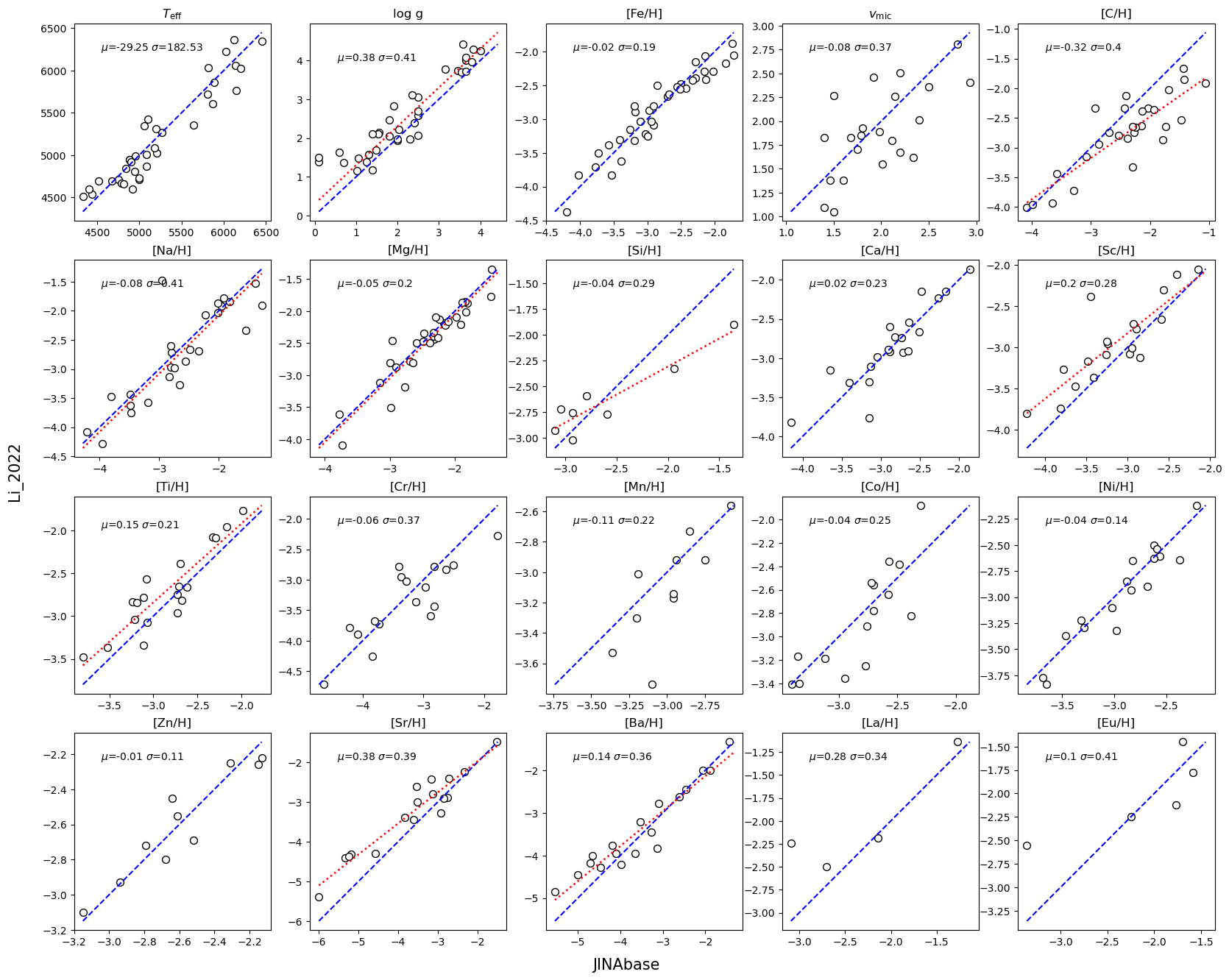}
    \caption{Comparison of atmospheric parameters and chemical abundances for stars common to \cite{li_four-hundred_2022} and JINAbase \citep{abohalima_jinabasedatabase_2018}. The dashed line in each panel represents the 1:1 relationship, with the mean and standard deviation of differences between the two sources shown. In the panels for $\rm{log}\,\textsl{g}$, [C/H], [Na/H], [Mg/H], [Si/H], [Sc/H], [Ti/H], [Sr/H], and [Ba/H], a red dotted line shows the linear fit between values from JINAbase and \cite{li_four-hundred_2022}.}
    \label{figA2}
\end{figure*}

\begin{figure*}
    \centering
    \includegraphics[width=1.0\textwidth]{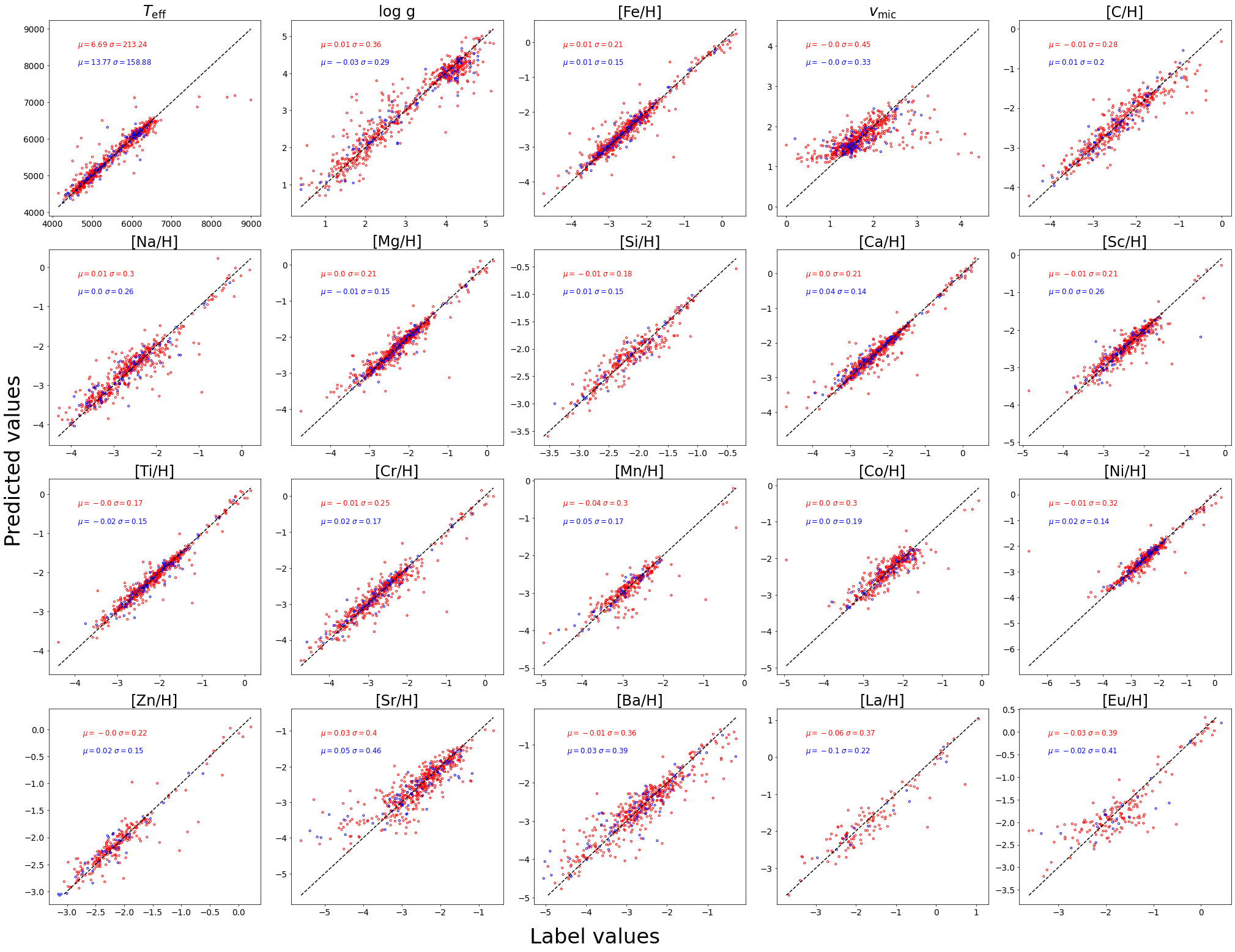}
    \caption{Comparison of predicted values and labels for training (red dots) and test stars (blue dots). The dashed line in each panel represents the 1:1 relationship, with the mean and standard deviation of differences shown in red for the training set and blue for the test set.}
    \label{figA3}
\end{figure*}

\begin{figure*}
    \centering
    \includegraphics[width=1.0\textwidth]{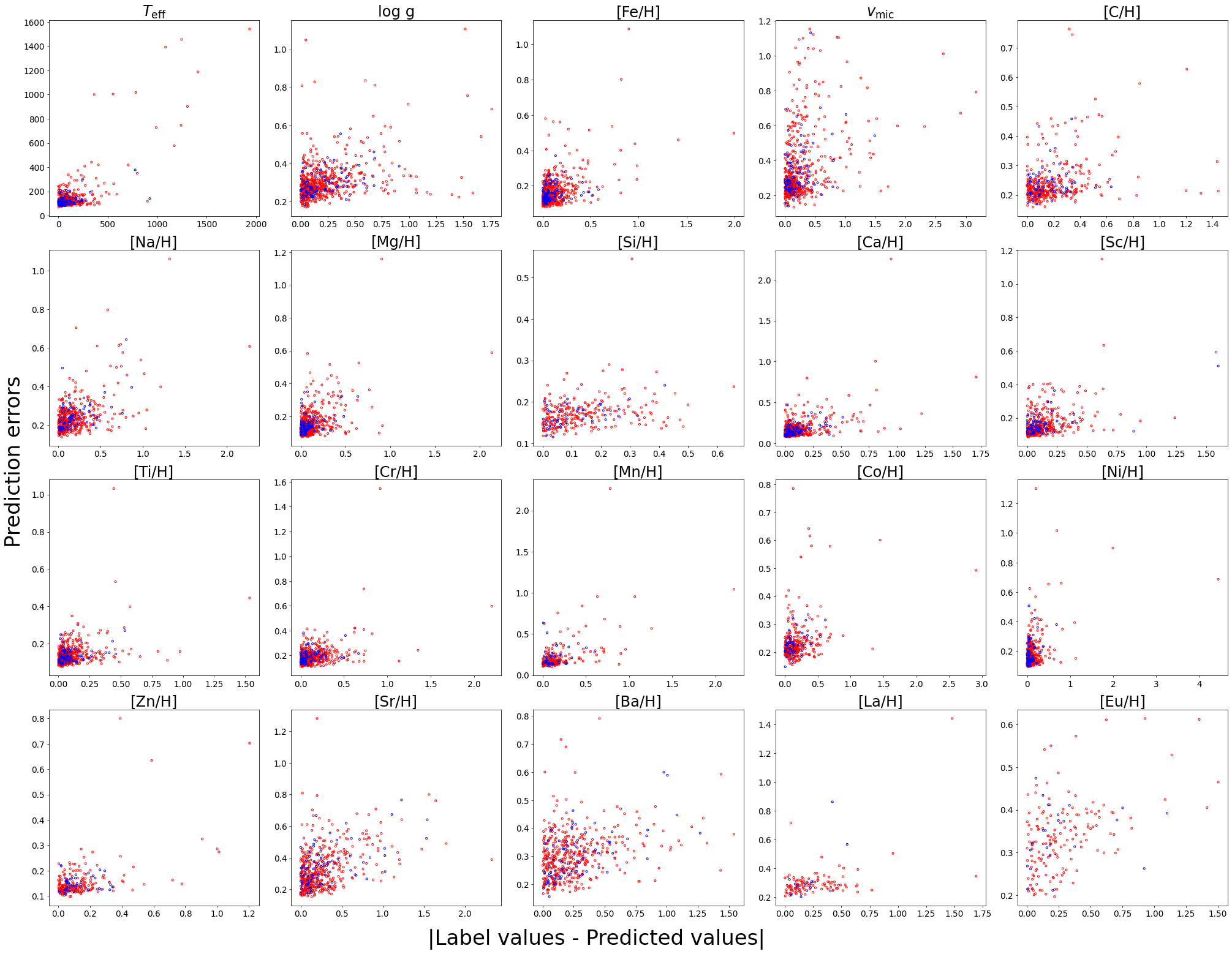}
    \caption{Correlation between prediction errors and the absolute difference between predicted and label values. Red and blue dots represent training and test stars, respectively.}
    \label{figA4}
\end{figure*}

\begin{figure*}
    \centering
    \includegraphics[width=1.0\textwidth]{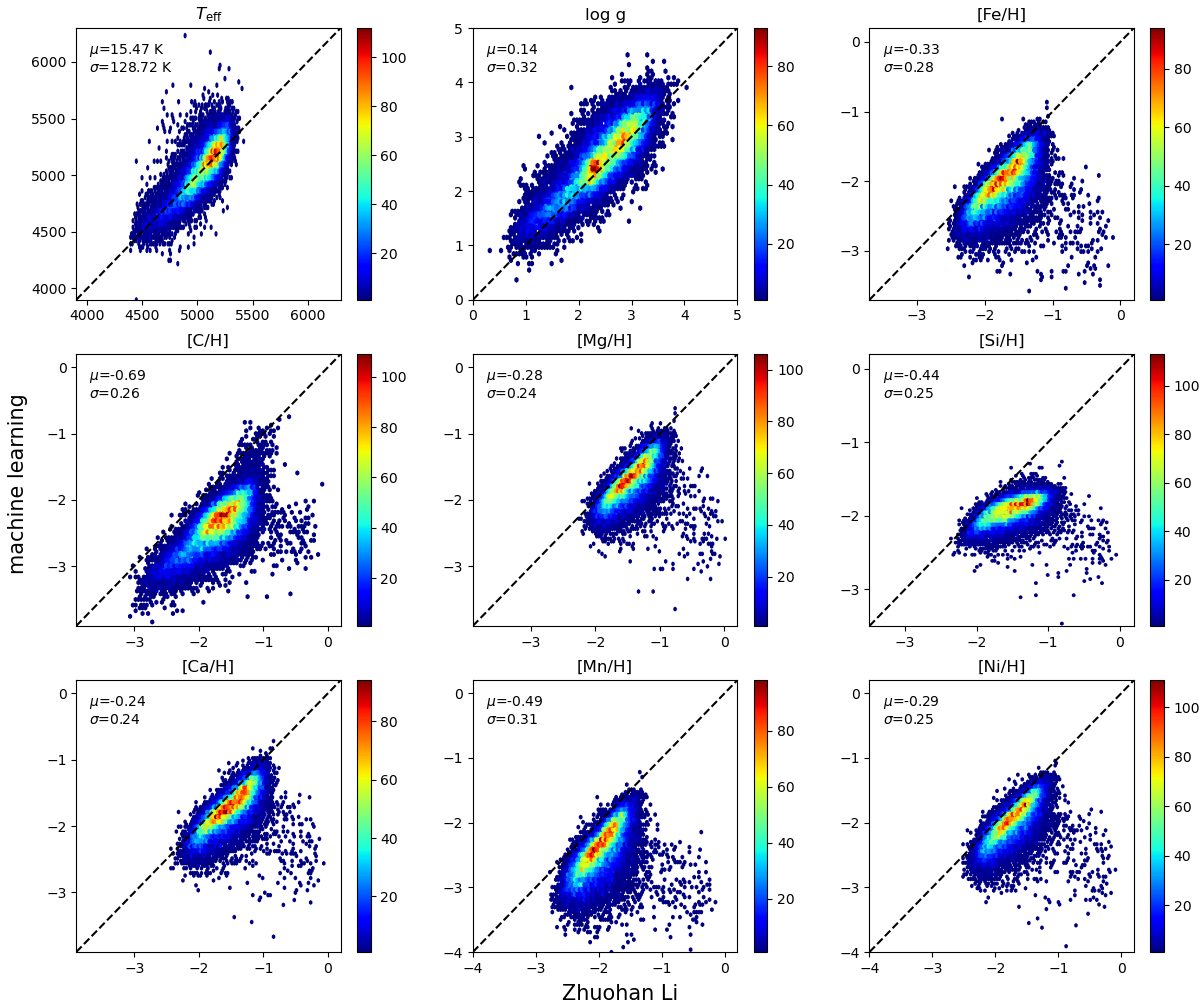}
    \caption{Comparison of $T_\mathrm{eff}$, $\rm{log}\,\textsl{g}$, [Fe/H], and the abundances of C, Mg, Si, Ca, Mn, and Ni obtained in this work using machine learning and those from \cite{li_stellar_2022} for $\sim20,000$ common stars. These elements were chosen because both studies include them. Dashed lines mark the 1:1 correspondence, and the color scale represents the stellar density distribution.
}
    \label{figA5}
\end{figure*}

\begin{figure}
    \centering
    \includegraphics[width=1.0\textwidth]{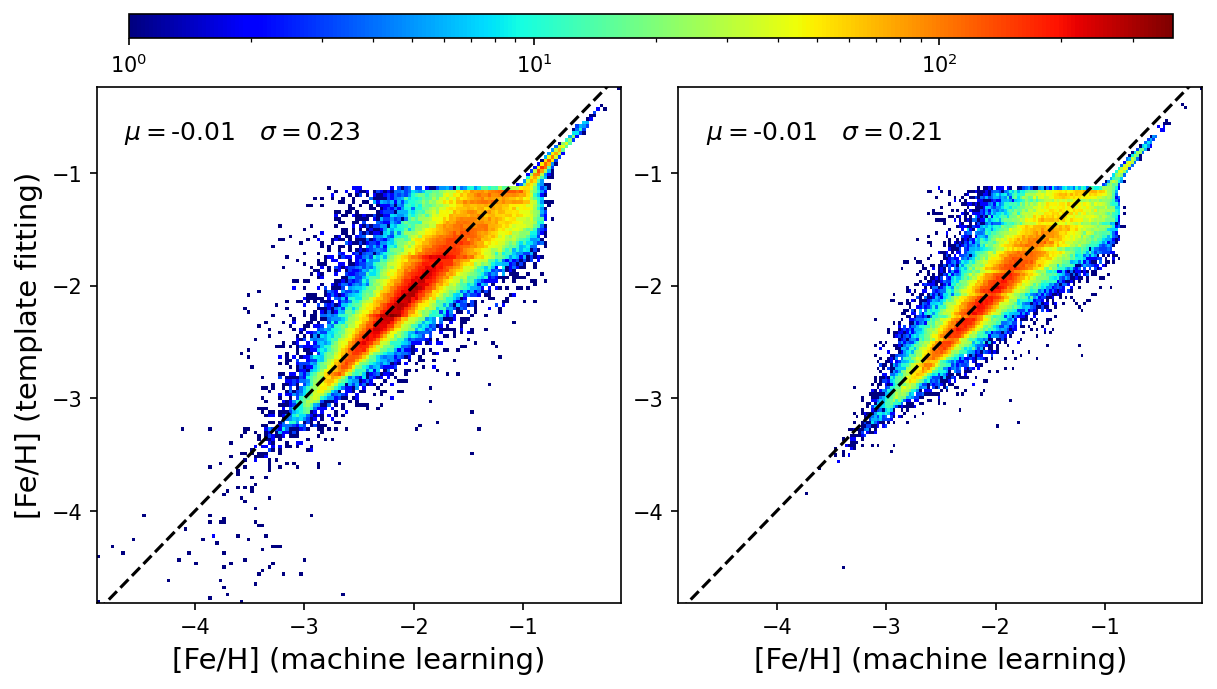}
    \caption{Comparison of iron abundances from template fitting and machine learning. The left panel shows the stars with $\rm{flag}=0$, while the right panel shows the stars after error filtering (criteria in Section \ref{sec2.1.2}). Stellar number density is indicated by color, with a colorbar at the top.}
    \label{figA6}
\end{figure}

\begin{figure}
    \centering
    \includegraphics[width=1.0\textwidth]{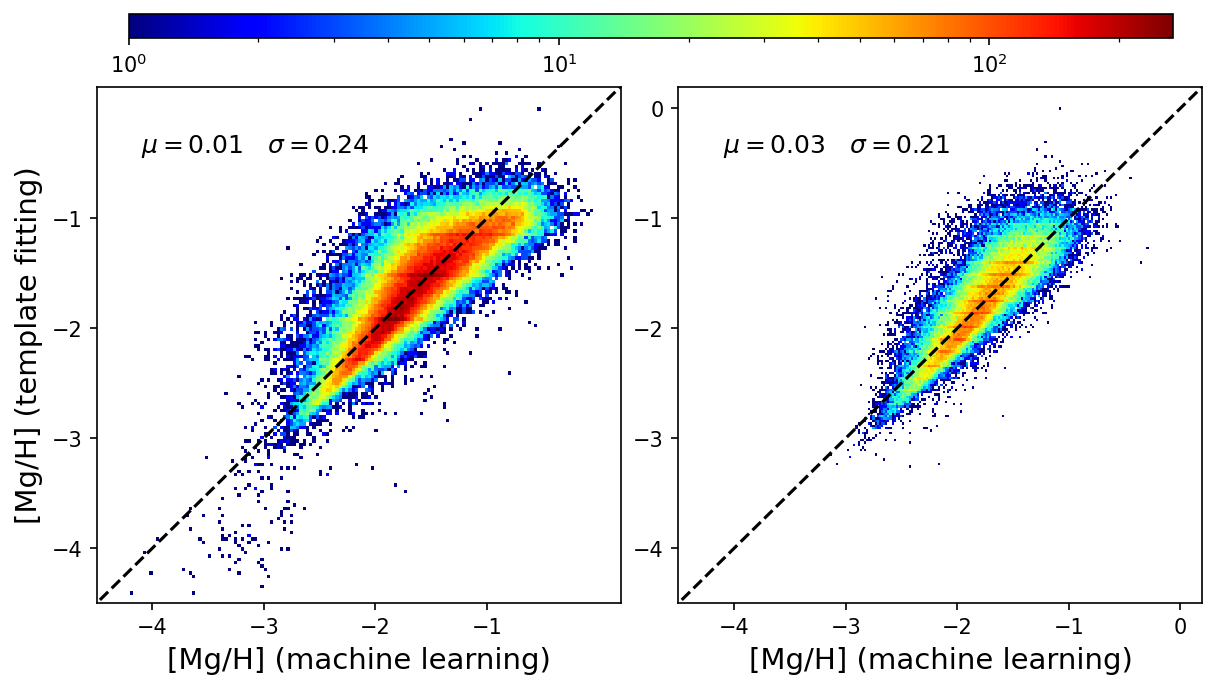}
    \caption{Same as Fig. \ref{figA6}, but for magnesium.}
    \label{figA7}
\end{figure}

\begin{figure}
    \centering
    \includegraphics[width=1.0\textwidth]{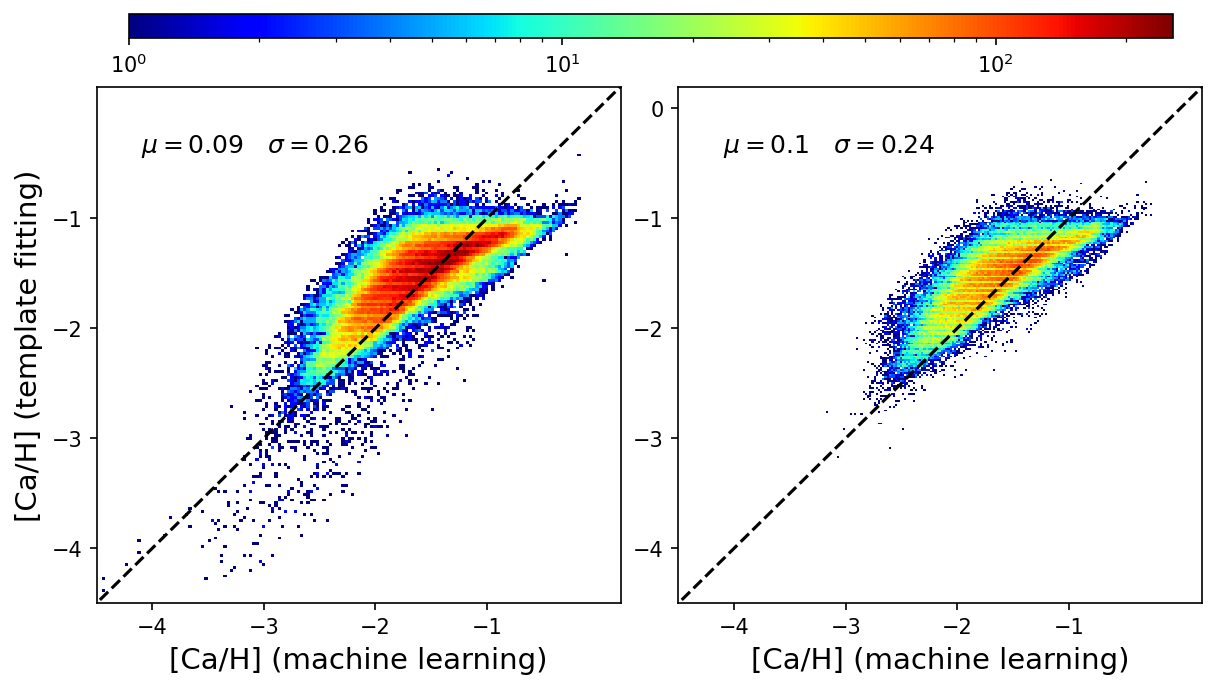}
    \caption{Same as Fig. \ref{figA6}, but for calcium.}
    \label{figA8}
\end{figure}

\begin{figure}
    \centering
    \includegraphics[width=1.0\textwidth]{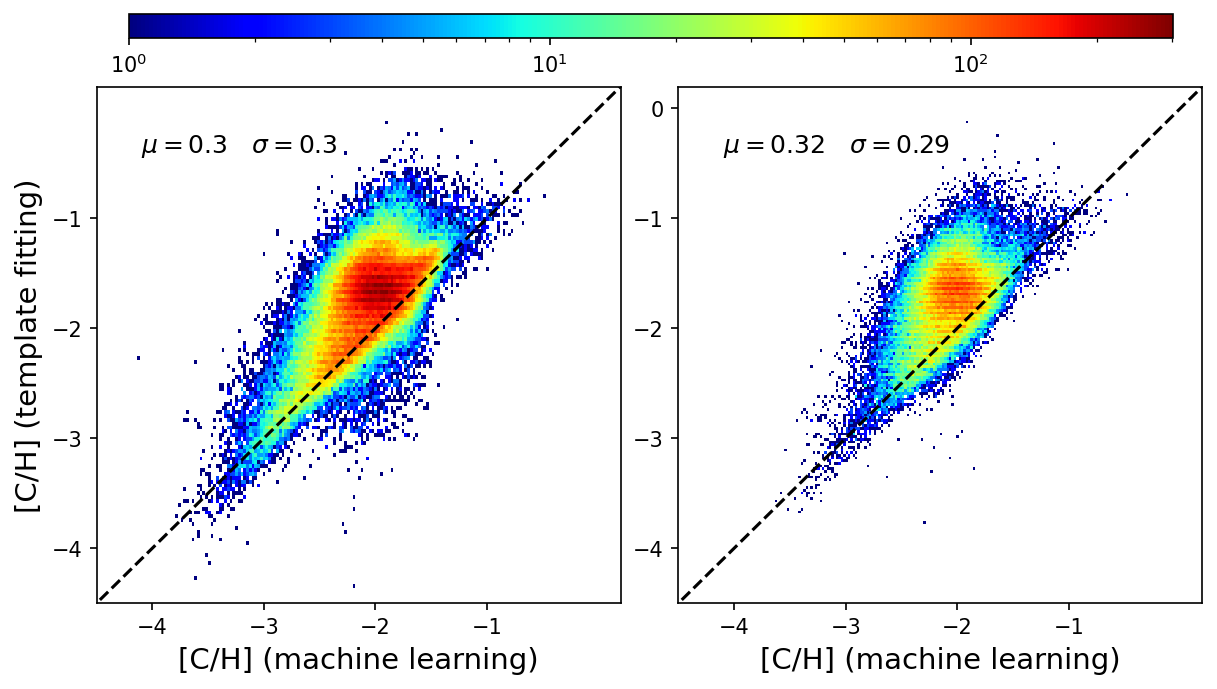}
    \caption{Same as Fig. \ref{figA6}, but for carbon.}
    \label{figA9}
\end{figure}

\begin{figure*}
    \centering
    \includegraphics[width=1.0\textwidth]{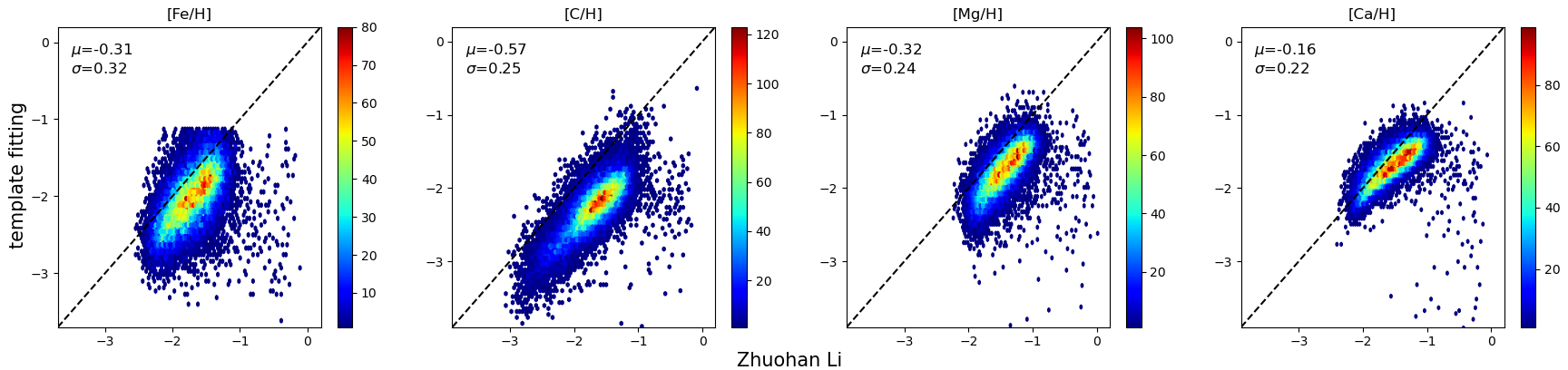}
    \caption{Similar to Fig. \ref{figA5}, but comparing [Fe/H] and the abundances of C, Mg, and Ca from this work via template fitting with those from \cite{li_stellar_2022}.   
}
    \label{figA10}
\end{figure*}

This appendix provides details on the estimation of chemical abundances. For the machine learning reference set, we used stars from \cite{li_four-hundred_2022} and JINAbase \cite{abohalima_jinabasedatabase_2018}, Fig. \ref{figA1} presents the distribution of stellar parameters for the reference set and target stars. Noting systematic biases between the two sources. Fig. \ref{figA2} compares atmospheric parameters and chemical abundances for stars common to both sources, and Table \ref{tabA1} lists parameters for revising abundances from JINAbase \cite{abohalima_jinabasedatabase_2018}, with details in Section \ref{sec2.1.1}. Table \ref{tabA2} specifies the wavelength ranges used to estimate Fe, Mg, Ca, and C abundances via template fitting. Fig. \ref{figA3} shows machine learning accuracy on training and test sets, while Fig. \ref{figA4} illustrates the relationship between errors and prediction accuracy. Fig. \ref{figA6}, \ref{figA7}, \ref{figA8}, and \ref{figA9} compare Fe, Mg, Ca, and C abundances measured by machine learning and template fitting, with subplots showing results with and without error filtering. We estimated the atmospheric parameters and chemical abundances of $\sim100,000$ metal-poor stars using machine learning. For Fe, C, Mg, and Ca, we also conducted measurements through template fitting. To evaluate the accuracy of our results, we compared our values with those from \cite{li_stellar_2022}. Fig. \ref{figA5} presents the comparison between our machine learning results and those from \cite{li_stellar_2022}, while Fig. \ref{figA10} shows the comparison of template fitting results with values from \cite{li_stellar_2022}.

\begin{table}
    \centering
    \caption{Parameters $a$ and $b$ of the linear functions representing the relationships between values of $\rm{log}\,\textsl{g}$ and [X/H] (X = C, Na, Mg, Si, Sc, Ti, Sr, Ba) from JINAbase \citep{abohalima_jinabasedatabase_2018} and \cite{li_four-hundred_2022}. See Section \ref{sec2.1.1} for details.}
    \setlength{\tabcolsep}{2.1mm}{
    \begin{tabular}{lcc}
    \toprule
        &a &b \\
    \midrule
        $\rm{log}\,\textsl{g}$ & 1.0 & 0.3 \\
        $[$C/H$]$ & 0.697 & -1.058 \\
        $[$Na/H$]$ & 1.0 & -0.08 \\
        $[$Mg/H$]$ & 1.0 & -0.05 \\
        $[$Si/H$]$ & 0.545 & -1.218 \\
        $[$Sc/H$]$ & 0.794 & -0.447 \\
        $[$Ti/H$]$ & 0.921 & -0.077 \\
        $[$Sr/H$]$ & 0.781  & -0.417 \\
        $[$Ba/H$]$ & 0.823 & -0.488 \\
    \bottomrule
    \end{tabular}}
    \label{tabA1}
\end{table}

\begin{table}
    \centering
    \caption{Wavelength ranges used to estimate Fe, Mg, Ca, and C abundances via template fitting.}
    \setlength{\tabcolsep}{2.1mm}{
    \begin{tabular}{lc}
    \toprule
        Element & Wavelength range \\
    \midrule
        Fe&  4953 - 4962 Å \\
        Mg&  5179 - 5188 Å\\
        Ca&  3915 - 3955 Å\\
        C&  4283 - 4321 Å\\
    \bottomrule
    \end{tabular}}
    \label{tabA2}
\end{table}

\bibliographystyle{raa}
\bibliography{bibtex}

\end{CJK*}
\end{document}